%% file: main.tex
\PassOptionsToPackage{dvipsnames}{xcolor}

\documentclass[runningheads]{llncs}

\usepackage{color}
\usepackage{amsmath}
\usepackage{graphicx}
\usepackage{stmaryrd}
\usepackage[all]{xy}
\usepackage{multirow}
\usepackage{paralist}
\usepackage{hhline}
\usepackage{bm}
\usepackage{longtable}
\usepackage{makecell}
\usepackage{mdframed}

\usepackage{wasysym}
\usepackage{extarrows}
\usepackage{tikz}
\usetikzlibrary{tikzmark}
\usepackage{annotate-equations}

\usepackage{amssymb}
\usepackage{amsmath}
\usepackage{stmaryrd}
\usepackage{hyperref}

\usepackage{braket}
\usepackage{cleveref}

\usepackage{algorithm}
\usepackage{algpseudocode}

\usepackage{tikz}
\usetikzlibrary{quantikz2}

\input{lstlisting}

\usepackage{tikz}
\usetikzlibrary{shapes.geometric, arrows}
\usetikzlibrary{trees}
\tikzstyle{arrow} = [thick,->,>=stealth]

\usepackage{quantikz}

\newenvironment{ruletable}[1]
{
    \begin{longtable}{p{3cm} l}
    \caption{#1}\\
    \hline
    \textbf{Rule} & \textbf{Description} \\
    \hline
    \endfirsthead

    \hline
    \textbf{Rule} & \textbf{Description} \\
    \hline
    \endhead

    \hline
    \endfoot

    \hline
    \endlastfoot
}
{
    \end{longtable}
}

\newcommand{\CC}{C\nolinebreak\hspace{-.05em}\raisebox{.4ex}{\tiny\bf +}\nolinebreak\hspace{-.10em}\raisebox{.4ex}{\tiny\bf +}}
\def\CC{{C\nolinebreak[4]\hspace{-.05em}\raisebox{.4ex}{\tiny\bf ++}}}

\def\>{\ensuremath{\rangle}}
\def\<{\ensuremath{\langle}}

\newcommand {\cH } {{\mathcal{H}}}
\newcommand {\bu } {{\mathbf{u}}}
\newcommand {\bv } {{\mathbf{v}}}
\newcommand {\bC } {{\mathbb{C}}}
\newcommand {\ghz } {{\mathrm{GHZ}}}
\newcommand {\cnot } {{\mathsf{CNOT}}}
\newcommand {\ccnot } {{\mathsf{CCNOT}}}
\newcommand {\Swap } {{\mathsf{SWAP}}}
\renewcommand{\matrix}[1]{\begin{bmatrix}#1\end{bmatrix}}

\usepackage[misc]{ifsym}
\usepackage[T1]{fontenc}
\usepackage{graphicx}
\usepackage{color}

\begin{document}
\title{D-Hammer: Efficient Equational Reasoning for Labelled Dirac Notation}
\author{}
\institute{}

\author{Yingte Xu\inst{1}\orcidID{0000-0001-9071-7862} \and
Li Zhou\inst{2}\orcidID{0000-0002-9868-8477} \and
\\
Gilles Barthe\inst{1,3}\orcidID{0000-0002-3853-1777} \Letter}
\authorrunning{Yingte Xu, Li Zhou, and Gilles Barthe}
\institute{MPI-SP, Germany
\email{\{yingte.xu, gilles.barthe\}@mpi-sp.org}
\and
Key Laboratory of System Software (Chinese Academy of Sciences) and State Key Laboratory of Computer Science, Institute of Software, Chinese Academy of Sciences, China.
\email{zhouli@ios.ac.cn}
\and
IMDEA Software Institute, Spain}
%%%%%%%%%%%%%%%%%%%%%%%%%%%%%%%%%%%%%%%%%%%%%%%%%%%%%%%%%%%%%%%%%%%%%%%%

%
\maketitle              % typeset the header of the contribution
\begin{abstract}
    Labelled Dirac notation is a formalism commonly used by physicists to represent many-body quantum systems and by computer scientists to assert properties of quantum programs. It is supported by a rich equational theory for proving equality between expressions in the language. These proofs are typically carried on pen-and-paper, and can be exceedingly long and error-prone. We introduce D-Hammer, the first tool to support automated equational proof for labelled Dirac notation. The salient features of D-Hammer include: an expressive, higher-order, dependently-typed language for labelled Dirac notation; an efficient normalization algorithm; and an optimized \CC\ implementation. We evaluate the implementation on representative examples from both plain and labelled Dirac notation. In the case of plain Dirac notation, we show that our implementation significantly outperforms DiracDec (Xu et al., POPL'25).

% \keywords{First keyword  \and Second keyword \and Another keyword.}
\end{abstract}
%
%
%

%%%%%%%%%%%%%%%%
\newcommand*{\sem}[1]{{\llbracket #1 \rrbracket}}
\newcommand{\DiracDec}{\textsf{DiracDec}}

\newcommand{\reduce}{\triangleright}

\newcommand{\Sort}{\mathsf{Sort}}
\newcommand{\WF}{\mathcal{WF}}

\newcommand{\Index}{\mathsf{Index}}
\newcommand{\Type}{\mathsf{Type}}
\newcommand{\Basis}{\mathsf{Basis}}

\newcommand{\SType}{\mathcal{S}}
\newcommand{\KType}{\mathcal{K}}
\newcommand{\BType}{\mathcal{B}}
\newcommand{\OType}{\mathcal{O}}
\newcommand{\SET}{\mathsf{Set}}

\newcommand{\ZEROK}{\mathbf{0}_\mathcal{K}}
\newcommand{\ZEROB}{\mathbf{0}_\mathcal{B}}
\newcommand{\ZEROO}{\mathbf{0}_\mathcal{O}}

\newcommand{\PAIR}{\mathsf{PAIR}}

\newcommand{\ZERO}{\mathsf{0}}
\newcommand{\ONE}{\mathsf{1}}
\newcommand{\ADDS}{\mathsf{ADDS}}
\newcommand{\ADD}{\mathsf{ADD}}
\newcommand{\MULS}{\mathsf{MULS}}
\newcommand{\MUL}{\mathsf{MUL}}
\newcommand{\CONJ}{\mathsf{CONJ}}
\newcommand{\CJG}{\mathsf{CJG}}
\newcommand{\ADJ}{\mathsf{ADJ}}
\newcommand{\DELTA}{\mathsf{DELTA}}
\newcommand{\DOT}{\mathsf{DOT}}
\newcommand{\SCR}{\mathsf{SCR}}
\newcommand{\TSR}{\mathsf{TSR}}
\newcommand{\KET}{\mathsf{KET}}
\newcommand{\BRA}{\mathsf{BRA}}
\newcommand{\ONEO}{\mathbf{1}_\mathcal{O}}
\newcommand{\OUTER}{\mathsf{OUTER}}
\newcommand{\MULK}{\mathsf{MULK}}
\newcommand{\MULB}{\mathsf{MULB}}
\newcommand{\MULO}{\mathsf{MULO}}

\newcommand{\var}{\mathsf{var}}
\newcommand{\reg}{\mathsf{Reg}}
\newcommand{\DType}{\mathcal{D}}
\newcommand{\cR}{\mathcal{R}}
\newcommand{\cN}{\mathcal{N}}
\newcommand{\tD}{\tilde{D}}
\newcommand{\te}{\tilde{e}}
\newcommand{\tT}{\tilde{T}}
\newcommand{\tADD}{\widetilde{ADD}}
\newcommand{\bU}{\mathbf{U}}
\renewcommand{\<}{\langle}
\newcommand{\simp}{\mathsf{Simp}}
\newcommand{\List}{\mathsf{list}}
\renewcommand{\>}{\rangle}
\newcommand {\cD } {{\mathcal{D}}}
\newcommand {\cl } {{\mathit{cl}}}

\newenvironment{hproof}{%
  \renewcommand{\proofname}{Proof Sketch}\proof}{\endproof}

%%%%%%%%%%%%%%%%%%%%%%%%%%%

\input{01intro_preliminary.tex}

\input{02language_design.tex}

\input{03labelled.tex}
\input{05tool_eval.tex}

\input{06related_conclusion.tex}

%%%%%%%%%%%%%%%%%%%%%%%%%%

% \begin{credits}
%     \subsubsection{\ackname} A bold run-in heading in small font size at the end of the paper is
%     used for general acknowledgments, for example: This study was funded
%     by X (grant number Y).
% \end{credits}

\bibliographystyle{splncs04}
\bibliography{ref}

\appendix
\include{appendix}

\end{document}

%% file: lstlisting.tex
\usepackage{listings}
\usepackage[dvipsnames]{xcolor}

\lstdefinestyle{dhammer}{
    columns=fullflexible,
    mathescape = true,
    sensitive=true,
    basicstyle=\ttfamily\small, % Set sans-serif font as basic font
    breaklines=true,
    backgroundcolor=\color{lightgray!50!white}, % Light gray background
    morekeywords=[1]{Def, Var, CheckEq, with},
    morekeywords=[2]{INDEX, OTYPE, REG, USET}, 
    keywordstyle=[1]\bfseries\color{NavyBlue},
    keywordstyle=[2]\bfseries\color{black}, 
    morecomment=[l]{//},
    commentstyle=\itshape\color{black!50!white},
    string=[b]"
}

%% file: 01intro_preliminary.tex
\section{Introduction}

Dirac notation~\cite{dirac1939new}, also known as bra-ket notation, is a mathematical formalism for representing quantum states using linear algebra notation. For example, Dirac notation uses the linear combination \( a\ket{\psi} + b\ket{\phi} \) to represent the superposition of the quantum states \( \ket{\psi} \) and \( \ket{\phi} \). Another essential ingredient of Dirac notation is the tensor product $\otimes$, which is used to describe composite states. For instance, the tensor expression \( \ket{\psi} \otimes \ket{\phi} \) denotes the composition of the two quantum states \( \ket{\psi} \) and \( \ket{\phi} \). A variant of Dirac notation, called labelled Dirac notation, is often used to describe composite quantum states. In labelled Dirac notation, bras and kets are tagged with labels to identify the subsystems they operate on. For example, the labelled tensor \( \ket{\psi}_{S'} \otimes \ket{\phi}_{S} \) indicates that \( \ket{\psi}_{S} \) and \( \ket{\phi}_{S'} \) describe two quantum states over subsystem $S$ and $S'$, respectively. By considering the relationship between $S$ and $S'$, one can obtain identities for free, e.g.\, 
$$\ket{\phi}_S \otimes \ket{\psi}_{T} = \ket{\psi}_{T} \otimes \ket{\phi}_S \quad \text{if}~S \cap S' = \emptyset$$
In turn, commutativity of the tensor product ensures that one can
reason locally about quantum systems, and contributes to making labelled Dirac notation a convenient, compositional formalism for reasoning about quantum states, akin to how bunched logics support compositional reasoning about mutable states.

Labelled Dirac notation is also widely used to express assertions in quantum programs. Specifically, many quantum Hoare logics rely on labelled Dirac notation and its variants to express program assertions (see, for example, \cite{DBLP:conf/lics/ZhouBHYY21} \cite{Zhou2023} \cite{qRHL_unruh2019} \cite{QSL_Le_2022} \cite{Zhong2024-np} \cite{incorrectness_2022} \cite{qafny2024}). These logics also employ implicit equational reasoning between labelled Dirac expressions to glue applications of proof rule--similar to the rule of consequence in the setting of classical program verification. Consequently, it is
essential for the verification of quantum programs to have automated
means of proving equality of two complex expressions based on labelled
Dirac notation.

\subsubsection*{Contributions}
This paper presents D-Hammer, an automated tool for reasoning about
labelled Dirac notation. D-Hammer uses a rich, dependently typed
language to formalize labelled Dirac notation and supports common
idioms for describing quantum systems, including big operators of the
form \( \sum_{i \in I} a_i \ket{\phi_i}_S \) to represent indexed
superpositions of states. The semantics of typable expressions are
given in terms of Hilbert spaces, tailored to interpret the tensor
product as an AC symbol. We leverage this interpretation to define a
rich equational theory for labelled Dirac notation and prove its
soundness with respect to our denotational semantics. Finally, we
define an efficient normalization procedure to prove equivalence
between two expressions. We then evaluate our procedure with respect
to examples from the literature. Our evaluation covers examples in
plain and labelled Dirac notation. The main conclusions are that our
approach outperforms DiracDec~\cite{diracdec} to reason about plain
Dirac notation and is able of proving complex examples from the
literature on labelled Dirac notation, including examples from prior
work on quantum separation logic~\cite{DBLP:conf/lics/ZhouBHYY21}.
For completeness, we also evaluate D-Hammer on examples from the
literature on equivalence checking of (parametrized) quantum
circuits.

\section{Motivation and Preliminaries}
\subsection{Plain Dirac Notation}

Dirac notation, also known as bra-ket notation, provides an intuitive
and concise mathematical framework for describing quantum states and
operations in quantum mechanics.  We write $\cH$ for a Hilbert space,
i.e., a vector space equipped with an standard inner product
$\langle\bu,\bv\rangle \in \bC$ for $\bu,\bv\in\cH$.  Dirac notation
consists of the following components that reflect the basic postulates
of quantum mechanics:
\begin{itemize}
  \item Ket $|u\rangle$ is a column vector that denotes a quantum state $\bu$ in the state Hilbert space $\cH$. For example, the computational bases of qubit system are commonly written as $|0\rangle = \matrix{1 \\ 0}$ and $|1\rangle = \matrix{0 \\ 1}$.
  \item Bra $\langle u|$ is a row vector, the conjugate transpose of $|u\rangle$, that denotes the dual state of $|u\rangle$. 
  %It is alternative to interpret as a linear mapping from $\cH$ to $\bC$ defined as $\langle u| : |v\rangle \mapsto \langle \bu,\bv\rangle$. It is the conjugate transpose of $|u\rangle$.
  For example, $\langle 0| = \matrix{1 & 0}$ and $\langle 1| = \matrix{0 & 1}$.
  \item Inner product $\langle u|v\rangle \triangleq \langle \bu,\bv\rangle$ which indicates the probability amplitude for $|u\rangle$ to collapse into $|v\rangle$. 
  By convention, it is computed by matrix multiplication of two states, e.g., 
  $\langle 0|1\rangle = \matrix{1 & 0} \matrix{0 \\ 1} = 0.$
  \item Outer product $|u\rangle\langle v| \triangleq |w\rangle \mapsto (\langle v|w\rangle) |u\rangle$. Any linear map, such as unitary tranformation, measurement operator, etc, can be decomposed as the sum of outer products. It is also computed by matrix multiplication, e.g., 
  $|1\rangle\langle 0| = \matrix{0 \\ 1} \matrix{1 & 0}  = \matrix{0 & 0 \\ 1 & 0}.$
  \item Tensor product $|u\rangle\otimes|v\rangle$ (or simply $|u\rangle|v\rangle$ or $|uv\rangle$), $\langle u|\otimes\langle v|$ (or simply $\langle u|\langle v|$ or $\langle uv|$) for describing the state, dual state and linear map of composite systems respectively. It is computed by the the Kronecker product of matrices, e.g., $\langle 0|\langle 1| = {\color{blue}\matrix{1 & 0}}\otimes {\color{red}\matrix{1 & 0}} = \matrix{({\color{blue}1}*{\color{red}1}) & ({\color{blue}1}*{\color{red}0}) & ({\color{blue}0}*{\color{red}1}) & ({\color{blue}0}*{\color{red}0})} = \matrix{1&0&0&0}$.
\end{itemize}

\subsection{Labelled Dirac Notation and Motivating Example}
Labelled Dirac notation is a generalization of Dirac notation for
describing many-body quantum systems. The following example shows the necessity of labels:

\begin{example}
  \label{example1}
  Let $p,q,r$ be three qubits and initially in the (unnormalized)
  $\ghz$ state $|\ghz\rangle \triangleq |000\rangle+|111\rangle$.
  Applying the 3-qubit Toffoli gate ($\ccnot$) with control qubits
  $p,r$ and target qubit $q$ to GHZ is equivalent to applying 2-qubit
  $\cnot$ gate with control qubit $r$ or $p$ and target qubit $q$.
  Using Dirac notation, the identity is written as:
\begin{align*}
  \ccnot|\ghz\rangle &= (I\otimes \cnot)|\ghz\rangle \\
  &= (\Swap\otimes I)(I\otimes \cnot)(\Swap\otimes I)|\ghz\rangle,
\end{align*}
which might be illustrated by the following circuit models.
\begin{figure}[h]
  % \begin{minipage}{0.48\linewidth}
      \centering
      \vspace{-0.5cm}
      \begin{quantikz}[column sep=0.1em, row sep=0em]
      \lstick{$p\qquad\qquad $} & \lstick{} &[0.2cm] & \ctrl{1} & [0.2cm] &  \quad \ \ \;\! \quad\qquad\quad & 
      &[0.2cm] & & [0.2cm] & \quad \ \ \;\! \quad\qquad\quad & 
      &[0.2cm] & \gate[2,swap,style={dashed}]{} & [0.2cm] & [0.2cm]\gate[2,swap,style={dashed}]{} & [0.2cm] & \\[-0.1cm]
      \lstick{$r\qquad\qquad $} & \lstick{$|\ghz\>$} & & \ctrl{1} & & \quad = \quad\qquad\quad & \lstick{$|\ghz\>$} 
      & & \ctrl{1} & & \quad = \quad\qquad\quad & \lstick{$|\ghz\>$}
      & & & \ctrl{1} & & & \\
      \lstick{$q\qquad\qquad $} & \lstick{} & & \targ{} & & \quad \ \ \;\! \quad\qquad\quad & 
      & & \targ{} & & \quad \ \ \;\! \quad\qquad\quad &
      & & & \targ{} & & & \\
      \end{quantikz}
      \vspace{-0.5cm}
      % \vspace{-0.7em}
      % \caption{Circuit computing the last bit of $a+(011)_2$.}
      % \label{fig:addition}
  %  \end{minipage}
  \end{figure}
\end{example}

Formalizing the statement in Dirac notation requires the following
steps: 1. Arrange the qubits in a conventional order, here we choose
$p,r,q$ to simplify the representation of $\ccnot$; 2. Lift the local
operation $\cnot$ to the global system. When $\cnot$ acts on $r,q$, it is
straightforward, as $r$ and $q$ are consistent with the chosen order. We only need
to tensor it with an identity operator $I$ on $p$, i.e., $(I\otimes
\cnot)$.  For $\cnot$ acting on $p,q$, note that $p,q$ are not
adjacent in the chosen order. Thus, we additionally need the $\Swap$ gate to
temporarily exchange the qubits $p$ and $r$, i.e., globally, we apply
$(\Swap\otimes I)$ before and after $(I\otimes \cnot)$ to lift $\cnot$ on $r,q$.

Roughly speaking, encoding in plain Dirac notation requires
tensoring identity operators and using additional $\Swap$ gates,
since the conventional order does not generally guarantee the following: 1. the order
of all local operations is consistent with it, 2. the local operations
only involve adjacent qubits in the conventional order.

To address the limitations of Dirac notations, physicists routinely
use labels (or subscripts) to indicate the systems on which quantum
states or operations are applied, thereby avoiding unnecessary lifting
and swap gates. For example, rewriting the previous example using labels,
we obtain:
\begin{align*}
  \ccnot_{prq}|\ghz\rangle_{pqr} = 
  \cnot_{rq}|\ghz\rangle_{pqr}
  = \cnot_{pq}|\ghz\rangle_{pqr}
\end{align*}
This formalization avoids determining and maintaining the conventional
order of qubits, nor lifting using additional $I$ and $\Swap$s. In
this setting, the tensor products become associative and
commutative, allowing us to rearrange qubits as needed for
calculations. For our example, we can perform the calculation as
follows:
\begin{align*}
  &\ccnot_{prq}|\ghz\rangle_{pqr}
  = (\ccnot|000\rangle + \ccnot|111\rangle)_{prq}
  = (|000\rangle + |110\rangle)_{prq}\\
  &\cnot_{rq}|\ghz\rangle_{pqr}
  = (\cnot|00\rangle)_{rq}|0\rangle_p + (\cnot|11\rangle)_{rq}|1\rangle_p
  =(|000\rangle + |101\rangle)_{rqp}\\
  &\cnot_{pq}|\ghz\rangle_{pqr}
  = (\cnot|00\rangle)_{pq}|0\rangle_r + (\cnot|11\rangle)_{pq}|1\rangle_r
  = (|000\rangle + |101\rangle)_{pqr}
\end{align*}
The right-hand side (RHS) of each line is equivalent, as shown.  
In addition, labelled
Dirac notation can conveniently describe local measurements, partial traces (representing the state or evolution of subsystems in many-body systems), and partial inner products (which correspond to partial traces in pure states). These capabilities are sufficient for handling the mathematical formulas of quantum mechanics in many-body systems.

Labelled Dirac notation is not only pervasive in the description of many-body systems but also plays a crucial role in quantum program logic. Just as classical program logic uses variable names to construct logical formulas, avoiding the need for global memory functions, quantum program logic similarly uses variable names to label the subsystems on which quantum gates act, rather than lifting them to the global system. Actually, lifting operations would lead to an exponential increase in formula length relative to the number of variables, as discussed in \cite{QSL_Le_2022}.

In the following, we will use this motivating example as the primary focus in our demonstrations. Instead of GHZ states, we use a simpler example involving Bell states:
\begin{example}
    \label{ex: motivating}
    Let \( q \) and \( r \) represent two quantum systems in the Hilbert space \( \mathcal{H}_T \). Let \( M \) be a quantum operation acting on \( \mathcal{H}_T \), and let \( \ket{\Phi} = \sum_{i \in T} \ket{i} \otimes \ket{i} \) be the maximally entangled state. Then, it holds that
    \[
    M_q \ket{\Phi}_{(q, r)} = M_r^T \ket{\Phi}_{(q, r)}.
    \]
\end{example}
As explained earlier regarding labels, we can consider the global system $(q,r)$ and transform the equation above into the plain Dirac notation:
\begin{align}
    (M \otimes I) \ket{\Phi} = (I \otimes M^T) \ket{\Phi}.
    \label{eq: motivating plain}
\end{align}
The following sections introduce the formal language and labelled Dirac notation, and present a systematic approach for reducing labels. We will also demonstrate how an automated system can be built and used to solve similar equalities.

%% file: 02language_design.tex
\section{Dirac notation}
This section introduces the language of Dirac notation, its
denotational and axiomatic semantics, and describes D-Hammer approach
to equational reasoning. Three main ingredients of our language are:
\begin{itemize}
\item a rich typing discipline that distinguishes between scalars,
  kets, bras and operators, but supports sufficient overloading to
  remain close to standard Dirac notation;
\item higher-order, indexed (a.k.a.\, weakly dependent) types. It allows
  to formally encode defined symbols like transpose or trace, which
  are usually used to represent the term in an abstract manner;
\item operators with indefinite arities. Indefinite arities are
  instrumental for reasoning efficiently about associative and
  commutative (AC) symbols have indefinite arities, as they enable
  normalization by sorting.
\end{itemize}

\subsection{Language}
Since a Hilbert space $\mathcal{H}_V$ is dependent on the basis set
$V$, types for Dirac notation also depends on the type index.
Therefore, the language is organized into three layers: the index, the
type, and the term.  Terms represent concrete instances such as kets,
bras, and operators, which will be typed and checked. The index
represents classical data types and appears in type expressions to
differentiate between various Hilbert spaces and sets.

\begin{definition}[Index Syntax]
    The syntax for type indices is:
    \begin{align*}
        \sigma ::=\ & x \mid \sigma_1 \times \sigma_2.
    \end{align*}
\end{definition}
Here, \( x \) is a variable, and \( \sigma_1 \times \sigma_2 \) represents the product type for tensor product spaces or Cartesian product sets.

\begin{definition}[Type Syntax]
    The syntax for Dirac notation types is:
    \begin{align*}
        T ::=\ & \Basis(\sigma) \mid \SType \mid \KType(\sigma) \mid \BType(\sigma) \mid \OType(\sigma_1, \sigma_2) \mid T_1 \to T_2 \mid \forall x.T \mid \SET(\sigma).
    \end{align*}
\end{definition}
\( \Basis(\sigma) \) denotes the type for basis elements in the index
\( \sigma \).  \( \SType \) represents scalars, while \(
\KType(\sigma) \) and \( \BType(\sigma) \) refer to ket and bra types
in the Hilbert space \( \sigma \), respectively.  \( \OType(\sigma_1,
\sigma_2) \) represents linear operators with \( \sigma_2 \) as the
domain and \( \sigma_1 \) as the codomain.  \( \SET(\sigma) \) refers
to the type of subsets of \( \sigma \), used to denote the values of
bound variables in summations.  The remaining two constructs define
function types: \( T_1 \to T_2 \) represent the set of functions that
take a \( T_1 \)-type argument and return a \( T_2 \)-type term, while
\( \forall x. T \) represents the dependently typed functions that
take an index argument \( x :\Index \) and produce a \( T \)-type
term, where $T$ may depend on \( x \).  Index functions are essential
for defining polymorphic transformations over Hilbert spaces.

\begin{definition}[Term Syntax]
    The syntax for Dirac notation terms is:
    \begin{align*}
        e ::=\ & x \mid \lambda x : T.e \mid \lambda x : \Index.e \mid e_1\ e_2 \mid (e_1, e_2) \\
        & |\ 0 \mid 1 \mid e_1 \times \cdots \times e_n \mid e^* \mid \delta_{e_1, e_2} \\
        & |\ \ZEROK(\sigma) \mid \ZEROB(\sigma) \mid \ZEROO(\sigma_1, \sigma_2) \mid \ONEO(\sigma) \\
        & |\ \ket{e} \mid \bra{t} \mid e^\dagger \mid e_1.e_2 \mid e_1 + \cdots + e_n \mid e_1 \otimes e_2 \mid e_1 \cdot e_2 \\
        & |\ \mathbf{U}(\sigma) \mid e_1 \star e_2 \mid \sum_{e_1} e_2.
    \end{align*}
\end{definition}
The terms above are explained in five lines.
\begin{enumerate}
    \item \textbf{function and basis}: \( \lambda x : T.e \) represents the abstraction for normal functions, and \( \lambda x : \Index.e \) represents the abstraction for index functions.
    \( e_1\ e_2 \) denotes function application.
    \( (e_1, e_2) \) is the basis pair for product types.
    \item \textbf{scalar}: \( 0 \), \( 1 \), \( e_1 \times \cdots \times e_n \) and \( e^* \) are symbols for scalars.
    \( \delta_{e_1, e_2} \) compares whether two basis are the same and evaluates to $1$ or $0$ accordingly.
    \item \textbf{Dirac constant}: zero ket, zero bra, zero operator and identity operator.
    \item \textbf{Dirac function}: \( \ket{e} \) is a ket, \( \bra{t} \) is a bra, and \( e^\dagger \) denotes the conjugate transpose of \( e \). \( e_1.e_2 \) represents scaling the term \( e_2 \) by scalar \( e_1 \). \(e_1 + \cdots + e_n\) is the addition. \( e_1 \otimes e_2 \) represents tensor product, and \( e_1 \cdot e_2 \) represents the multiplication.
    \item \textbf{summation}: \( \mathbf{U}(\sigma) \) denotes the universal set with index \( \sigma \). \( e_1 \star e_2 \) represents the Cartesian product of \( e_1 \) and \( e_2 \). \( \sum_{e_1} e_2 \) is the big operator sum, modeled by folding the function \( e_2 \) over the value set \( e_1 \). Typically, the sum's body is given by an abstraction. For convenience, we also use the notation \( \sum_{x \in s} X \) to represent \( \sum_{s} \lambda x : T . X \).
\end{enumerate}

The scalar multiplication $\times$ and addition $+$ are AC symbols,
and they have indefinite arity.  We use letters like $a, b, c$ to
represent scalar variables, $K$ and $B$ to represent ket and bra
variables, and $O$ for operators.  Therefore, $O \cdot K$ is
interpreted as the operator-ket multiplication, and scalars can also
be constructed from inner products $B \cdot K$.

\subsection{Typing System}
The typing system is responsible for classifying terms within a proof system, according to the types of variables and definitions. 
We use a context \( \Gamma \) to preserve the assumptions \( x : T \) and definitions \( x := t : T \).
\begin{definition}[Context]
    The syntax for context \( \Gamma \) is:
    \begin{align*}
        \Gamma ::= &\ [] \mid \Gamma; x : \Index \mid \Gamma; x : T \mid \Gamma; x := t : T.
    \end{align*}
\end{definition}
Definitions refer to symbols that can be expanded or unfolded, and typically represent abstract concepts. such as transpose or trace in Dirac notation. Assumptions, on the other hand, define the types of variables.
We say an expression \( t \) has type \( X \) in context \( \Gamma \) if the typing judgment \( \Gamma \vdash t : X \) can be proven through the rules in~\Cref{sec: full typing rules}. These are two instances:
\begin{gather*}
    \frac{\Gamma \vdash t : \Basis(\sigma)}{\Gamma \vdash \ket{t} : \KType(\sigma)},
    \qquad
    \frac{\Gamma \vdash B : \BType(\sigma) \qquad \Gamma \vdash K : \KType(\sigma)}{\Gamma \vdash B \cdot K : \SType}.
\end{gather*}
The ket \( \ket{t} \) will have the type \( \KType(\sigma) \) if \( t \) is a basis term of index \( \sigma \). Similarly, the inner product between a bra and a ket of the same index \( \sigma \) is typed as a scalar. It corresponds to the constraint of inner product that vectors should be from the same Hilbert space.
Especially, the big operator sum is modeled by folding a function over a set, with the typing rule as follows:
\[
    \frac{\Gamma \vdash s : \SET(\sigma) \qquad \Gamma \vdash f : \Basis(\sigma) \to \KType(\tau)}{\Gamma \vdash \sum_{s} f : \KType(\tau)}.
\]

% \begin{lemma}
%     The typing of expressions is both decidable and unique.
% \end{lemma}

% \begin{proof}
%     The type of an expression can be determined recursively. For any given function symbol and argument types, there is at most one typing rule, ensuring the uniqueness of typing.
% \end{proof}

\subsection{Semantics}

The semantics of a language define the meaning of its expressions. In this context, the objective of our algorithm is to determine whether two expressions are semantically equivalent. We define the semantics in a denotational manner, mapping syntax to set-theoretic objects.

\subsubsection{Denotational Semantics}
Denotational semantics maps types to sets, and expressions and
indices to values in the interpretation of types and indices,
respectively. As in other dependently typed systems, all
interpretations are parametrized by a valuation mapping \( v \), which
assigns values to variables and indices. We let \( \sem{e}_v \) denote
the interpretation of an expression $e$ w.r.t.\, a valuation \( v \),
and use similar notations for types and indices. As usual, we say that
a valuation \( v \) is valid w.r.t.\, a context $\Gamma$ if for every
variable declaration $x:T$, we have \( \sem{x}_v \in\sem{T}_v \) and
for every definition $x:= t: T$, we have  \( \sem{x}_v=\sem{t}_v \).

In more detail, variables typed with \( \Index \) are interpreted as
finite sets, and the product of two indices \( \sem{\sigma_1 \times
  \sigma_2} \) is defined as the Cartesian product of the sets \(
\sem{\sigma_1} \) and \( \sem{\sigma_2} \). More generally, each type
is interpreted as a set. For example, the scalar type \( \sem{\SType}
\) is interpreted as the set of complex numbers \( \mathbb{C} \), and
the ket and bra types \( \sem{\KType(\sigma)} \) and \(
\sem{\BType(\sigma)} \) are interpreted as the Hilbert space \(
\mathcal{H}_{\sem{\sigma}} \) and its dual \(
\mathcal{H}_{\sem{\sigma}}^* \), respectively. Terms are explained as
the set elements. For example, the semantics of ket tensor product
$\sem{K_1 \otimes K_2} \equiv \sem{K_1} \otimes \sem{K_2}$, is
obtained by first calculating the semantics $\sem{K_1}$ and
$\sem{K_2}$ as vectors, and then take the vector tensor product as
result.  The complete interpretation of terms and types is provided
in~\Cref{sec: full denotational sem}.

% One special case is the delta function \( \delta_{s,t} \), which is interpreted as:
% \[
%     \sem{\delta_{s,t}} =
%     \begin{cases}
%         1, & \text{if } \sem{s} = \sem{t}, \\
%         0, & \text{if } \sem{s} \neq \sem{t}.
%     \end{cases}
% \]d

The type system is sound w.r.t.\, the denotational semantics of
expressions. Specifically, for a well-formed context \( \Gamma \), term \( t \), and type \( T \), if \( \Gamma \vdash t : T \), then for any valuation \( v \) that is valid for $\Gamma$, the interpretation of \( t \)  w.r.t.\, $v$ is an element of the interpretation of \( T \) w.r.t.\, $v$.

\begin{lemma}[Soundness of type system]
  \label{lemma:sound type system}
  If \( \Gamma \vdash t : T \), then for all valuations \( v \) valid w.r.t.\,
  $\Gamma$, we have \( \sem{t}_v \in \sem{T}_v \).
\end{lemma}
% \begin{proof}
% By induction on the derivation.
% \end{proof}
This interpretation formalizes the standard understanding of Dirac
notation and provides the foundation for the algorithm. However,
computers cannot directly reason about equivalence through
mathematical interpretations. We proceed by defining a proof system
that abstracts these concepts.

\subsubsection{Axiomatic semantics} 

The proof system for equivalence is based on equational logic, together with axioms that describe the properties of Dirac notation. A full list of these axioms can be found in~\Cref{sec: full axioms}. The axioms cover fundamental aspects of linear spaces, as well as other structures like the tensor and inner products. For example, we have the absorption law for zero symbols:
\(X \cdot \mathbf{0} = \mathbf{0},\)
and the bilinearity of the dot product:
\begin{align*}
(a.X) \cdot Y = a \cdot (X \cdot Y), \quad X \cdot (Y_1 + Y_2) = X \cdot Y_1 + X \cdot Y_2, \\
X \cdot (a.Y) = a \cdot (X \cdot Y), \quad (X_1 + X_2) \cdot Y = X_1 \cdot Y + X_2 \cdot Y.
\end{align*}
The entire axioms are separated into two sets $R$ and $E$.
$R$ contains the axioms normalized by term rewriting. Other axioms requiring special algorithms, which are collected in the set $E$.
\begin{definition}[axiom set E]
\label{def: axiom E}
\begin{align*}
    \textup{(AC-equivalence)} &\ \text{e.g.,} \quad X + Y = Y + X, \quad (X + Y) + Z = X + (Y + Z), \\
    \textup{($\alpha$-equivalence)} &\ \lambda x . A = \lambda y . A\{x/y\},
    \quad
    \textup{(SUM-SWAP)} \ \sum_{i \in s_1} \sum_{j \in s_2} A = \sum_{j \in s_2} \sum_{i \in s_1} A, \\
    \textup{(scalar theories)} &\ \text{e.g.,} \quad a + 0 = a, \quad a \times (b + c) = a \times b + a \times c.
\end{align*}
\end{definition}

% These equational axioms provide an operable theory for the proof automation algorithm. Denotational semantics can be seen as one model for this theory, meaning that equivalences derived from the axioms always imply equivalence in the interpretations.

We say an equation $e_1 = e_2$ is provable, denoted as $\Gamma \vdash e_1 = e_2 : T$, if $\Gamma \vdash e_1 : T$ and $\Gamma \vdash e_2 : T$ are provable, and $e_1 = e_2$ can be deduced in $\Gamma$ using the axioms and equational logic.
An equation \( e_1 = e_2 \) is valid in context $\Gamma$, written as \(
\Gamma \vDash e_1 = e_2 \), if \( \sem{e_1}_v = \sem{e_2}_v \) for
all valuations \( v \) that are valid w.r.t.\, \( \Gamma \).

\begin{theorem}[Soundness of equational theory]\label{lem: axiom sound}
If \( \Gamma \vdash e_1 = e_2 :T \) then \( \Gamma \vDash e_1 = e_2\).
\end{theorem}
The proof of soundess is standard: we prove that all axioms are sound,
and that all proof rules are sound.

% \begin{proof}
% By induction on the structure of the derivation
% \end{proof}
Next, we formalize the motivating example~\Cref{ex: motivating} in
Dirac notation.
\begin{example}[Motivating Example Formalization]
    \label{ex: formalizing motivating}
    Definitions and assumptions in the context \( \Gamma \) are formalized as follows:
    \begin{align*}
        & \text{TPO} && := \lambda T_1 : \Index. \lambda T_2 : \Index. \lambda O : \OType(T_1, T_2). \sum_{i \in \mathbf{U}(T_1)} \sum_{j \in \mathbf{U}(T_2)} \bra{i} O \ket{j} . \ket{j}\bra{i} \\
        & &&\quad : \forall T_1. \forall T_2. \OType(T_1, T_2) \to \OType(T_2, T_1); \\
        &\text{phi} &&:= \lambda T : \Index. \sum_{i \in \mathbf{U}(T)} \sum_{j \in \mathbf{U}(T)} \ket{(i, j)} : \forall T.\KType(T \times T); \\
        & T && : \Index;  \qquad \qquad \qquad \qquad \qquad M \qquad : \OType(T, T).
    \end{align*}
    Notice how the functions and higher-order typing helps to formalize the abstract concepts here.
    The symbol \( \text{TPO} \) represents the transpose of an operator, polymorphic on the Hilbert spaces \( T_1 \) and \( T_2 \). 
    The symbol \( \text{phi} \) takes the index \( T \) and defines the maximally entangled states, summing over all basis elements in \( T \), as indicated by the universal set \( \mathbf{U}(T) \).
    With the assumption of the index \( T \) and operator \( M \), we can express the equivalence in the plain Dirac notation as:
    \[
    (\textrm{M} \otimes \mathbf{1}_\mathcal{O}(\textrm{T})) \cdot (\textrm{phi T}) = (\mathbf{1}_\mathcal{O}(\textrm{T}) \otimes (\textrm{TPO T T M})) \cdot (\textrm{phi T}).
    \]
\end{example}

\subsection{Normalization}

The equivalence of Dirac notations is established through normalization, which transforms equivalent expressions into the same syntax under a set of axioms. We employ an efficient algorithm to perform the normalization fully on $R \cup E$.

\begin{enumerate}
    \item \textbf{Rule based term rewriting}: Expand definitions and simplify expressions.
    \item \textbf{Variable expansion}: Convert to abstract element-wise representation.
    \item \textbf{Rule based term rewriting}: Normalize terms on \( R \) modulo \( E \).
    \item \textbf{Sorting without bound variables}: Normalize AC-equivalence.
    \item \textbf{Swapping successive summations}: Normalize SUM-SWAP equivalence.
    \item \textbf{Use de Bruijn index}: Normalize \( \alpha \)-equivalence.
\end{enumerate}

Step 1 through 3 involve term rewriting for $R$. 
Term rewriting is the process of repeatedly reducing a term using a set of rules in the form of $l\ \reduce\ r$. The reduction works by matching the subterms with the left-hand side of a rule and replacing it with the right-hand side. 
For example, the term $(x\times y) . \ket{t} + \ket{t}$ is matched by the rule $a.K + K\ \reduce\ (a + 1).K$, and is rewritten into $(x\times y + 1) . \ket{t}$.
Step 1 and 3 use the same set of rewriting rules in~\Cref{sec: rewriting rules}.
Step 2 expands variables to their abstract element-wise representation, e.g., $K \ \reduce\ \sum_{i} (\bra{i} \cdot K). \ket{i}$, which is useful when reasoning about sums.

Steps 4 through 6 are specialized algorithms designed to further normalize the axiom set $E$.
The main challenge here is the coexistence of AC-equivalence and SUM-SWAP, which means that naive sorting cannot alwasy convert equivalent terms into the same form.
For step 4 and 5, the key observation is that in a successive sum expression \( \sum_{i \in s_1} \cdots \sum_{j \in s_n} A \), the names and order of the bound variables \( i, \dots, j \) can be freely permuted. Therefore we first ignore bound variables and normalize AC-equivalence by sorting. Afterward, the order of summation can be established accordingly. The final step uses de Bruijn indices~\cite{deBruijn1972lambda} to resolve $\alpha$-equivalence. For further details, refer to~\Cref{sec: decide}.

\Cref{fig: normalization demo} shows the normalization outline for $(M \otimes \mathbf{1}_\mathcal{O}(T)) \cdot (\text{phi}\ T)$.
\begin{figure}[h]
    \scriptsize
    \begin{align*}
        & (\eqnmarkbox[blue]{mvar}{M} \otimes \eqnmarkbox[red]{identity}{\mathbf{1}_\mathcal{O}(T)})(\eqnmarkbox[red]{phi-definition}{\textrm{phi}\ T})
        \Rightarrow
        (\sum_{k,l\in T} \bra{k} M \ket{l}.\ket{k}\bra{l}
        \eqnmarkbox[red]{sumpush1}{\otimes \sum_{i \in T}} \ket{i}\bra{i})(\eqnmarkbox[red]{sumpush2}{\sum_{j\in T}}\ket{j, j}) 
        \\
        & \Rightarrow
        \sum_{i,j,k,l\in T} \bra{k} M \ket{l} . \ket{k, i}\eqnmarkbox[red]{braket-dot}{\bra{l, i} \cdot \ket{j,j}}
        \\
        & \Rightarrow
        \eqnmarkbox[red]{delta-elim-sum}{\sum_{i,j,k,l\in T}} (\eqnmarkbox[red]{delta-elim-delta}{\delta_{l,j} \times \delta_{i,j}} \times \bra{k} M \ket{l}) . \ket{k, i}
        \Rightarrow
        \eqnmarkbox[brown]{swapsum}{\sum_{j, k \in T}} \bra{k} M \ket{j} . \ket{k, j}
        \\[1em]
        & \Rightarrow
        \sum_{k, j \in T} \bra{k} M \ket{j} . \ket{k, j}
        \ \eqnmarkbox[brown]{deBruijn}{\xRightarrow{\text{de Bruijn}}}\ 
        \sum_T\sum_T\bra{\$1} M \ket{\$0} . \ket{\$1, \$0}
    \end{align*}
    \annotate[yshift=1em]{right}{mvar}{\scriptsize{variable expansion}}
    \annotate[yshift=0.5em, xshift=1cm]{right}{identity}{\scriptsize{identity operator}}
    \annotate[yshift=0em]{below,left}{phi-definition}{\scriptsize{definition}}
    \annotatetwo[yshift=0em]{below}{sumpush1}{sumpush2}{\scriptsize{sum lifting}}
    \annotate[yshift=0em]{below}{braket-dot}{\scriptsize{inner product}}
    \annotatetwo[yshift=0em,xshift=2cm]{below}{delta-elim-sum}{delta-elim-delta}{\scriptsize{$\delta$-elimination}}
    \annotate[yshift=0em]{below}{swapsum}{\scriptsize{sum swapping}}
    \caption{A normalization outline for the left-hand side of~\Cref{eq: motivating plain}. Matched subterms are marked with colors. Blue marking represents variable expansion, red marking represents rule applications, and brown marking represents normalization step 4-6.}
    \label{fig: normalization demo}
\end{figure}

In contrast, previous work performs normalization only partially on $R$, and proves equivalence by checking all possible permutations according to $E$.
Our algorithm fully normalize the term, as illustrated below, and is more efficient as a result.
\[
\text{(partial)}\qquad e_1 \mathop{\twoheadrightarrow}^R e_1' \mathop{=}^E e_2' \mathop{\twoheadleftarrow}^R e_2
\qquad
\text{(full)}\qquad e_1 \mathop{\twoheadrightarrow}^{R \cup E} e \equiv e \mathop{\twoheadleftarrow}^{R \cup E} e_2
\]
% While the permutation checking algorithm has factorial complexity on the number of AC symbol arguments, our algorithm uses normalization by sorting, and the complexity is polynomial on the term size. 
% This improvement in efficiency is also confirmed by experiments.

% \begin{gather*}
%     \eqnmarkbox[blue]{node1}{GEBV} = \sum_{i}^{n} \eqnmarkbox[red]{node2}{X_{i}} b_{i} \\
%     \eqnmarkbox[green]{node3}{ssssss}
% \end{gather*}
    
%     \annotate[yshift=1em]{left}{node1}{Genomic Estimated Breeding Value}
%     \annotate[yshift=-1em]{below}{node2}{Genotype at locus i}
%     \annotatetwo[yshift=2em]{above}{node2}{node3}{var 1}

% Lastly, we can prove that the equivalence established by this normalization procedure is sound with respect to the semantics.

% \begin{theorem}[Soundness]
%     For any well-formed context \( E \) and well-typed expressions \( e_1 \) and \( e_2 \), if  $e_1$ and $e_2$ have the same normal form, then \( \sem{e_1} = \sem{e_2} \).
% \end{theorem}
% \begin{proof}
%     Because of~\Cref{lem: axiom sound}, we only need to prove soundness with respect to axioms. 
%     This is because the term rewriting procedure follows from the fact that each rewriting rule preserves equivalence. Furthermore, the operations in the sort and swap transformations respect the AC-equivalence and SUM-SWAP axioms. Finally, the de Bruijn normalization ensures soundness for \( \alpha \)-equivalence.
% \end{proof}

%% file: 03labelled.tex
% Labelled Dirac Notation

\section{Labelled Dirac Notation}
\label{sec: labelled}

% what is the labelled dirac notation?

% the semantics -- cylinder extension -- uniqueness

% \subsection{Syntax, Typing and Rewriting Rules}

% syntax -- typing -- rules

% \subsection{Normalization}
% normalization -- algorithm that eliminate labels

In this section, we extend the language by allowing quantum variables to indicate the quantum system on which vectors and operators act. 
As discussed, this enables us to express and reason about the states and operations locally, without referring to the entire system. 
We further demonstrate how to transform the equivalence problem into one involving the plain Dirac notation studied earlier.

\subsection{Syntax, Typing and Semantics}
We begin by introducing the notation of quantum registers for structured variable combinations. This is necessary because, unlike assignments for classical variables, unitary transformations on composite systems--a quantum version of assignments--cannot generally be decomposed into separate unitary transformations on individual subsystems.
Let $\cR$ be the set of quantum variables.
\begin{definition}[Quantum Register] Register $R$ is inductively generated by
  \begin{align*}
    R ::= r\in\cR \mid (R, R).
  \end{align*}
\end{definition}
For simplicity, we restrict register formation to pairings, which corresponds to the structure of tensor product. In this context, $\cH_{(R_1,R_2)}$ is isomorphic to $\cH_{R_1}\otimes \cH_{R_2}$, which allows us to view the tensor product space as the space of paired registers.

The no-cloning theorem, a fundamental property of quantum computing, prevents us from copying an unknown quantum state. This requires an additional check on the valid registers--they should not include repeated quantum variables--which is often handled in programming languages via linear types. As such, we define the \emph{order-free} variable set of a register as all quantum variables appearing in the register; we let $\var(R)$ denote the variable set of register $R$.
We use the variable set to establish side conditions of typing valid registers and employ it as the type parameter for annotating labelled Dirac terms.

Now, we are ready to extend the type syntax and term syntax as follows:
\begin{definition}[Labelled Dirac Notation]
  The \textbf{labelled Dirac notation} includes all plain Dirac notation symbols and the generators defined below.
  Here, $s\subseteq \cR$ is a quantum variable set.
  \begin{align*}
    T & ::= \DType(s,s) \mid \mathsf{Reg}(\sigma) \\
    e & ::= R \mid |i\>_r \mid {}_r\<i| \mid e_R \mid e_{R;R} \mid
    e \otimes e \otimes \cdots \otimes e \mid e \cdot e.
  \end{align*}
\end{definition}
$\DType(s_1, s_2)$ is the unified type for all labelled Dirac notation, where $s_1$ indicates the codomain systems and $s_2$ indicates the domain systems.
Roughly speaking, we define Hilbert space $\cH_s\triangleq\bigotimes_{r\in s}\cH_r$ for each set $s$, so that $\cH_{\emptyset}$ is a one-dimensional space isomorphic to complex numbers. Then the function view of ket and bra~\cite{Zhou2023} provides an alternative way, i.e., a ket on subsystem $s$ as a linear map from $\cH_{\emptyset}$ to $\cH_s$, a bra as a linear map from $\cH_s$ to $\cH_{\emptyset}$, to unify the type of kets, bras, and operators. For instance, labelled ket $|i\>_r$ has type $\DType(\{r\}, \emptyset)$, and labelled bra ${}_r\<i|$ has type $\DType(\emptyset, \{r\})$.
$\reg(\sigma)$ are types for registers $R$, and the index $\sigma$ indicates the type of Hilbert space represented by the register.
It is allowed to lift a plain Dirac notation associate with corresponding quantum variables or registers, e.g., $\ket{i}_r$ and ${}_r\bra{i}$ are labelled basis, $e_R$ for bra, ket, and $e_{R;R'}$ for operators which additionally allows different domain $R'$ and codomain $R$.
We further introduce new $\cdot$ for generalized composition (unified for all kinds of multiplications between kets, bras and operators) and $\otimes$ for labelled tensor product since they do not share the same properties v.s. its counterpart in plain Dirac notation, i.e., generalized composition is not associative and labelled tensor is indeed an AC symbol.
% In labelled Dirac notation, the structure of tensor product does not matter, therefore $\otimes$ is an AC symbol.
% Following the unified type $\DType(s,s)$, all kinds of multiplications are represented by the same dot product $e \cdot e$.
% Finally, $\ket{i}_r$ and ${}_r\bra{i}$ are labelled basis for the normal form of labelled Dirac notation, where $r$ are registers symbols in $\mathcal{R}$. 

\paragraph*{Typing rules.}
There are various rules for computing types and checking vadility of
registers and labelled terms. Here we display some of the rules and
refer the reader to~\Cref{sec: full typing rules} for the full set of rules.
\begin{gather*}
  \frac{
      \Gamma \vdash R : \reg(\sigma) \quad
      \Gamma \vdash Q : \reg(\tau)
      \quad \var(R) \cap \var(Q) = \emptyset
  }{\Gamma \vdash (R,Q) : \reg(\sigma \times \tau)} \\[0.2cm]
  \frac{
          \Gamma \vdash r : \reg(\sigma) \quad
          \Gamma \vdash i : \Basis(\sigma)}
  {\Gamma \vdash |i\>_r : \DType(\{r\}, \emptyset)}
  \qquad
  \frac{\Gamma \vdash R : \reg(\sigma)\qquad \Gamma \vdash K : \KType(\sigma)}{\Gamma \vdash K_R : \DType(\var(R), \emptyset)} \\[0.2cm]
    \frac{
        \Gamma \vdash D_i : \DType(s_i,s_i') \qquad
        \forall\,i\neq j.\ s_i\cap s_j = \emptyset \qquad
        \forall\,i\neq j.\ s'_i\cap s'_j = \emptyset
    }
    {\Gamma \vdash D_1 \otimes \cdots \otimes D_i : \DType(\bigcup_i s_i, \bigcup_i s_i')}.
\end{gather*}
The first rule states that a paired register is of the product type and its components must be disjoint.
To lift a plain Dirac notation into the labelled version (line 2), we enforce that the term and register share the same indices, reflecting the fact the state should be consistant to the corresponding subsystems.
The third line provides the typing of labelled tensor product, with a check to ensure that the component subsystems are disjoint from each other.

\paragraph*{Semantics.} 
The labelled Dirac notation handles lifting and ordering for us, and its semantics accurately capture these details. The key points are: 1. cylindrical extension, which lift a ket or bra or operator to larger domain and codomain; 2. general composition, which further employs cylindrical extension that obeys the principle of ``localizing objects as much as possible''~\cite{Zhou2023}.
Since $\cR$ is given, we let $\sigma_r : \Index$ denote type of $r$, i.e.,  $\Gamma\vdash r : \reg(\sigma_r)$, for simplicity.

\begin{definition}[Cylindrical Extension]
  For any $D : \cD(s_1,s_2)$ and $s$ that disjoint with both $s_1$ and $s_2$, we define $\cl(D,s) \triangleq D\otimes \mathbf{1}_{s}$ of type $\cD(s_1\cup s, s_2\cup s)$.
\end{definition}
Formally, we equip $\cR$ with a default order, such as the alphabetical order of names. For any valid register $R$, there exists the operator $\Swap_R$ that sorts $R$; for example, $\Swap_{(q,p)} \triangleq \sum_{i\in \bU(\sigma_p)}\sum_{j\in \bU(\sigma_q)}|i\>\<j|\otimes |j\>\<i|$. We can further define $\Swap_{s_1,s_2,\cdots}$ for merging disjoint sets orderly. See Appendix \ref{sec: full denotational sem} for details.

For given context $\Gamma$ and any valuation $v$,
we interpret 
$$\mbox{$\sem{\cD(s,s')}_v \triangleq \mathcal{L}(\bigotimes_{j}\cH_{\sem{\sigma_{r'_j}}_v}, \bigotimes_{i}\cH_{\sem{\sigma_{r_i}}_v})$}$$ where $\mathcal{L}$ denotes the set of linear maps, 
$s = \{r_1,\cdots,r_n\}$ and $s' = \{r'_1,\cdots,r'_m\}$ ($r_i$ and $r'_j$ are sorted).
% we define $\sigma_s\triangleq \sigma_1\times\cdots\times\sigma_n$ if $s$ is non-empty. \label{context: sigma s}
% For any valuation $v$, we further write $\cH_{\sem{s}_v} \triangleq \bigotimes_{i}\cH_{\sem{\sigma_i}_v}$ which equals to $\cH_{\sem{\sigma_s}_v}$ for non-empty $s$.
% We interpret $\sem{\cD(s,s')} \triangleq \mathcal{L}(\cH_{\sem{s'}_v}, \cH_{\sem{s}_v})$ where $\mathcal{L}$ denotes the set of linear maps.
% for any $s\subseteq \cR$, where $\Gamma\vdash r : \reg(\sigma_r)$ for all $r\in\cR$. 
We interpret labelled Dirac notations inductively as (assume $\Gamma\vdash D_i : \cD(s_i,s'_i)$) :
\begin{itemize}
  \item $\sem{|i\>_r}_v = |\sem{i}_v\>$;\quad 
        $\sem{\<i|_r}_v = \<\sem{i}_v|$;\quad  
        $\sem{K_R}_v = \sem{\Swap_R}_v \cdot \sem{K}_v$; \\
        $\sem{B_R}_v = \sem{B}_v\cdot \sem{\Swap_R}_v^\dagger$;\quad
        $\sem{O_{R_1,R_2}}_v = \sem{\Swap_{R_1}}_v\cdot \sem{O}_v\cdot \sem{\Swap_{R_2}}_v^\dagger$;
  \item $\sem{D_1\otimes \cdots \otimes D_n}_v = \sem{\Swap_{s_1,\cdot,s_n}}_v\cdot (\sem{D_1}_v\otimes \cdots \otimes \sem{D_1}_v)\cdot \sem{\Swap_{s'_1,\cdot,s'_n}}_v^\dagger$;
  \item $\sem{D_1\cdot D_2}_v = \sem{\cl(D_1, s_2\backslash s'_1)}_v\cdot \sem{\cl(D_2, s'_1\backslash s_2)}_v$. Note that $s_2\backslash s'_1$ and $s'_1\backslash s_2$ are the minimal extension that make it interpretable. E.g., to interpret ${}_p\<i|\cdot |j\>_{p,q}$, we at least need to extend ${}_p\<i|$ to ${}_p\<i|\otimes I_{q}$.
\end{itemize}
It can be shown that Lemma \ref{lemma:sound type system} also holds for labelled terms, i.e., $\sem{D}_v \in \sem{\cD(s,s')}_v$ given $\Gamma\vdash D : \cD(s,s')$. Following the semantics, labelled tensor is independent of its order, i.e., $\sem{D_1\otimes D_2}_v = \sem{D_2\otimes D_1}_v$ and $\sem{D_1\otimes (D_2\otimes D_3)}_v = \sem{(D_1\otimes D_2)\otimes D_3}_v$, which ensures the soundness of treating labelled tensor as an AC symbol.

\subsection{Elimination of labels}
It is possible to eliminate labels from labelled Dirac expressions and
thus to transform any equation in Labelled Dirac notation into an
equation in plain Dirac notation. This is achieved by the following
three steps:

\paragraph*{1. Elimination of $e_R$ or $e_{R;R}$.}
We decompose all $e_R$ or $e_{R;R}$ to the labelled basis with scalar coefficients. 
%We add rules to the term rewriting system, which in general try to represent labelled Dirac notation with labelled basis and scalar coefficients. The first step is the label elimination. 
Take operator as an example:
\begin{align*}
    & O_{R,R'} \ \reduce\ \sum_{i_{r_1}\in\bU(\sigma_{r_1})}\cdots \sum_{i_{r_n}\in\bU(\sigma_{r_n})}
    \sum_{i_{r'_1}\in\bU(\sigma_{r'_1})}\cdots \sum_{i_{r'_{n'}}\in\bU(\sigma_{r'_{n'}})} \\
    & \qquad (\<i_R|\cdot O\cdot |i_{R'}\>).(|i_{r_1}\>_{r_1}\otimes\cdots\otimes|i_{r_n}\>_{r_n} \otimes {}_{r'_1}\<i_{r'_1}|\otimes\cdots\otimes{}_{r'_{n'}}\<i_{r'_{n'}}|).
\end{align*}
where $|i_R\>$ and $\<i_{R'}|$ are constructed by tensoring the basis according to the structure of $R$ (see Appendix \ref{sec: full denotational sem}). The rules for $e_R$ (ket and bra) are similar.
%This step reduces all labelled terms $e_R$ or $e_{R;R}$.

\paragraph*{2. Rewriting to normal form.}
We add three types of rules for dealing with operators on labelled terms, 1) recursively applying them to subterms, 2) pushing big operators out and 3) eliminating generalized composition and bra-ket pairs.
%Other symbols on labelled Dirac notation are also 
Take the rule for conjugate $(D_1 \cdot D_2)^\dagger \ \reduce\ D_2^\dagger \cdot D_1^\dagger$ as an example of 1).
For 2), we use distributivity rules for scaling, labelled tensor and generalized composition. For example, rule
\[
% \textrm{(R-SUM-PUSHD0)}
X_1 \otimes \cdots (\sum_{i \in M} D) \cdots \otimes X_2\ \reduce\ \sum_{i \in M} (X_1 \otimes \cdots D \cdots \otimes X_n)
\]
will lift summation to the outside.
Extra rules for 3) are established including:
% The final step operates on sum and dot product. They will lift summation to the outside, and eliminate the bra-ket pairs whenever possible.
\begin{align*}
    % & \textrm{(R-SUM-PUSHD0)}
    % && X_1 \otimes \cdots (\sum_{i \in M} D) \cdots \otimes X_2\ \reduce\ \sum_{i \in M} (X_1 \otimes \cdots D \cdots \otimes X_n) \\
    % %
    % & \textrm{(R-SUM-PUSHD1)}
    % && (\sum_{i \in M} D_1) \cdot D_2 \ \reduce\ \sum_{i \in M} (D_1 \cdot D_2) \\
    %
    & \textrm{(R-L-SORT0)}
    && A : \DType(s_1, s_2), B : \DType(s_1', s_2'), s_2 \cap s_1'=\emptyset \Rightarrow A \cdot B \ \reduce\ A \otimes B \\
    & \textrm{(R-L-SORT1)}
    && {}_r\bra{i}\cdot\ket{j}_r \ \reduce\ \delta_{i, j} \\
    & \textrm{(R-L-SORT2)}
    && {}_r\bra{i}\cdot(Y_1 \otimes \cdots \otimes \ket{j}_r \otimes \cdots \otimes Y_m) \ \reduce\ \delta_{i, j}.(Y_1  \otimes \cdots \otimes Y_m)
\end{align*}
% These rules are added to the rewriting system in~\Cref{sec: decide} and executed together.
Assuming no variables of $\DType(s_1, s_2)$, 
repeating the application of the above rules yield the normal form of both sides of the equality--the addition of big operators: 
\begin{equation}
  \label{eqn:Labelled normal}
  \sum_{i}\cdots\sum_{j} a_1 . (\ket{i}_{p} \otimes \cdots \otimes \bra{j}_{q})
  + \cdots +
  \sum_{k}\cdots\sum_{l} a_m . (\ket{k}_{r} \otimes \cdots \otimes \bra{l}_{s})
\end{equation}
where each sum body is a tensor of labelled basis with scalar coefficients in plain Dirac notation.
\paragraph*{3. Ordering and elimination of quantum variables.} 
We further sort tensors of labelled basis in every sum body of Eqn. (\ref{eqn:Labelled normal}) by 1. ket first and 2. the default order of variables. This yield the same shape of every subterm on both sides of the equality, e.g., 
\begin{equation}
  \label{eqn:Labelled normal1}
  \sum_{i}\cdots\sum_{j}\cdots \sum_{i'}\cdots\sum_{j'} a_1 . ((\ket{i}_p \otimes \cdots \otimes \ket{j}_q) \otimes (\bra{i'}_{p'}\otimes\cdots\otimes \bra{j'}_{q'}))
\end{equation}
and thus it is equivalent to prove the equivalence of additions of subterms of:
\begin{equation}
  \label{eqn:Labelled normal2}
  \sum_{i}\cdots\sum_{j}\cdots \sum_{i'}\cdots\sum_{j'} a_1 . ((\ket{i} \otimes \cdots \otimes \ket{j}) \cdot (\bra{i'}\otimes\cdots\otimes \bra{j'}))
\end{equation}
which do not involve any labels. 

%Define $T_{s,s'}$ as $\SType$ if $s=s'=\emptyset$, $\KType(\sigma_s)$ if $s'=\emptyset$, $\BType(\sigma_{s'})$ if $s=\emptyset$ and $\OType(\sigma_s,\sigma_{s'})$ otherwise (see Sec. \ref{context: sigma s}). We claim the soundness of our normalization as:
% Then if the original equality is $\Gamma\vdash D_1 = D_2 : \cD(s,s')$, what remains to be check $\Gamma\vdash e_1 = e_2 : T_{s,s'}$.
\paragraph*{}

The procedure to eliminate labels is sound and complete, in the sense
that expressions in Labelled Dirac notation are equivalent iff their
translations to plain Dirac notation are equivalent.
\begin{theorem}[Label Elimination]
  \label{thm: normalization}
  Assume $\Gamma\vdash D_1 : \cD(s,s')$, $\Gamma\vdash D_2 :
  \cD(s,s')$ and no variables of $\DType(\cdot,\cdot)$ appear in
  $D_1,D_2$.  Let $e_1 = e_2$ be obtained by above normalization
  procedure on $D_1 = D_2$.  Then $e_1=e_2$ is an equation in plain
  Dirac notation and $\Gamma\vDash D_1 = D_2$ if and only if
  $\Gamma\vDash e_1 = e_2$.
\end{theorem}
The idea of the proof is as follows: first, we define a set of proof
rules to rewrite every labelled Dirac notation into the form of
Eqn. (1). We prove that each rule is sound w.r.t.\, semantics, and
that every labelled Dirac expression can be rewritten to an expression
of the form of Eqn. (1); the proof is by induction on the structure of
the expression. Next, we show that reordering of labelled basis
preserves the type and semantics and thus yield expressions of the
form of Eqn. (4), as desired. Details appear in~\Cref{app:label_elim}.

%% file: 05tool_eval.tex
\section{Implementation and Case Study}
We present \textbf{D-Hammer}, an open-source
and publicly available\footnote{\texttt{https://github.com/LucianoXu/D-Hammer}}
implementation of our approach.  \textbf{D-Hammer} is an
equational prover for Labelled Dirac notation written in \CC.  It
features a parser built using ANTLR4, and scalar reasoning is powered
by the Mathematica Engine. Users can use commands to make definitions
and assumptions in the maintained context, conduct the normalization
and equivalence checking, and obtain the rewriting trace output.
D-Hammer can be run interactively from the command line or integrated
into other \CC\ projects as a library.

\subsubsection{Structure and Mechanism} 
The project consists of the following components:
\begin{itemize}
    \item \texttt{antlr4}: A third-party library for building the parser.
    \item \texttt{WSTP interface}: A wrapper to link with Mathematica Engine.
    \item \texttt{ualg}: The framework module for universal algebra, defining basic concepts like terms and substitutions.
    \item \texttt{dhammer}: The main module containing symbols definitions, type checking, rewriting rules, normalization algorithm and the prover.
    \item \texttt{example}: An example benchmark for evaluation.
    \item \texttt{toplevel}: The command line application.
\end{itemize}
% The project structure is illustrated in~\Cref{fig: dhammer structure}.
% \texttt{ualg} is the module for universal algebra, defining basic concepts like terms and substitutions. It serves as the library for \texttt{dhammer}, which are then utilized in the example benchmarks and the toplevel command line application. Important components of \texttt{dhammer} are:

% \input{fig/proj_structure.tex}

The internal data structure for terms follows a pointer-based syntax tree, using the function application style:
\[
    \texttt{
        term ::= ID | ID [term (, term)*].
    }
\]
The syntax tree can either be an identifier, or an application with an identifier as the function head, and several syntax trees as arguments. Below are several examples of Dirac notation terms and their corresponding syntax trees.
\footnotesize{
\begin{align*}
    & X_1 + X_2 + X_3 && \texttt{ADD[X1, X2, X3]} 
    \\
    & \lambda x: \OType(T_1,  T_2). x^\dagger && \texttt{FUN[x, OTYPE[T1, T2], ADJ[x]]}
    \\
    & \sum_{i \in \mathbf{U}(T)} \ket{i} \bra{i} && \texttt{SUM[USET[T], FUN[i, BASIS[T], OUTER[KET[i], BRA[i]]]]}
\end{align*}
}
% The syntax tree structure is also compatible with the datatype of Mathematica. This improves the interoperability between D-Hammer and the Mathematica system, enabling them to work interleavingly.
To improve usability, D-Hammer also supports many special notations for terms, and most Dirac notation terms is encoded in the natural, intuitive way.
Here are some examples for the parsing syntax.

\begin{figure}
    \center
\begin{tabular}{c >{\centering\arraybackslash}p{4cm} l}
    \hline
    syntax & parsing result & explanation \\
    \hline
    \texttt{|e>} & \texttt{KET[e]} & the ket basis\\
    \texttt{e1 + ... + en} & \texttt{ADD[e1, ..., en]} & the addition\\
    \texttt{e1\ e2} & \texttt{COMPO[e1, e2]} & composition in Dirac notation \\
    \texttt{e1\^{}*} & \texttt{CONJ[e1]} & scalar conjugation \\
    \texttt{fun i : T => X} & \texttt{FUN[i, T, X]} & lambda abstraction \\
    \hline
\end{tabular}
\end{figure}

Finally, D-Hammer uses a prover to host the computation. The prover maintains a well-formed context $\Gamma$, and processes commands to modify the context and conduct calculations. The commands are listed below.
\begin{itemize}
    \item \texttt{\textcolor{NavyBlue}{Def} ID := term.} It defines the \texttt{ID} as the \texttt{term}, using the \textbf{W-Def} typing rule.
    \item \texttt{\textcolor{NavyBlue}{Var} ID := term.} It make an assumption of \texttt{ID} with the \texttt{term} as type, using the \textbf{W-Assume} typing rules.
    \item \texttt{\textcolor{NavyBlue}{Check} term.} Type checking the \texttt{term} and output the result.
    \item \texttt{\textcolor{NavyBlue}{Normalize} term.} Normalize the \texttt{term} using the algorithm introduced in~\Cref{sec: decide}.
    \item \texttt{\textcolor{NavyBlue}{CheckEq} term \textcolor{NavyBlue}{with} term.} Check the equivalence of the two terms calculating and comparing their normal forms.
\end{itemize}
The prover will type check the terms for each command. We can also use \texttt{\textcolor{NavyBlue}{Normalize} term \textcolor{NavyBlue}{with trace}.} to output the proof trace during normalization. The proof trace is a sequence of records, including the rule or transformation appied, the position of application, and the pre- and post-transformation terms. The record helps understand the normalization procedure better, and can be turned into verified proofs in theorem provers in the future.

\subsubsection{Use Case}
As a tutorial, we encode the motivating~\Cref{ex: motivating}, examine and explain how to check it using D-Hammer. The encoding is shown below.

    \begin{lstlisting}[style=dhammer]
Var T : INDEX. Var M : OTYPE[T, T].
Def phi := idx T => Sum nv in USET[T], |(nv, nv)>.
Var r1 : REG[T]. Var r2 : REG[T].
CheckEq M_r1 (phi T)_(r1, r2) with (TPO T T M)_r2 (phi T)_(r1, r2).
    \end{lstlisting}        

The first three lines use the \texttt{\textcolor{NavyBlue}{Var}} and \texttt{\textcolor{NavyBlue}{Def}} commands to set up the context for the Dirac notation.
\texttt{T} is a type index, representing arbitrary Hilbert space types. \texttt{M} is assumed to be an operator in the Hilbert space with type \texttt{T}. \texttt{phi} is defined as the maximally entangled state, depending on the bound variable \texttt{T} as index.
\texttt{r1} and \texttt{r2} are register names for the two subsystems.

In the left-hand side of \texttt{\textcolor{NavyBlue}{CheckEq}} command, \texttt{M\_r1} denotes the labelled notation $M_{r_1}$, and \texttt{(phi T)\_(r1, r2)} denotes the entangled state $\ket{\Phi}_{(r_1, r_2)}$. They are connected by a white space, which is parsed into the composition of Dirac notation, and will be reduced into the operator-ket multiplication after typing. The right hand side is interpreted similarly, except the defined symbol \texttt{TPO} in the context:

\begin{lstlisting}[style=dhammer]
Def TPO := idx sigma => idx tau => fun O : OTYPE[sigma, tau] => Sum i in USET[sigma], Sum j in USET[tau], (<i| O |j>).(|j> <i|).
\end{lstlisting}

The \texttt{TPO} symbol represents the transpose of operators, and encodes the formalization in~\Cref{ex: formalizing motivating}. Other commonly used concepts in Dirac notation are encoded and provided as defined symbols in D-Hammer.

Within one second, the prover reports the result of equivalence with their common normal form:
    \begin{lstlisting}[style=dhammer]
The two terms are equal.
[Normalized Term] SUM[USET[T], FUN[BASIS[T], SUM[USET[T], FUN[BASIS[T], SCR[DOT[BRA[$\texttt{\$1}$], MULK[M, KET[$\texttt{\$0}$]]], LTSR[LKET[$\texttt{\$1}$, r1], LKET[$\texttt{\$0}$, r2]]]]]]] : DTYPE[RSET[r1, r2], RSET]
    \end{lstlisting}

The normal form is in the internal syntax tree format mentioned above. A more readable interpretation is:
\[
\sum_{\mathbf{U}(T)} \sum_{\mathbf{U}(T)} \bra{\$1}M\ket{\$ 0} . \ket{\$1}_{r_1} \otimes \ket{\$0}_{r_2} : \DType(\{r_1, r_2\}, \emptyset).
\]
Here $\$0$ and $\$1$ are de Bruijn indices. The result is a ket on the $\{r_1, r_2\}$ system as expected, and follows pattern proposed in~\Cref{sec: labelled}.

\section{Evaluation}\label{sec:eval}
We evaluate D-Hammer on several example sets, and make a comparison with the previous tool DiracDec~\cite{diracdec}.
The experiments are carried out using a MacBook Pro with M3 Max chip. Results are summarized as follows, which indicates
significant performance improvements.

\begin{center}
    \begin{tabular}{c|c c c|c c c}
        \hline
        \multirow{2}{*}{source} & \multicolumn{3}{c|}{DiracDec} & \multicolumn{3}{c}{D-Hammer} \\
        \cline{2-7}
                                 & expressable & success & time(s)           & expressable & success & time(s)                 \\
        \hline
        textbook(QCQI)          & 18          & 18        &    1.02        &    18      & 18          &   0.82      \\
        CoqQ                    & 162          & 156       &    48.69       &   158     &  158   &     9.74     \\
        % others                  & 7          &  6         &   77.20    &    7        &   6     &  2.13     \\
        % circuits                 & 2          & 2       &    17.67       &   3     &  2   &     1.4     \\
        % research paper                & 4          & 4         &  59.53       &   4    & 4       &  0.73     \\
        % labelled Dirac notation      &   -         &   -          &     -           &      -      &        18       &     22.72    \\
        \hline
    \end{tabular}        
\end{center}

\subsubsection{Textbook (QCQI)}
As a warm-up, we consider 18 examples from Nielsen and Chuang's
classic texbook~\cite{nielsen2010quantum}. All examples can be encoded
in DiracDec and D-Hammer and are solved very efficiently.

\begin{figure}
    \centering
    \includegraphics[width=0.65\textwidth]{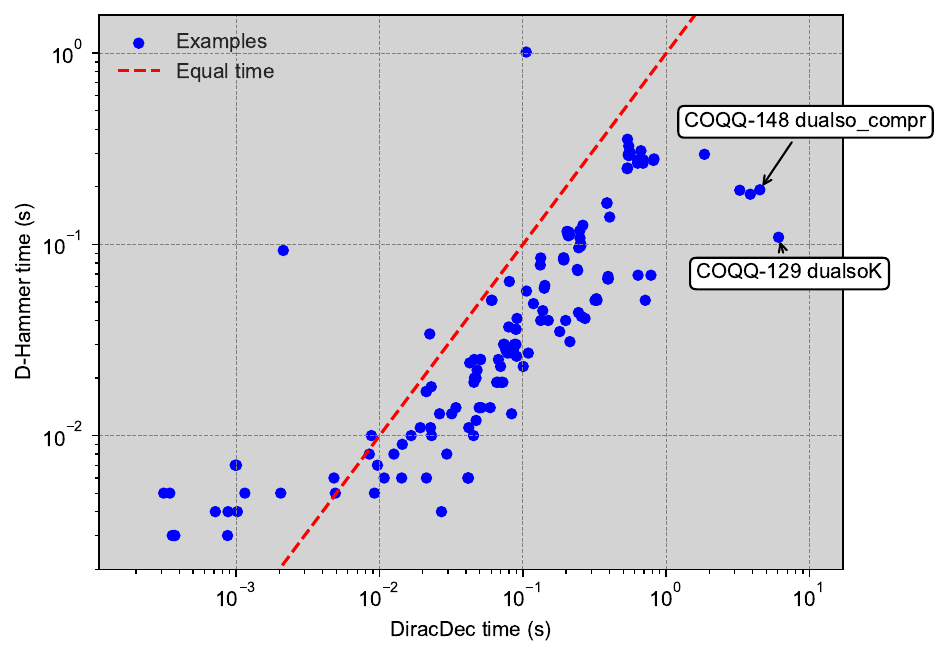}
    \caption{Time comparison between DiracDec and D-Hammer on the CoqQ benchmark.}
    \label{fig: CoqQ plot}
    \vspace{-0.4cm}
\end{figure}

\subsubsection{CoqQ}
As a more substantial example, we consider the examples from
CoqQ~\cite{Zhou2023}, an extensive formalization of quantum
information theory and quantum programming languages in the
Coq proof assistant.

CoqQ has been used as the main benchmark for evaluating DiracDec.
Specifically, \cite{diracdec} isolates 162 statements in CoqQ that
are in scope of the DiracDec language.

We have ported 158 out of 162 examples to D-Hammer. The remaining 4
examples uses projectors $\texttt{fst}$ and $\texttt{snd}$ on basis
pairs, where $\texttt{fst} (s, t) = s$ and $\texttt{snd} (s, t) =
t$. However, we found that this feature is rarely used and removed the
support in D-Hammer. Note that the omission of the projection rules
has a limited influence on the performance evaluation, because the
efficiency of the 158 examples is not affected by projection rules.

D-Hammer verifies the whole 158 examples in less than 10 seconds,
whereas DiracDec verifies 156/162 examples in more than 45 seconds.
\Cref{fig: CoqQ plot} shows a direct comparison of the efficiency of
the two tools. We observe that D-Hammer is slower than DiracDec on
small examples, due to marginal overhead, but becomes faster by an
average factor of 2 to 40 times as the running time of examples
increases. This non-linear growth suggests that efficiency gains
result from algorithmic improvements rather than the shift to a C++
implementation. Furthermore, examples with great improvements,
e.g. COQQ-129 and COQQ-148 shown in~\Cref{fig: CoqQ plot}, tend to use
deeply nested sums, for which our algorithm is more efficient.

% \subsubsection{Others}
% Other examples are taken from quantum circuits and one recent paper [TODO]. Because these examples involve decomposition on concrete $\ket{0}$ and $\ket{1}$ qubit basis, resulting a lot of addition elements, our algorithm dealing with AC symbols brings along significant improvement.
% One typical example is
% \begin{gather*}
%     (I \otimes P) \cdot U \cdot (I \otimes P) \cdot U^\dagger = \ket{0}\bra{0} \otimes I + \ket{1}\bra{1}\otimes (P \cdot P), \\
%     \textrm{where } P \triangleq e^{-i\theta/2} \ket{0}\bra{0} + e^{i\theta/2}\ket{1}\bra{1}, \qquad U \triangleq \ket{0}\bra{0} \otimes X + \ket{1}\bra{1} \otimes I.
% \end{gather*}
% It takes DiracDec about one minute, but D-Hammer solves it within one second.

\subsubsection{Labelled Dirac Notation}
We present a new set of examples for labelled Dirac notation (LDN), as illustrated in~\Cref{fig: labelled examples}. These examples include six representative cases drawn from various sources, such as well-established theorems, research paper results and quantum circuit equivalence.
D-Hammer successfully normalizes these examples and checks their equivalence using the algorithm outlined in~\Cref{sec: labelled}.
Among the examples, LDN-16 is a generalization of Example \ref{example1} and LDN-4 for Example \ref{ex: motivating}. LDN-12 shows the flexibility in combining labelled Dirac notations.
LDN-14 shows how to calculate controlled-not gate in different ways. 

A particularly noteworthy result is D-Hammer's solution to LDN-10, a highly complex and lengthy example. It is a theorem on quantum separation logic from~\cite{DBLP:conf/lics/ZhouBHYY21}, and proving it is challenging even for experts. Notably, it involves 7 registers, making it practically impossible to organize and referring to the subsystems without using labels.

\begin{figure}[h]
    \center
    % \scriptsize
    \setlength{\extrarowheight}{2pt}
    \begin{tabular}{l c c l}
        \hline
        example & source & time(s) & equation \\
        \hline
        LDN-4 & theorem & 0.03 & \( M_{r_1}\sum_{i}\ket{(i,i)}_{(r_1,r_2)} = M^T_{r_2}\sum_{i}\ket{(i,i)}_{(r_1, r_2)} \)\\
        LDN-10 & paper~\cite{DBLP:conf/lics/ZhouBHYY21} & 5.17 & \( \mathrm{tr}_{((a',(b,b')),c')}\Big[\mathrm{tr}_{r}\Big(U_{(r,(a,b))} \cdot\Big(|s\>_r\<s|\otimes 
        \Big[V_{((a',(b,b')),c')} \cdots \)\\
        LDN-11 & paper~\cite{PALSBERG2024206} & 0.12 & \( U_{(a,b)}\cdot W_{(b,c)}\cdot V_{(a,c)} = 
        \sum_i |i\>_a\<i|\otimes \big((P_i)_c\cdot W_{(b,c)}\cdot (Q_i)_c\big)\) \\
        LDN-12 & circuit & 0.07 & \( \ket{i}_{a;b}\bra{j} \cdot C_{(b,c)} \cdot D_{(c,d)} = {}_b\bra{j} \cdot C_{(b,c)} \cdot D_{(c,d)} \cdot \ket{i}_{a}\) \\
        LDN-14 & circuit & 7.37 & \scriptsize{\( \textsf{CNOT}_{rq}\ket{\textrm{GHZ}}_{pqr} = (\textsf{CNOT}\ket{00})_{rq}\ket{0}_p + (\textsf{CNOT}\ket{11})_{rq}\ket{1}_p \)} \\
        LDN-16 & theorem & 0.08 & \scriptsize{\( M_{prq}\ket{\textrm{GHZ}}_{prq} = N_{rq} \ket{\textrm{GHZ}}_{pqr}, M \triangleq \sum_{ij}\ket{ij}\bra{ij}\otimes U_{ij} \cdots \)} \\
        \hline
    \end{tabular}
    \caption{Part of examples for labelled Dirac notations. See~\Cref{sec: examples for labelled} for the full list.}
    \vspace{-0.2cm}
    \label{fig: labelled examples}
\end{figure}

\subsubsection*{Quantum circuits}
Quantum circuits is a prominent model of quantum computation. This
model is adopted by numerous tools, which can evaluate, optimize, or
prove equivalence of quantum circuits. These tools are based on a
variety of approaches, based on ZX-calculus~\cite{9868772}, or
decision diagrams~\cite{10.1145/3394885.3431590}, or other formalisms
discussed in \Cref{sec:relwork}. In general, these tools are aggressively
optimized to achieve scalability.

\begin{figure}[h]
    \begin{align*}
        & \textbf{(circuit)}\ &&
        \begin{quantikz}[scale=0.8]
            \qw & \gate{R_x(\frac{\pi}{2})} & \gate{R_x(\frac{\pi}{2})} & \qw \\
            \qw & \gate{T} & \qw & \qw \\
        \end{quantikz}
        =
        \begin{quantikz}[scale=0.8]
            \qw & \gate{H} & \gate{Z} & \gate{H} & \qw \\
            \qw & \gate{T} & \gate{H} & \gate{H} & \qw \\
        \end{quantikz} \\
        & \textbf{(notation)}\ &&
        \begin{aligned}
            & (\textsf{Rx}(\pi/2) \otimes \ONEO) \cdot (\ONEO \otimes \textsf{T}) \cdot (\textsf{Rx}(\pi/2) \otimes \ONEO) \\
            = & (\ONEO \otimes \textsf{H}) \cdot (\ONEO \otimes \textsf{H}) \cdot (\textsf{H} \otimes \ONEO) \cdot (\textsf{Z} \otimes \ONEO) \cdot (\textsf{H} \otimes \ONEO) \cdot (\ONEO \otimes \textsf{T})
        \end{aligned}
    \end{align*}
    \caption{The quantum circuit and Dirac notation encoding for example QC-5.}
    \label{fig: QC-5}
\end{figure}
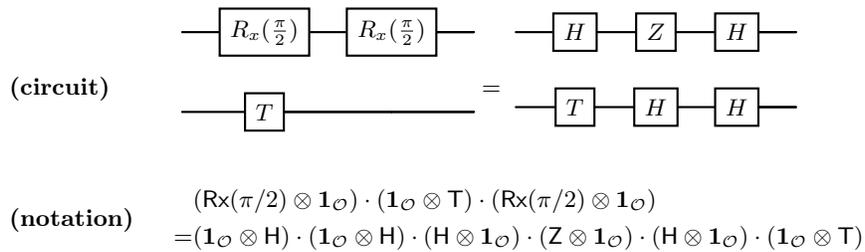

Quantum circuits can also be described as unitary operators in Dirac
notation. Therefore, D-Hammer can check the equivalence of quantum
circuits through their Dirac notation representations. Although it is
not the intended application of D-Hammer, we include an evaluation of
D-Hammer on some simple examples. An evaluation of some examples is
given in~\Cref{fig: circuit eval}, and~\Cref{fig: QC-5} shows one of
them. As expected, D-Hammer does not perform well in comparison with
specialized tools: it is always outperformed, often by several order
of magnitude, on small quantum circuit examples, and it always times
out on larger quantum circuit examples. In the future it would be of
interest to make D-Hammer more competitive on quantum circuits by
adopting some of the aggressive optimizations for other works.

\begin{figure}[h]
    \center
    % \scriptsize
    \setlength{\extrarowheight}{2pt}
    \begin{tabular}{c @{\hspace{2em}} c @{\hspace{2em}} c @{\hspace{2em}} c}
        \hline
        example & D-Hammer & ZX-Calculus~\cite{9868772} & Decision Diagrams (simulation)~\cite{10.1145/3394885.3431590} \\
        \hline
        QC-1 & 0.029 & 0.0 & 0.0 \\
        QC-2 & 0.16 & 0.00039 & 0.0038 \\
        QC-3 & 0.29 & 5.7e-5 & 0.0057 \\
        QC-4 & 0.013 & 4.3e-5 & 0.0017 \\
        QC-5 & 15 & 6.4e-5 & 0.0044 \\
        QC-6 & timeout & 0.00014 & 0.016 \\
        \hline
    \end{tabular}
    \caption{Time consumptions (in seconds) for quantum circuit equivalence checking using different tools.}
    \vspace{-0.2cm}
    \label{fig: circuit eval}
\end{figure}

\vspace{-2em}

\vspace{-2em}

%% file: 06related_conclusion.tex
\section{Related Work}\label{sec:relwork}

\paragraph*{Comparison with DiracDec}
Xu \emph{et al}~\cite{diracdec} define a language and an associate and
commutative rewriting system for Dirac notation, and implement their
language and rewriting system in Mathematica. Our approach follows a
similar pattern. However, there are significant differences in terms
of scope, expressiveness, and efficiency. 
% We review these differences below.

The most obvious difference is that D-Hammer targets labelled Dirac
notation, whereas DiracDec targets plain Dirac notation. As already
mentioned, the use of labelled Dirac notation is essential in many
applications; in particular, labelled Dirac notation can simplify
notation and proofs, and is in general better suited for writing and
reasoning about complex, many-body, quantum states. However, there are
several other important differences. First, our language leverages
higher-order functions to provide a compact and expressive
representation of big operators. Second, our language adopts AC
symbols with indefinite arities, which leads to compact representation
of terms, and eases AC reasoning. Third, the relation with Mathematica
is fundamentally different. DiracDec is implemented in Mathematica; as
a consequence, its behavior, and in particular, its typing rules are
constrained by the lack of typing in Mathematica. In contrast,
D-Hammer is designed as a separate tool; as a consequence, it benefits
from an improved representation of terms (e.g.\, AC operators with
indefinite arity), a more expressive type system (e.g.\, dependent
types), and a more efficient rewriting engine. D-Hammer still relies
on Mathematica to reason about functions that are not natively
supported by rewriting rules, but these interactions are constrained
and do not have negative effects on the overall efficiency of the
system. This is reflected in our experimental comparison of D-Hammer
and DiracDec, which shows how the former outperforms the latter.

\paragraph*{Comparison with ZX Calculus}
The ZX calculus~\cite{DBLP:conf/icalp/CoeckeD08,coecke2011interacting}
is a graphical calculus for quantum states. A main appeal of the ZX
calculus is that its foundations are grounded in categorical quantum
mechanics~\cite{DBLP:conf/lics/AbramskyC04}, a powerful framework for
modeling quantum physics. Another main appeal of the ZX calculus is
that it has a natural operational interpretation based on graph
rewriting. There is a large body of work that defines rewriting
systems for fragments/extensions/variants of the ZX calculus, and
studies their theoretical properties, in particular completeness (a
set of rules is complete if it can prove all valid identities) and
minimality (a complete set of rules is minimal if removing a rule
leads to incompleteness). Two main proof techniques for completeness
are via termination, or via interpretation. In the first case, one
shows that a rewriting system has unique normal forms and that two
expressions are semantically equivalent iff they have the same normal
form---there exists many relaxations of this result---whereas in the
second case one shows completeness by exhibiting a well-behaved
translation to another system for which completeness holds. Both proof
techniques have been used to prove completeness for multiple settings,
including the Clifford fragment~\cite{backens2014zx}, the
Toffoli+Hadamard fragment~\cite{hadzi15zx}, the Clifford+T
fragment~\cite{jeandel2018complete} and the (qubit) universal
fragment~\cite{HNW18,JeandelPV18beyond}. Subsequent works generalize
completeness to qudits~\cite{Poor2023}, quantum
circuits~\cite{DBLP:conf/lics/ClementHMPV23} or optical quantum
circuits~\cite{clement_et_al:2022}. We refer the interested reader
to~\cite{vandewetering2020zx} for a historical and technical account
of completeness results up to 2020.

There are many similarities between the ZX calculus and our formalism
for Labelled Dirac notation. However, there are also some differences
between our formalism and the ones used in ZX calculus. In particular,
our language supports big operators, tensors, and various operations
on Hilbert spaces. These features are typically not considered in
prior work on the ZX calculus, and as a consequence many examples
handled by D-Hammer lack an immediate translation into ZX-calculus.
This additional generality comes with a price. On the one hand, our
theoretical results are weaker: we do not claim completeness or
minimality. Similarly, practical implementations of the ZX-calculus~\cite{kissinger2020pyzx,9868772} outperform D-Hammer on examples
that can be handled by both tools, as shown in \Cref{sec:eval}.

\paragraph*{Comparison with other tools}
Beyond ZX calculus, there exists many other tools for simplifying and
proving equivalence of quantum circuits; we refer the reader to recent
surveys~\cite{QPV-survey2021,CharetonLVVBX23} for detailed accounts.
Notable works include~\cite{amy2018towards}, which uses the path-sums
formalism to check circuits with 1,000 of T gates,
and~\cite{AutoQ_pldi_2023,AutoQ2023,AutoQ_popl2025}, which uses
automata-based approaches to verify quantum circuits at scale---their
tool \textsc{autoq} is able to verify circuits with over 100,00 gates,
and was recently extended to support parametrized verification.

\paragraph*{Canonical forms in multi-body quantum physics}
Canonical forms play a fundamental role in quantum physics. For
instance, \cite{perezgarcia2007,RevModPhys.93.045003,Acuaviva_2023}
discuss canonical forms of Matrix Product States (MPS) and Tensor
Networks respectively. An exciting direction for further work is to
further develop automated deduction techniques for quantum physics.

\paragraph*{Comparison with egraphs}
Our algorithm is based on term rewriting. However, it is a challenge
to device well-behaved and efficient term rewriting for (labelled)
Dirac notation. An alternative to term rewriting would be to use
equality saturation~\cite{DBLP:journals/pacmpl/WillseyNWFTP21}, a
powerful equational reasoning technique that does not require
existence of normal forms. Equality saturation may be particularly
useful when considering further extensions of labelled Dirac notation.

\section{Conclusion and Future Work}
We have designed and implemented D-Hammer, a dependently typed
higher-order language and proof system for labelled Dirac
notation. D-Hammer benefits from an optimized implementation in
\CC\ and a tight integration with Mathematica to reason about a broad
range of mathematical functions, including trigonometric and
exponential functions, that are commonly used in quantum physics.
There are two important directions for future work. 
% The first
% direction is to enhance D-Hammer with support for big tensor notation,
% i.e.\, to support expressions of the form $\bigotimes_{i\in I}
% e_i$. 
The first direction is to extend D-Hammer with a mechanism to
generate independently verifiable certificates.  There is a large body
of work on producing certificates for automated tools, in particular
SMT solvers; see e.g.~\cite{DBLP:journals/cacm/BarbosaBCDKLNNOPRTZ23}
for a recent overview. One potential option would be to integrate
D-Hammer with the Coq or Lean proof assistants; in the first case, one
would benefit from the formalization of labelled Dirac notation in
CoqQ~\cite{Zhou2023}, whereas in the second case, one would benefit
from powerful mechanisms to integrate rewriting procedures into the
Lean proof assistant. The second direction is to connect
D-Hammer with quantum program verifiers. Two potential applications
are automating equational proofs for tools that already use Dirac
notation, and to substitute numerical methods for tools that use
matrices instead of symbolic assertions.

\subsection*{Acknowledgements}
\small
We thank the anonymous reviewers for their constructive feedback that helped improve this paper. We also extend our gratitude to Jam Kabeer Ali Khan and Ivan Ariel Renison for their work of testing of D-Hammer.

\normalsize

%% Disclosure of Interests
\subsubsection*{Disclosure of Interests}
\small
This research was supported by the National Key R\&D Program of China under Grant No. 2023YFA1009403.
\normalsize

%% Based on the first equational reasoning tool for Dirac notation called DiracDec, this work improves and extends the theory for practical applications, and provides the solver D-Hammer. Experiments show that the tool demonstrates advantages in decidability, efficiency and usability. 

%% We expect D-Hammer to have applications in areas like quantum program verification or proofs of post-quantum cryptography protocols in the future.
%% One promising following up is to connect D-Hammer with theorem provers like Coq. It involves transforming theorem prover expressions into D-Hammer, and verify the proof trace of D-Hammer in theorem provers. Besids, most quantum program verifiers nowadays depend on matrix calculations. D-Hammer can serve as the replacement for matrix methods to enable symbolic deductions.

%% file: appendix.tex
\begin{center} {\Large \textbf{Appendix}} \end{center}

\section{Full Typing Rules}

\label{sec: full typing rules}

This section includes the full list of typing rules.

\begin{itemize}
    \item Rules for a well-formed context.
    \[
        \textbf{W-Empty} \qquad
        \frac{}{\WF([])}
    \]
    \[
        \textbf{W-Assum-Index} \qquad
        \frac{\WF(\Gamma) \qquad}{\WF(\Gamma; x : \Index)}
    \]
    \[
        \textbf{W-Assum-Term} \qquad
        \frac{\Gamma \vdash T : \Type}{\WF(\Gamma; x:T)}
    \]
    \[
        \textbf{W-Def-Term} \qquad
        \frac{\Gamma \vdash t:T \qquad x \notin \Gamma}{\WF(\Gamma; x:=t:T)}
    \]

    \item Rules for type indices.
    \[
        \textbf{Index-Var} \qquad
        \frac{\WF(\Gamma) \qquad x : \Index \in \Gamma}{\Gamma \vdash x : \Index}
    \]
    \[
        \textbf{Index-Prod} \qquad
        \frac{\Gamma \vdash \sigma : \Index \qquad \Gamma \vdash \tau : \Index}{\Gamma \vdash \sigma \times \tau : \Index}
    \]
    \[
        \textbf{Index-Qudit} \qquad
        \frac{\WF(\Gamma)}{\Gamma \vdash \mathsf{bool} : \Index}
    \]

    \item Rules for types.
    \[
        \textbf{Type-Lam} \qquad
        \frac{\Gamma \vdash T : \Type \qquad \Gamma \vdash U : \Type}{\Gamma \vdash T \to U : \Type}
    \]
    \[
        \textbf{Type-Index} \qquad
        \frac{\Gamma; x : \Index \vdash U : \Type}{\Gamma \vdash \forall x.U : \Type}
    \]
    \[
        \textbf{Type-Basis} \qquad
        \frac{\Gamma \vdash \sigma : \Index}{\Gamma \vdash \Basis(\sigma) : \Type}
    \]
    \[
        \textbf{Type-Ket} \qquad
        \frac{\Gamma \vdash \sigma : \Index}{\Gamma \vdash \KType(\sigma) : \Type}
    \]
    \[
        \textbf{Type-Bra} \qquad
        \frac{\Gamma \vdash \sigma : \Index}{\Gamma \vdash \BType(\sigma) : \Type}
    \]
    \[
        \textbf{Type-Opt} \qquad
        \frac{\Gamma \vdash \sigma : \Index \qquad \Gamma \vdash \tau : \Index}{\Gamma \vdash \OType(\sigma, \tau) : \Type}
    \]
    \[
        \textbf{Type-Scalar} \qquad
        \frac{\WF(\Gamma)}{\Gamma \vdash \SType : \Type}
    \]
    \[
        \textbf{Type-Set} \qquad
        \frac{\Gamma \vdash \sigma : \Index}{\Gamma \vdash \SET(\sigma) : \Type}
    \]
    \[
        \textbf{Type-Register} \qquad
        \frac{\Gamma \vdash  \sigma : \Index}{\Gamma \vdash \reg(\sigma) : \Type}
    \]
    \[
        \textbf{Type-Labelled} \qquad
        \frac{\Gamma \vdash r : \reg(\sigma_r) \text{ for all $r$ in $s_1$ and $s_2$} }{\Gamma \vdash \DType(s_1, s_2) : \textsf{Type}} 
    \]

    \item Rules for variable and function typings. Here $U\{x/u\}$ means replacing the bound variable $x$ with $u$ in $U$.
    \[
        \textbf{Term-Var} \qquad
        \frac{
            \begin{aligned}
                & \WF(\Gamma) \\
                & (x:T) \in \Gamma \text{ or } (x := t : T) \in \Gamma \text{ for some $t$}
            \end{aligned}
        }
        {\Gamma \vdash x : T}
    \]
    \[
        \textbf{Lam} \qquad
        \frac{\Gamma; x : T \vdash t : U}{\Gamma \vdash (\lambda x : T . t) : T \to U}
    \]
    \[
        \textbf{Index} \qquad
        \frac{\Gamma; x : \Index \vdash t : U}{\Gamma \vdash (\lambda x : T.t) : \forall x.U}
    \]
    \[
        \textbf{App-Lam} \qquad
        \frac{\Gamma \vdash t:U \to T \qquad \Gamma \vdash u:U}{\Gamma \vdash (t\ u):T}
    \]
    \[
        \textbf{App-Index} \qquad
        \frac{\Gamma \vdash t:\forall x.U \qquad \Gamma \vdash u : \Index}{\Gamma \vdash (t\ u):U\{x/u\}}
    \]
    
    \item Basis term typing rules. 
    \[  \textbf{Basis-0} \qquad
        \frac{\WF(\Gamma)}{\Gamma \vdash 0 : \Basis(\mathsf{bool})}
    \]
    \[  \textbf{Basis-1} \qquad
        \frac{\WF(\Gamma)}{\Gamma \vdash 1 : \Basis(\mathsf{bool})}
    \]
    \[
        \textbf{Basis-Pair} \qquad
        \frac{\Gamma\vdash s : \Basis(\sigma) \qquad \Gamma\vdash t : \Basis(\tau)} {\Gamma\vdash (s, t) : \Basis(\sigma \times \tau)}
    \]

    \item Composition typing rules. 
    \[  \textbf{Compo-SS} \qquad
        \frac{\Gamma \vdash x : \SType \qquad \Gamma \vdash y : \SType}{\Gamma \vdash x \circ y : \SType}
    \]
    \[  \textbf{Compo-SK} \qquad
        \frac{\Gamma \vdash x : \SType \qquad \Gamma \vdash y : \KType(\sigma)}{\Gamma \vdash x \circ y : \KType(\sigma)}
    \]
    \[  \textbf{Compo-SB} \qquad
        \frac{\Gamma \vdash x : \SType \qquad \Gamma \vdash y : \BType(\sigma)}{\Gamma \vdash x \circ y : \BType(\sigma)}
    \]
    \[  \textbf{Compo-SO} \qquad
        \frac{\Gamma \vdash x : \SType \qquad \Gamma \vdash y : \OType(\sigma, \tau)}{\Gamma \vdash x \circ y : \OType(\sigma, \tau)}
    \]
    \[  \textbf{Compo-KS} \qquad
        \frac{\Gamma \vdash x : \KType(\sigma) \qquad \Gamma \vdash y : \SType}{\Gamma \vdash x \circ y : \KType(\sigma)}
    \]
    \[  \textbf{Compo-KK} \qquad
        \frac{\Gamma \vdash x : \KType(\sigma) \qquad \Gamma \vdash y : \KType(\tau)}{\Gamma \vdash x \circ y : \KType(\sigma \times \tau)}
    \]
    \[  \textbf{Compo-KB} \qquad
        \frac{\Gamma \vdash x : \KType(\sigma) \qquad \Gamma \vdash y : \BType(\tau)}{\Gamma \vdash x \circ y : \OType(\sigma, \tau)}
    \]
    \[  \textbf{Compo-BS} \qquad
        \frac{\Gamma \vdash x : \BType(\sigma) \qquad \Gamma \vdash y : \SType}{\Gamma \vdash x \circ y : \BType(\sigma)}
    \]
    \[  \textbf{Compo-BK} \qquad
        \frac{\Gamma \vdash x : \BType(\sigma) \qquad \Gamma \vdash y : \KType(\sigma)}{\Gamma \vdash x \circ y : \SType}
    \]
    \[  \textbf{Compo-BB} \qquad
        \frac{\Gamma \vdash x : \BType(\sigma) \qquad \Gamma \vdash y : \BType(\tau)}{\Gamma \vdash x \circ y : \BType(\sigma \times \tau)}
    \]
    \[  \textbf{Compo-BO} \qquad
        \frac{\Gamma \vdash x : \BType(\sigma) \qquad \Gamma \vdash y : \OType(\sigma, \tau)}{\Gamma \vdash x \circ y : \BType(\tau)}
    \]
    \[  \textbf{Compo-OS} \qquad
        \frac{\Gamma \vdash x : \OType(\sigma, \tau) \qquad \Gamma \vdash y : \SType}{\Gamma \vdash x \circ y : \OType(\sigma, \tau)}
    \]
    \[  \textbf{Compo-OK} \qquad
        \frac{\Gamma \vdash x : \OType(\sigma, \tau) \qquad \Gamma \vdash y : \KType(\tau)}{\Gamma \vdash x \circ y : \KType(\sigma)}
    \]
    \[  \textbf{Compo-OO} \qquad
        \frac{\Gamma \vdash x : \OType(\sigma, \tau) \qquad \Gamma \vdash y : \OType(\tau, \rho)}{\Gamma \vdash x \circ y : \OType(\sigma, \rho)}
    \]
    \[
        \textbf{Compo-DD} \qquad
        \frac{
            \begin{aligned}
                \Gamma \vdash x : \DType(s_1,s_1') \\
                \Gamma \vdash y : \DType(s_2,s_2')
            \end{aligned}
            \qquad 
            \begin{aligned}
                s_1 \cap s_2 \backslash s_1' = \emptyset \\
                s_2' \cap s_1' \backslash s_2 = \emptyset
            \end{aligned}
        }
        {\Gamma \vdash x\circ y : \DType(s_1 \cup (s_2\backslash s_1'), s_2' \cup (s_1'\backslash s_2))}
    \]

    \item Scalar term typing rules.
    \[
        \textbf{Sca-0} \qquad
        \frac{\WF(\Gamma)}{\Gamma \vdash 0 : \SType}
    \]
    \[
        \textbf{Sca-1} \qquad
        \frac{\WF(\Gamma)}{\Gamma \vdash 1 : \SType}
    \]
    \[
        \textbf{Sca-Delta} \qquad
        \frac{ \Gamma\vdash s : \Basis(\sigma) \qquad \Gamma\vdash t : \Basis(\sigma) } {\Gamma \vdash \delta_{s, t} : \SType}
    \]
    \[
        \textbf{Sca-Add} \qquad
        \frac{\Gamma\vdash a_i : \SType \text{ for all $i$}}{\Gamma\vdash a_1 + \cdots + a_n : \SType}
    \]
    \[
        \textbf{Sca-Mul} \qquad
        \frac{\Gamma\vdash a_i : \SType \text{ for all $i$}}{\Gamma\vdash a_1 \times \cdots \times a_n : \SType}
    \]
    \[
        \textbf{Sca-Conj} \qquad
        \frac{\Gamma \vdash a : \SType}{\Gamma \vdash a^*:\SType}
    \]
    \[
        \textbf{Sca-Dot} \qquad
        \frac{\Gamma\vdash B : \BType(\sigma) \qquad \Gamma\vdash K : \KType(\sigma)}{\Gamma \vdash B \cdot K : \SType}
    \]
    \[
        \textbf{Sca-Sum} \qquad
        \frac{\Gamma \vdash s : \SET(\sigma) \qquad \Gamma \vdash f : \Basis(\sigma) \to \SType}{\Gamma \vdash \sum_{s} f : \SType}
    \]

    \item Ket term typing rules.
    \[
        \textbf{Ket-0} \qquad
        \frac{\Gamma \vdash \sigma : \Index}{\Gamma \vdash \ZEROK(\sigma) : \KType(\sigma)}
    \]
    \[
        \textbf{Ket-Basis} \qquad
        \frac{\Gamma\vdash t : \Basis(\sigma)}{\Gamma \vdash \ket{t} : \KType(\sigma)}
    \]
    \[
        \textbf{Ket-Adj} \qquad
        \frac{\Gamma \vdash B : \BType(\sigma)}{\Gamma \vdash B^\dagger : \KType(\sigma)}
    \]
    \[
        \textbf{Ket-Scr} \qquad
        \frac{\Gamma \vdash a : \SType \qquad \Gamma \vdash K : \KType(\sigma)}{\Gamma \vdash a.K : \KType(\sigma)}
    \]
    \[
        \textbf{Ket-Add} \qquad
        \frac{\Gamma \vdash K_i : \KType(\sigma) \text{ for all $i$}}{\Gamma \vdash K_1 + \cdots + K_n : \KType(\sigma)}
    \]
    \[
        \textbf{Ket-MulK} \qquad
        \frac{\Gamma \vdash O : \OType(\sigma, \tau) \qquad \Gamma \vdash K : \KType(\tau)}{\Gamma \vdash O \cdot K : \KType(\sigma)}
    \]
    \[
        \textbf{Ket-Tsr} \qquad
        \frac{\Gamma \vdash K_1 : \KType(\sigma) \qquad \Gamma \vdash K_2 : \KType(\tau)} {\Gamma \vdash K_1 \otimes K_2 : \KType(\sigma \times \tau)}
    \]
    \[
        \textbf{Ket-Sum} \qquad
        \frac{\Gamma \vdash s : \SET(\sigma) \qquad \Gamma \vdash f : \Basis(\sigma) \to \KType(\tau)}{\Gamma \vdash \sum_{s} f : \KType(\tau)}
    \]

    \item Bra term typing rules.
    \[
        \textbf{Bra-0} \qquad
        \frac{\Gamma \vdash \sigma : \Index}{\Gamma \vdash \ZEROB(\sigma) : \BType(\sigma)}
    \]
    \[
        \textbf{Bra-Basis} \qquad
        \frac{\Gamma\vdash t : \Basis(\sigma)}{\Gamma \vdash \bra{t} : \BType(\sigma)}
    \]
    \[
        \textbf{Bra-Adj} \qquad
        \frac{\Gamma \vdash K : \KType(\sigma)}{\Gamma \vdash K^\dagger : \BType(\sigma)}
    \]
    \[
        \textbf{Bra-Scr} \qquad
        \frac{\Gamma \vdash a : \SType \qquad \Gamma \vdash B : \BType(\sigma)}{\Gamma \vdash a.B : \BType(\sigma)}
    \]
    \[
        \textbf{Bra-Add} \qquad
        \frac{\Gamma \vdash B_i : \BType(\sigma) \text{ for all $i$}}{\Gamma \vdash B_1 + \cdots + B_n : \BType(\sigma)}
    \]
    \[
        \textbf{Bra-MulB} \qquad
        \frac{\Gamma \vdash B : \KType(\sigma) \qquad \Gamma \vdash O : \OType(\sigma, \tau)}{\Gamma \vdash B \cdot O : \BType(\tau)}
    \]
    \[
        \textbf{Bra-Tsr} \qquad
        \frac{\Gamma \vdash B_1 : \BType(\sigma) \qquad \Gamma \vdash B_2 : \BType(\tau)} {\Gamma \vdash B_1 \otimes B_2 : \BType(\sigma \times \tau)}
    \]
    \[
        \textbf{Bra-Sum} \qquad
        \frac{\Gamma \vdash s : \SET(\sigma) \qquad \Gamma \vdash f : \Basis(\sigma) \to \BType(\tau)}{\Gamma \vdash \sum_{s} f : \BType(\tau)}
    \]

    \item Operator term typing rules.
    \[
        \textbf{Opt-0} \qquad
        \frac{\Gamma \vdash \sigma : \Index \qquad \Gamma \vdash \tau : \Index}{\Gamma \vdash \ZEROO(\sigma, \tau) : \OType(\sigma, \tau)}
    \]
    \[
        \textbf{Opt-1} \qquad
        \frac{\Gamma \vdash \sigma : \Index}{\Gamma \vdash \ONEO(\sigma) : \OType(\sigma, \sigma)}
    \]
    \[
        \textbf{Opt-Adj} \qquad
        \frac{\Gamma \vdash O : \OType(\sigma, \tau)}{\Gamma \vdash O^\dagger : \OType(\tau, \sigma)}
    \]
    \[
        \textbf{Opt-Scr} \qquad
        \frac{\Gamma \vdash a : \SType \qquad \Gamma \vdash O : \OType(\sigma, \tau)}{\Gamma \vdash a.O : \OType(\sigma, \tau)}
    \]
    \[
        \textbf{Opt-Add} \qquad
        \frac{\Gamma \vdash O_i : \OType(\sigma, \tau) \text{ for all $i$}}{\Gamma \vdash O_1 + \cdots + O_n : \OType(\sigma, \tau)}
    \]
    \[
        \textbf{Opt-Outer} \qquad
        \frac{\Gamma\vdash K : \KType(\sigma) \qquad \Gamma\vdash B : \BType(\tau)}{\Gamma\vdash K \cdot B : \OType(\sigma, \tau)}
    \]
    \[
        \textbf{Opt-Mulo} \qquad
        \frac{\Gamma \vdash O_1 : \OType(\sigma, \tau) \qquad \Gamma \vdash O_2 : \OType(\tau, \rho)}{\Gamma \vdash O_1 \cdot O_2 : \OType(\sigma, \rho)}
    \]
    \[
        \textbf{Opt-Tsr} \qquad
        \frac{\Gamma \vdash O_1 : \OType(\sigma_1, \tau_1) \qquad \Gamma \vdash O_2 : \OType(\sigma_2, \tau_2)} {\Gamma \vdash O_1 \otimes O_2 : \OType(\sigma_1 \times \sigma_2, \tau_1 \times \tau_2)}
    \]
    \[
        \textbf{Opt-Sum} \qquad
        \frac{\Gamma \vdash s : \SET(\sigma) \qquad \Gamma \vdash f : \Basis(\sigma) \to \OType(\tau, \rho)}{\Gamma \vdash \sum_{s} f : \OType(\tau, \rho)}
    \]

    \item Set term typing rules.
    \[
        \textbf{Set-U} \qquad
        \frac{\Gamma \vdash \sigma : \Index}{\Gamma \vdash \mathbf{U}(\sigma) : \SET(\sigma)}
    \]
    \[
        \textbf{Set-Prod} \qquad
        \frac{\Gamma \vdash A : \SET(\sigma) \qquad \Gamma \vdash B : \SET(\tau)}{\Gamma \vdash A \star B : \SET(\sigma \times \tau)}
    \]

    \item Register term typing rules.
    \[
        \textbf{Reg-Var} \qquad
        \frac{\WF(\Gamma) \qquad r : \reg(\sigma) \in \Gamma}{\Gamma \vdash r : \reg(\sigma)}
    \]
    \[
        \textbf{Reg-Pair}\qquad
        \frac{
            \begin{aligned}
                \Gamma \vdash R : \reg(\sigma) \\
                \Gamma \vdash Q : \reg(\tau)
            \end{aligned}
            \qquad \var(R) \cap \var(Q) = \emptyset
        }{\Gamma \vdash (R,Q) : \reg(\sigma \times \tau)}
    \]

    \item Typing rules for labelled Dirac notation.
    \[
        \textbf{L-Basis-Ket}\qquad 
        \frac{r : \reg(\sigma) \in \Gamma\qquad \Gamma \vdash i : \Basis(\sigma)}{\Gamma \vdash |i\>_r : \DType(\{r\}, \emptyset)}
    \]
    \[
        \textbf{L-Basis-Bra}\qquad 
        \frac{r : \reg(\sigma) \in \Gamma \qquad \Gamma \vdash i : \Basis(\sigma)}{\Gamma \vdash {}_r\<i| : \DType(\emptyset, \{r\})}
    \]
    \[
        \textbf{L-Ket}\qquad 
        \frac{\Gamma \vdash R : \reg(\sigma)\qquad \Gamma \vdash K : \KType(\sigma)}{\Gamma \vdash K_R : \DType(\var R, \emptyset)}
    \]
    \[
        \textbf{L-Bra}\qquad 
        \frac{\Gamma \vdash R : \reg(\sigma)\qquad \Gamma \vdash B : \BType(\sigma)}{\Gamma \vdash B_R : \DType(\emptyset, \var R)}
    \]
    \[
        \textbf{L-Opt}\qquad 
        \frac{
            \begin{aligned}
                \Gamma \vdash R_1 : \reg (\sigma_1) \\
                \Gamma \vdash R_2 : \reg (\sigma_2)
            \end{aligned}
            \qquad
            \Gamma \vdash O : \OType(\sigma_1, \sigma_2)
        }
        {\Gamma \vdash O_{R_1;R_2} : \DType(\var R_1, \var R_2)}
    \]
    \[
        \textbf{L-Conj}\qquad 
        \frac{\Gamma \vdash D : \DType(s_1,s_2)}{\Gamma \vdash D^\dagger : \DType(s_2,s_1)}
    \]
    \[
        \textbf{L-Scl}\qquad 
        \frac{\Gamma \vdash S : \SType\qquad \Gamma \vdash D : \DType(s_1,s_2)}{\Gamma \vdash S.D : \DType(s_1,s_2)}
    \]
    \[
        \textbf{L-Add}\qquad
        \frac{\Gamma \vdash D_i : \DType(s_1,s_2)\quad \text{forall } i}{\Gamma \vdash D_1+\cdots+D_n : \DType(s_1,s_2)}
    \]
    \[
        \textbf{L-Tsr}\qquad
        \frac{
            \Gamma \vdash D_i : \DType(s_i,s_i') \qquad
            \bigcap_i s_i = \emptyset \qquad
            \bigcap_i s_i' = \emptyset
        }
        {\Gamma \vdash D_1 \otimes \cdots \otimes D_i : \DType(\bigcup_i s_i, \bigcup_i s_i')}
    \]
    \[
        \textbf{L-Dot}\qquad
        \frac{
            \begin{aligned}
                \Gamma \vdash D_1 : \DType(s_1,s_1') \\
                \Gamma \vdash D_2 : \DType(s_2,s_2')
            \end{aligned}
            \qquad 
            \begin{aligned}
                s_1 \cap s_2 \backslash s_1' = \emptyset \\
                s_2' \cap s_1' \backslash s_2 = \emptyset
            \end{aligned}
        }
        {\Gamma \vdash D_1\cdot D_2 : \DType(s_1 \cup (s_2\backslash s_1'), s_2' \cup (s_1'\backslash s_2))}
    \]
    \[
        \textbf{L-Sum}\qquad
        \frac{\Gamma \vdash s : \SET(\sigma) \qquad \Gamma \vdash f : \Basis(\sigma) \to \DType(s_1, s_2)}{\Gamma \vdash \sum_{s} f : \DType(s_1, s_2)}
    \]

\end{itemize}

\section{Denotational Semantics}
\label{sec: full denotational sem}

\begin{definition}[Interpretation of indices]
The interpretation $\sem{\sigma}$ of a index is defined inductively as follows:
    \begin{align*}
        & \textrm{(Basis types)} &&
        \sem{\sigma_1 \times \sigma_2} \equiv \sem{\sigma_1} \times \sem{\sigma_2}.
     \end{align*}
\end{definition}

\begin{definition}[Interpretation of types]
The interpretation $\sem{T}$ of a type is defined inductively as follows:
\begin{align*}
    & \text{(Basis types)} && \begin{aligned}
      & \sem{\Basis(\sigma)} \equiv \sem{\sigma},
 \end{aligned} \\
    & \text{(Dirac types)} &&
 \sem{\SType} \equiv \mathbb{C},
 \qquad
 \sem{\KType(\sigma)} \equiv \mathcal{H}_{\sem{\sigma}}, 
 \qquad
 \sem{\BType(\sigma)} \equiv \mathcal{H}^*_{\sem{\sigma}},
      \\ & &&
 \sem{\OType(\sigma, \tau)} \equiv \mathcal{L}(\mathcal{H}_\sem{\tau}, \mathcal{H}_\sem{\sigma})
    \\
    % & \text{(Function types)} &&
    % \sem{T_1 \to T_2} \equiv \sem{T_2}^\sem{T_1}
    % \qquad
    % \sem{\forall x.T} \equiv \sem{T}^I \\
    & \text{(Set types)} &&
    \sem{\SET(\sigma)} = \mathcal{P}(\sem{\sigma}) \\
    & \text{(Labelled Dirac types)} &&
    \sem{\cD(s,s')} = \mathcal{L}(\bigotimes_{j}\cH_{\sem{\sigma_{r'_j}}_v}, \bigotimes_{i}\cH_{\sem{\sigma_{r_i}}_v})
\end{align*}
where the sets $s = \{r_1,\cdots,r_n\}$ and $s' = \{r'_1,\cdots,r'_m\}$ ($r_i$ and $r'_j$ are sorted) and $\Gamma \vdash r_i : \reg(\sigma_{r_i})$, $\Gamma \vdash r'_j : \reg(\sigma_{r'_j})$.
\end{definition}

We now turn to the interpretation of expressions. As usual, the interpretation is parametrized by a valuation \(v\), which maps all variables $x$ to their value $v(x)$. 

For any set $s = \{r_1,r_2,\cdots,r_n\}$ ($r_i$ is ordered) with $\Gamma \vdash r_i : \reg(\sigma_{r_i})$, we define 
$$\mathbf{1}_s \triangleq (\mathbf{1}_{\mathcal{O}}(\sigma_{r_1}))_{r_1}\otimes\cdots(\mathbf{1}_{\mathcal{O}}(\sigma_{r_n}))_{r_n}.$$
For any register $R$ s.t. $\Gamma\vdash R : \reg(\sigma)$, suppose $\var(R) = \{r_1,r_2,\cdots,r_n\}$ ($r_i$ is ordered), 
we introduce variables $i_{r_k} : \Basis(\sigma_{r_k})$ with $\Gamma \vdash r_k : \reg(\sigma_{r_k})$ for $k = 1,\cdots, n$. 
We reconstruct the basis $|i_R\>$ (which is of type $\KType(\sigma)$) and $\<i_R|$ (which is of type $\BType(\sigma)$) of $R$ by:
\begin{itemize}
  \item $R = r_k$, $|i_R\> \triangleq |i_{r_k}\>$ and $\<i_R| \triangleq \<i_{r_k}|$;
  \item $R = (R_1,R_2)$: $|i_R\> \triangleq |i_{R_1}\> \otimes |i_{R_2}\>$ and $\<i_R| \triangleq \<i_{R_1}| \otimes \<i_{R_2}|$.
\end{itemize}
Now, we can define the operator $\Swap_R$ that sorts $R$ as: 
$$\Swap_R = \sum_{i_{r_1}}\cdots\sum_{i_{r_n}} (|i_{r_1}\>\otimes\cdots|i_{r_n}\>)\cdot(\<i_R|).$$
Similarly, we define $\Swap_{s_1,s_2,\cdots,s_n}$ for merging disjoint sets orderly. Suppose $s_i = \{r_{i1},\cdots r_{im_i}\}$ (ordered enumerated), and sorted $\bigcup_i\{r_{i1},\cdots r_{im_i}\}$ as $\{r_1,\cdots,r_k\}$, then  
$$\Swap_R = \sum_{i_{r_1}}\cdots\sum_{i_{r_k}} (|i_{r_1}\>\otimes\cdots|i_{r_n}\>)\cdot(\bigotimes_i(\<i_{r_{i1}}|\otimes\cdots \<i_{r_{im_i}}|)).$$

\begin{definition}[Semantics of expressions]
The interpretation of $e$ under valuation $v$, written as $\sem{e}_v$, is defined by the clauses of 
{\allowdisplaybreaks
 \begin{align*}
    & \text{(Scalars)} && 
 \sem{0} \equiv 0, 
 \qquad 
 \sem{1} \equiv 1,
 \qquad
 \sem{a + b} \equiv \sem{a} + \sem{b},
 \qquad
 \sem{a \times b} \equiv \sem{a} \times \sem{b}, 
    \\ & &&
 \sem{a^*} \equiv \sem{a}^*,
 \qquad
 \sem{\delta_{s, t}} \equiv \left\{
 \begin{array}{ll}
        1, & \text{where } \sem{s} = \sem{t}, \\
        0, & \text{where } \sem{s} \neq \sem{t}, 
 \end{array}
 \right.
 \qquad
 \sem{B \cdot K} \equiv \sem{B} \cdot \sem{K},
  \\[0.2cm]
    & \text{(Constants)} &&
 \sem{\mathbf{0}_\mathcal{K}(\sigma)} \equiv \mathbf{0}, 
 \qquad
 \sem{\mathbf{0}_\mathcal{B}(\sigma)} \equiv \mathbf{0},
 \qquad
 \sem{\mathbf{0}_\mathcal{O}(\sigma, \tau)} \equiv \mathbf{0}, 
 \qquad
 \sem{\mathbf{1}_\mathcal{O}(\sigma)} \equiv \mathbf{I},
  \\[0.2cm]
    & \text{(Basis)} &&
 \sem{\ket{t}} \equiv \ket{\sem{t}},
 \qquad
 \sem{\bra{t}} \equiv \bra{\sem{t}},
  \\[0.2cm]
    & \text{(Shared symbols)} &&
 \sem{D^\dagger} \equiv \sem{D}^\dagger,
 \qquad
 \sem{a.D} \equiv \sem{a} \sem{D},
 \qquad
 \sem{D_1 + D_2} \equiv \sem{D_1}+\sem{D_2}, \\
    & &&
 \sem{D_1 \cdot D_2} \equiv \sem{D_1} \cdot \sem{D_2},
 \qquad
 \sem{D_1 \otimes D_2} \equiv \sem{D_1} \otimes \sem{D_2}. \\[0.2cm]
       & \text{(set terms)} && 
    \sem{\mathbf{U}(\sigma)} \equiv \sem{\sigma},
    \qquad
    \sem{M_1 \times M_2} \equiv \sem{M_1} \times \sem{M_2},
    \\[0.2cm]
       & \text{(sum)} &&
    \llbracket \sum_{i \in M} X \rrbracket_v \equiv \sum_{m \in \sem{M}} \sem{X}_{v[i\mapsto m]} \\[0.2cm]
    & \text{(Labelled basis)} &&
      \sem{|i\>_r} = |\sem{i}\>,\qquad \sem{\<i|_r} = \<\sem{i}| \\[0.2cm]
    & \text{(Labelled lifting)} &&
    \sem{K_R} = \sem{\Swap_R} \cdot \sem{K}
 \qquad
 \sem{B_R} = \sem{B}\cdot \sem{\Swap_R} \\
 & && \sem{O_{R_1,R_2}} = \sem{\Swap_{R_1}}\cdot \sem{O}\cdot \sem{\Swap_{R_2}} \\[0.2cm]
 & \text{(Labelled tensor)} &&
 \sem{D_1\otimes \cdots \otimes D_n} = \sem{\Swap_{s_1,\cdot,s_n}}\cdot (\sem{D_1}\otimes \cdots \otimes \sem{D_1})\cdot \sem{\Swap_{s'_1,\cdot,s'_n}} \\[0.2cm]
 & \text{(Generalized dot)} &&
  \sem{D_1\cdot D_2} = \sem{\cl(D_1, s_2\backslash s'_1)}\cdot \sem{\cl(D_2, s'_1\backslash s_2)}
\end{align*}
}
where we assume $\Gamma\vdash D_i : \cD(s_i,s'_i)$.
\end{definition}
Denotational semantics of expressions. Symbol $D$ represents appropriate terms from the ket, bra, or operator sorts. States in $\mathcal{H}$ are represented by column vector, co-states in $\mathcal{H}^*$ by row vector, then all $\cdot$ above are interpreted as matrix multiplications, while $\otimes$ as Kronecker products.
We omit the semantics of functions.

\section{Axiomatic Semantics}
\label{sec: full axioms}
The full list of equational axioms are provided below.
{\allowdisplaybreaks
    \begin{align*}
        & \textsc{(Ax-Scalar)} &&
        (B \cdot K)^* = K^\dagger \cdot B^\dagger
        \\
            & \textsc{(Ax-Delta)} &&
        \delta_{s, t}^* = \delta_{s, t}
        \qquad
        \bra{s} \cdot \ket{t} = \delta_{s, t}
        \\ & &&
        \delta_{s, s} = 1
        \qquad
        s \neq t \vdash \delta_{s, t} = 0
        \qquad
        \delta_{s, t} = \delta_{t, s}
        \\
            & \textsc{(Ax-Linear)} &&
        \mathbf{0} + D = D
        \qquad
        D_1 + D_2 = D_2 + D_1
        \\ & &&
        (D_1 + D_2) + D_3 = D_1 + (D_2 + D_3)
        \\ & &&
        0.D = \mathbf{0}
        \qquad
        a.\mathbf{0} = \mathbf{0}
        \qquad
        1.D = D
        \\ & &&
        a.(b.D) = (a \times b).D
        \qquad
        (a + b).D = a.D + b.D
        \\ & &&
        a.(D_1 + D_2) = a.D_1 + a.D_2
        \\
        & \textsc{(Ax-Bilinear)} &&
        D \cdot \mathbf{0} = \mathbf{0} 
        \qquad
        D_1 \cdot (a.D_2) = a.(D_1 \cdot D_2)
        \\ & &&
        D_0 \cdot (D_1 + D_2) = D_0 \cdot D_1 + D_0 \cdot D_2
        \\ & &&
        \mathbf{0} \cdot D = \mathbf{0}
        \qquad
        (a.D_1) \cdot D_2 = a.(D_1 \cdot D_2)
        \\ & &&
        (D_1 + D_2) \cdot D_0 = D_1 \cdot D_0 + D_2 \cdot D_0
        \\ 
        & &&
        D \otimes \mathbf{0} = \mathbf{0}
        \qquad
        D_1 \otimes (a.D_2) = a.(D_1 \otimes D_2)
        \\ & &&
        D_0 \otimes (D_1 + D_2) = D_0 \otimes D_1 + D_0 \otimes D_2
        \\ & &&
        \mathbf{0} \otimes D = \mathbf{0} 
        \qquad
        (a.D_1) \otimes D_2 = a.(D_1 \otimes D_2)
        \\ & &&
        (D_1 + D_2) \otimes D_0 = D_1 \otimes D_0 + D_2 \otimes D_0
        \\ 
            & \textsc{(Ax-Adjoint)} &&
        \mathbf{0}^\dagger = \mathbf{0}
        \qquad
        (D^\dagger)^\dagger = D 
        \qquad
        (a.D)^\dagger = a^*.(D^\dagger)
        \\ & &&
        (D_1 + D_2)^\dagger = D_1^\dagger + D_2^\dagger
        \\
        & && (D_1 \cdot D_2)^\dagger = D_2^\dagger \cdot D_1^\dagger
        \qquad
        (D_1 \otimes D_2)^\dagger = D_1^\dagger \otimes D_2^\dagger
        \\
            & \textsc{(Ax-Comp)} &&
        D_0 \cdot (D_1 \cdot D_2) = (D_0 \cdot D_1) \cdot D_2
        \\ & &&
        (D_1 \otimes D_2) \cdot (D_3 \otimes D_4) = (D_1 \cdot D_3) \otimes (D_2 \cdot D_4)
        % \\ & &&
        % (K_1 \cdot B_1) \cdot (K_2 \cdot B_2) = (B_1 \cdot K_2) . (K_1 \otimes B_2)
        \\ & &&
        (K_1 \cdot B) \cdot K_2 = (B \cdot K_2).K_1
        \qquad
        B_1 \cdot (K \cdot B_2) = (B_1 \cdot K).B_2
        % \\ & &&
        % (K \cdot B) \cdot O = K \cdot (B \cdot O)
        % \qquad
        % O \cdot (K \cdot B) = (O \cdot  K) \cdot B
        % \\
        %     & &&
        % (K_1 \cdot B_1) \otimes (K_2 \cdot B_2) = (K_1 \otimes K_2) \cdot (B_1 \otimes B_2)
        \\ 
            & &&
        (B_1 \otimes B_2) \cdot (K_1 \otimes K_2) = (B_1 \cdot K_1) \times (B_2 \cdot K_2)
        \\ 
            & \textsc{(Ax-Ground)} &&
        \mathbf{1}_\mathcal{O}^\dagger = \mathbf{1}_\mathcal{O}
        \qquad
        \textbf{1}_\mathcal{O} \cdot D = D 
        \qquad
        \mathbf{1}_\mathcal{O} \otimes \mathbf{1}_\mathcal{O} = \mathbf{1}_\mathcal{O} 
        \\ & &&
        \ket{t}^\dagger = \bra{t}
        \qquad
        \ket{s} \otimes \ket{t} =\ket{(s, t)} 
        \\
        & \textsc{(Sum)}
        && \sum_{i \in s} \mathbf{0} = \mathbf{0}
        \qquad 
        \sum_{i \in \mathbf{U}(\sigma)} \ket{i} \cdot \bra{i} = \mathbf{1}_\mathcal{O}(\sigma) \\
        & && i \text{ free in } t \Rightarrow \sum_{i \in \mathbf{U}(\sigma)} \delta_{i, t} = 1 \\
        & && i \text{ free in } t \Rightarrow \sum_{i \in \mathbf{U}(\sigma)} \delta_{i, t}.A = A\{i/t\} \\
        & && \sum_{i \in M} \sum_{j \in M} \delta_{i, j} = \sum_{j \in M} 1 \\
        & && \sum_{i \in M} \sum_{j \in M} \delta_{i, j}.A = \sum_{j \in M} A\{i/j\} \\
        & &&  b_1 \times \cdots \times (\sum_{i \in M} a) \times \cdots \times b_n 
        \reduce\ \sum_{i \in M} (b_1 \times \cdots \times a \times \cdots \times b_n) \\
        & && (\sum_{i \in M}a)^* = \sum_{i \in M} a^*
        \qquad
        (\sum_{i \in M} A)^\dagger = \sum_{i \in M} A^\dagger \\
        & &&
        a.(\sum_{i \in M} A) = \sum_{i \in M} a.A
        \qquad
        (\sum_{i \in M} a).A = \sum_{i \in M} a.A \\
        & &&
        X \cdot (\sum_{i \in M} Y) = \sum_{i \in M} X \cdot Y
        \qquad
        (\sum_{i \in M} X) \cdot Y = \sum_{i \in M} X \cdot Y \\
        & &&
        X \otimes (\sum_{i \in M} Y) = \sum_{i \in M} X \otimes Y
        \qquad
        (\sum_{i \in M} X) \otimes Y = \sum_{i \in M} X \otimes Y \\
        & && \sum_{i \in M}(a_1 + \cdots + a_n) = (\sum_{i \in M} a_1) + \cdots + (\sum_{i \in M} a_n) \\
        & && \sum_{i \in M}(X_1 + \cdots + X_n) =(\sum_{i \in M} X_1) + \cdots + (\sum_{i \in M} X_n) \\
        & && \sum_{i \in \mathbf{U}(\sigma \times \tau)} A =\sum_{j \in \mathbf{U}(\sigma)} \sum_{k \in \mathbf{U}(\tau)} A\{i/(j, k)\} \\
        & && \sum_{i \in M_1 \star M_2} A =\sum_{j \in M_1} \sum_{k \in M_2} A\{i/(j, k)\} \\
        & \textrm{($\alpha$-equivalence)}
        && \lambda x . A = \lambda y . A\{x/y\} \\
        & \textsc{(Sum-Swap)}
        && \sum_{i \in s_1} \sum_{j \in s_2} A = \sum_{j \in s_2} \sum_{i \in s_1} A
    \end{align*}
}

\section{Rewriting Rules}

\label{sec: rewriting rules}

This section includes all the rewriting rules used in the system. Related rules are collected in the same table. 

\renewcommand{\arraystretch}{1.2} % Increases row height by 50%

\begin{ruletable}{Reductions for the definitions and function applications.}
    BETA-ARROW
    & $((\lambda x : T.t)\ u)\ \reduce\ t\{x/u\}$ \\
    BETA-INDEX
    & $((\lambda x : T.t)\ u)\ \reduce\ t\{x/u\}$ \\
    DELTA
    & $(c:=t:T) \in \Gamma \Rightarrow c\ \reduce\ t$
\end{ruletable}

\begin{ruletable}{The special to flatten all AC symbols within one call.}
    R-FLATTEN
    & $a_1 + \cdots + (b_1 + \cdots + b_m) + \cdots + a_n$ \\
    & $\reduce\ a_1 + \cdots + b_1 + \cdots + b_m + \cdots + a_n$ \\ 
    \\
    & $a_1 \times \cdots \times (b_1 \times \cdots \times b_m) \times \cdots \times a_n$ \\
    & $\reduce\ a_1 \times \cdots \times b_1 \times \cdots \times b_m \times \cdots \times a_n$ \\
    \\
    & $X_1 + \cdots + (X_1' + \cdots + X_m') + \cdots + X_n$ \\
    & $\reduce\ X_1 + \cdots + X_1' + \cdots + X_m' + \cdots + X_n$
\end{ruletable}

\begin{ruletable}{Rules for scalar symbols.}
    R-CONJ5
    & $ \delta_{s, t}^* \ \reduce\ \delta_{s, t}$ \\
    R-CONJ6
    & $ (B \cdot K)^* \ \reduce\ K^\dagger \cdot B^\dagger $ \\
    R-DOT0
    & $ \ZEROB(\sigma) \cdot K \ \reduce\ 0 $ \\
    R-DOT1
    & $ B \cdot \ZEROK(\sigma) \ \reduce\ 0 $ \\
    R-DOT2
    & $ (a.B) \cdot K \ \reduce\ a \times (B \cdot K) $ \\
    R-DOT3
    & $ B \cdot (a.K) \ \reduce\ a \times (B \cdot K) $ \\
    R-DOT4
    & $ (B_1 + \cdots + B_n) \cdot K \ \reduce\ B_1 \cdot K + \cdots + B_n \cdot K $ \\
    R-DOT5
    & $ B \cdot (K_1 + \cdots + K_n) \ \reduce\ B \cdot K_1 + \cdots + B \cdot K_n $ \\
    R-DOT6
    & $ \bra{s} \cdot \ket{t} \ \reduce\ \delta_{s, t} $ \\
    R-DOT7
    & $ (B_1 \otimes B_2) \cdot \ket{(s, t)} \ \reduce\ (B_1 \cdot \ket{s}) \times (B_2 \cdot \ket{t}) $ \\
    R-DOT8
    & $ \bra{(s, t)} \cdot (K_1 \otimes K_2) \ \reduce\ (\bra{s} \cdot K_1) \times (\bra{t} \cdot K_2) $ \\
    R-DOT9
    & $ (B_1 \otimes B_2) \cdot (K_1 \otimes K_2) \ \reduce\ (B_1 \cdot K_1) \times (B_2 \cdot K_2) $ \\
    R-DOT10
    & $ (B \cdot O) \cdot K \ \reduce\ B \cdot (O \cdot K) $ \\
    R-DOT11
    & $ \bra{(s, t)} \cdot ((O_1 \otimes O_2) \cdot K) \ \reduce\ ((\bra{s} \cdot O_1) \otimes (\bra{t} \cdot O_2)) \cdot K $ \\
    R-DOT12
    & $ (B_1 \otimes B_2) \cdot ((O_1 \otimes O_2) \cdot K) \ \reduce\ ((B_1 \cdot O_1) \otimes (B_2 \cdot O_2)) \cdot K $ \\
    R-DELTA0
    & $ \delta_{a, a} \ \reduce\ 1$ \\
    R-DELTA1
    & $ \delta_{(a, b), (c, d)} \ \reduce\ \delta_{a, c} \times \delta_{b, d}$ \\
\end{ruletable}

\begin{ruletable}{Rules for scaling.}
    R-SCR0
    & $ 1.X \ \reduce\ X $ \\
    R-SCR1
    & $ a.(b.X) \ \reduce\ (a \times b).X $ \\
    R-SCR2
    & $ a.(X_1 + \cdots + X_n) \ \reduce\ a.X_1 + \cdots + a.X_n $ \\
    R-SCRK0
    & $ K : \KType(\sigma) \Rightarrow 0.K \ \reduce\ \ZEROK(\sigma)$ \\
    R-SCRK1
    & $ a.\ZEROK(\sigma) \ \reduce\ \ZEROK(\sigma) $ \\
    R-SCRB0
    & $ B : \BType(\sigma) \Rightarrow 0.B \ \reduce\ \ZEROB(\sigma)$ \\
    R-SCRB1
    & $ a.\ZEROB(\sigma) \ \reduce\ \ZEROB(\sigma) $ \\
    R-SCRO0
    & $ O : \OType(\sigma, \tau) \Rightarrow 0.O \ \reduce\ \ZEROO(\sigma, \tau)$ \\
    R-SCRO1
    & $ a.\ZEROO(\sigma, \tau) \ \reduce\ \ZEROO(\sigma, \tau) $ \\
\end{ruletable}

\begin{ruletable}{Rules for addition. }
    R-ADDID
    & $+(X) \ \reduce\ X$ \\
    R-ADD0
    & $Y_1 + \cdots + X + \cdots + X + \cdots + Y_n \ \reduce\ Y_1 + \cdots + Y_n + \cdots + (1+1).X$ \\
    R-ADD1
    & $Y_1 + \cdots + X + \cdots + a.X + \cdots + Y_n \ \reduce\ Y_1 + \cdots + Y_n + (1+a).X$ \\
    R-ADD2
    & $Y_1 + \cdots + a.X + \cdots + X + \cdots + Y_n \ \reduce\ Y_1 + \cdots + Y_n + (a+1).X$ \\
    R-ADD3
    & $Y_1 + \cdots + a.X + \cdots + b.X + \cdots + Y_n \ \reduce\ Y_1 + \cdots + Y_n + (a+b).X$ \\
    R-ADDK0
    & $K_1 + \cdots + \ZEROK(\sigma) + \cdots + K_n\ \reduce K_1 + \cdots + K_n$ \\
    R-ADDB0
    & $B_1 + \cdots + \ZEROB(\sigma) + \cdots + B_n \ \reduce\ B_1 + \cdots + B_n$ \\
    R-ADDO0
    & $O_1 + \cdots + \ZEROO(\sigma, \tau) + \cdots + O_n \ \reduce\ O_1 + \cdots + O_n$ 
    \\
\end{ruletable}

\begin{ruletable}{Rules for adjoint.}
    R-ADJ0
    & $ (X^\dagger)^\dagger \ \reduce\ X $ \\
    R-ADJ1
    & $ (a.X)^\dagger \ \reduce\ (a^*).(X^\dagger) $ \\
    R-ADJ2
    & $ (X_1 + \cdots + X_n)^\dagger \ \reduce\ X_1^\dagger + \cdots + X_n^\dagger $ \\
    R-ADJ3
    & $ (X \otimes Y)^\dagger \ \reduce\ X^\dagger \otimes Y^\dagger$ \\
    R-ADJK0
    & $ \ZEROB(\sigma)^\dagger \ \reduce\ \ZEROK(\sigma) $ \\
    R-ADJK1
    & $ \bra{t}^\dagger \ \reduce\ \ket{t} $ \\
    R-ADJK2
    & $ (B \cdot O)^\dagger\ \reduce\ O^\dagger \cdot B^\dagger $ \\
    R-ADJB0
    & $ \ZEROK(\sigma)^\dagger \ \reduce\ \ZEROB(\sigma) $ \\
    R-ADJB1
    & $ \ket{t}^\dagger \ \reduce\ \bra{t} $ \\
    R-ADJB2
    & $ (O \cdot K)^\dagger\ \reduce\ K^\dagger \cdot O^\dagger $ \\
    R-ADJO0
    & $ \ZEROO(\sigma, \tau)^\dagger \ \reduce\ \ZEROO(\tau, \sigma) $ \\
    R-ADJO1
    & $ \ONEO(\sigma)^\dagger \ \reduce\ \ONEO(\sigma)$ \\
    R-ADJO2
    & $ (K \cdot B)^\dagger \ \reduce\ B^\dagger \cdot K^\dagger$ \\
    R-ADJO3
    & $ (O_1 \cdot O_2)^\dagger \ \reduce\ O_2^\dagger \cdot O_1^\dagger $ \\
\end{ruletable}

\begin{ruletable}{Rules for tensor product.}
    R-TSR0
    & $ (a.X_1) \otimes X_2 \ \reduce\ a.(X_1 \otimes X_2) $ \\
    R-TSR1
    & $ X_1 \otimes (a.X_2) \ \reduce\ a.(X_1 \otimes X_2) $ \\
    R-TSR2
    & $ (X_1 + \cdots + X_n) \otimes X' \reduce X_1 \otimes X' + \cdots + X_n \otimes X' $ \\
    R-TSR3
    & $ X' \otimes (X_1 + \cdots + X_n) \reduce X' \otimes X_1 + \cdots + X' \otimes X_n $ \\
    R-TSRK0
    & $ K : \KType(\tau) \Rightarrow \ZEROK(\sigma) \otimes K\ \reduce\ \ZEROK(\sigma \times \tau) $ \\
    R-TSRK1
    & $ K : \KType(\tau) \Rightarrow K \otimes \ZEROK(\sigma)\ \reduce\ \ZEROK(\tau \times \sigma) $ \\
    R-TSRK2
    & $\ket{s} \otimes \ket{t} \ \reduce\ \ket{(s, t)}$ \\
    R-TSRB0
    & $ B : \BType(\tau) \Rightarrow \ZEROB(\sigma) \otimes B\ \reduce\ \ZEROB(\sigma \times \tau) $ \\
    R-TSRB1
    & $ B : \BType(\tau) \Rightarrow B \otimes \ZEROB(\sigma)\ \reduce\ \ZEROB(\tau \times \sigma) $ \\
    R-TSRB2
    & $\bra{s} \otimes \bra{t} \ \reduce\ \bra{(s, t)}$ \\
    R-TSRO0
    & $ O : \OType(\sigma, \tau) \Rightarrow O \otimes \ZEROO(\sigma', \tau') \ \reduce\ \ZEROO(\sigma \times \sigma', \tau \times \tau') $ \\
    R-TSRO1
    & $ O : \OType(\sigma, \tau) \Rightarrow \ZEROO(\sigma', \tau') \otimes O \ \reduce\ \ZEROO(\sigma' \times \sigma, \tau' \times \tau) $ \\
    R-TSRO2
    & $\ONEO(\sigma) \otimes \ONEO(\tau)\ \reduce\ \ONEO(\sigma \times \tau)$ \\
    R-TSRO3
    & $ (K_1 \cdot B_1) \otimes (K_2 \cdot B_2)\ \reduce\ (K_1 \otimes K_2) \cdot (B_1 \otimes B_2) $ \\
\end{ruletable}

\begin{ruletable}{Rule for $O\cdot K$.}
    R-MULK0
    & $ \ZEROO(\sigma, \tau) \cdot K \ \reduce\ \ZEROK(\sigma) $ \\
    R-MULK1
    & $ O : \OType(\sigma, \tau) \Rightarrow O \cdot \ZEROK(\tau) \ \reduce\ \ZEROK(\sigma) $ \\
    R-MULK2
    & $ \ONEO(\sigma) \cdot K \ \reduce K $ \\
    R-MULK3
    & $ (a.O) \cdot K \ \reduce\ a.(O \cdot K) $ \\
    R-MULK4
    & $ O \cdot (a.K) \ \reduce\ a.(O \cdot K) $ \\
    R-MULK5
    & $ (O_1 + \cdots + O_n) \cdot K \ \reduce\ O_1 \cdot K + \cdots + O_n \cdot K $ \\
    R-MULK6
    & $ O \cdot (K_1 + \cdots + K_n) \ \reduce\ O \cdot K_1 + \cdots + O \cdot K_n $ \\
    R-MULK7
    & $ (K_1 \cdot B) \cdot K_2 \ \reduce\ (B \cdot K_2).K_1 $ \\
    R-MULK8
    & $ (O_1 \cdot O_2) \cdot K \ \reduce\ O_1 \cdot (O_2 \cdot K) $ \\
    R-MULK9
    & $ (O_1 \otimes O_2) \cdot ((O_1' \otimes O_2') \cdot K)\ \reduce\ ((O_1 \cdot O_1') \otimes (O_2 \cdot O_2')) \cdot K $ \\
    R-MULK10
    & $ (O_1 \otimes O_2) \cdot \ket{(s, t)}\ \reduce\ (O_1 \cdot \ket{s}) \otimes (O_2 \cdot \ket{t}) $ \\
    R-MULK11
    & $ (O_1 \otimes O_2) \cdot (K_1 \otimes K_2)\ \reduce\ (O_1 \cdot K_1) \otimes (O_2 \cdot K_2) $
\end{ruletable}

\begin{ruletable}{Rule for $B\cdot O$.}
    R-MULB0
    & $ B \cdot \ZEROO(\sigma, \tau) \ \reduce\ \ZEROB(\tau) $ \\
    R-MULB1
    & $ O : \OType(\sigma, \tau) \Rightarrow \ZEROB(\sigma) \cdot O\ \reduce\ \ZEROB(\tau) $ \\
    R-MULB2
    & $ B \cdot \ONEO(\sigma) \ \reduce B $ \\
    R-MULB3
    & $ (a.B) \cdot O \ \reduce\ a.(B \cdot O) $ \\
    R-MULB4
    & $ B \cdot (a.O) \ \reduce\ a.(B \cdot O) $ \\
    R-MULB5
    & $ (B_1 + \cdots + B_n) \cdot O \ \reduce\ B_1 \cdot O + \cdots + B_n \cdot O $ \\
    R-MULB6
    & $ B \cdot (O_1 + \cdots + O_n) \ \reduce\ B \cdot O_1 + \cdots + B \cdot O_n $ \\
    R-MULB7
    & $ B_1 \cdot (K \cdot B_2) \ \reduce\ (B_1 \cdot K).B_2 $ \\
    R-MULB8
    & $ B \cdot (O_1 \cdot O_2) \ \reduce\ (B \cdot O_1) \cdot O_2 $ \\
    R-MULB9
    & $ (B \cdot (O_1' \otimes O_2')) \cdot (O_1 \otimes O_2) \ \reduce\ B \cdot ((O_1' \otimes O_2') \cdot (O_1 \otimes O_2)) $ \\
    R-MULB10
    & $ \bra{(s, t)} \cdot (O_1 \otimes O_2)\ \reduce\ (\bra{s} \cdot O_1) \otimes (\bra{t} \cdot O_2) $ \\
    R-MULB11
    & $ (B_1 \otimes B_2) \cdot (O_1 \otimes O_2)\ \reduce\ (B_1 \cdot O_1) \otimes (B_2 \cdot O_2) $
\end{ruletable}

\begin{ruletable}{Rules for $K \cdot B$.}
    R-OUTER0
    & $ B : \BType(\tau) \Rightarrow \mathbf{0}_\mathcal{K}(\sigma) \cdot B\ \reduce\ \mathbf{0}_\mathcal{O}(\sigma, \tau) $ \\
    R-OUTER1
    & $ K : \KType(\sigma) \Rightarrow K \cdot \mathbf{0}_\mathcal{B}(\tau)\ \reduce\ \mathbf{0}_\mathcal{O}(\sigma, \tau) $ \\
    R-OUTER2
    & $ (a.K) \cdot B\ \reduce\ a.(K \cdot B) $ \\
    R-OUTER3
    & $ K \cdot (a.B)\ \reduce\ a.(K \cdot B) $ \\
    R-OUTER4
    & $ (K_1 + \cdots + K_n) \cdot B\ \reduce\ K_1 \cdot B + \cdots + K_n \cdot B $ \\
    R-OUTER5
    & $ K \cdot (B_1 + \cdots + B_n)\ \reduce\ K \cdot B_1 + \cdots + K \cdot B_n $ \\
\end{ruletable}

\begin{ruletable}{Rules for $O_1 \cdot O_2$.}
    R-MULO0
    & $ O : \OType(\tau, \rho) \Rightarrow \ZEROO(\sigma, \tau) \cdot O\ \reduce\ \ZEROO(\sigma, \rho) $ \\
    R-MULO1
    & $ O : \OType(\sigma, \tau) \Rightarrow O \cdot \ZEROO(\tau, \rho)\ \reduce\ \mathbf{0}_\mathcal{O}(\sigma, \rho) $ \\
    R-MULO2
    & $ \ONEO(\sigma) \cdot O \ \reduce\ O $ \\
    R-MULO3
    & $ O \cdot \ONEO(\sigma) \ \reduce\ O $ \\
    R-MULO4
    & $ (K \cdot B) \cdot O \ \reduce\ K \cdot (B \cdot O) $ \\
    R-MULO5
    & $ O \cdot (K \cdot B) \ \reduce\ (O \cdot K) \cdot B $ \\
    R-MULO6
    & $ (a.O_1) \cdot O_2 \ \reduce\ a.(O_1 \cdot O_2) $ \\
    R-MULO7
    & $ O_1 \cdot (a.O_2) \ \reduce\ a.(O_1 \cdot O_2) $ \\
    R-MULO8
    & $ (O_1 + \cdots + O_n) \cdot O'\ \reduce\ O_1 \cdot O' + \cdots + O_n \cdot O' $ \\
    R-MULO9
    & $ O' \cdot (O_1 + \cdots + O_n)\ \reduce\ O' \cdot O_1 + \cdots + O' \cdot O_n $ \\
    R-MULO10
    & $ (O_1 \cdot O_2) \cdot O_3\ \reduce\ O_1 \cdot (O_2 \cdot O_3) $ \\
    R-MULO11
    & $ (O_1 \otimes O_2) \cdot (O_1' \otimes O_2')\ \reduce\ (O_1 \cdot O_1') \otimes (O_2 \cdot O_2') $ \\
    R-MULO12
    & $ (O_1 \otimes O_2) \cdot ((O_1' \otimes O_2') \cdot O_3)\ \reduce\ ((O_1 \cdot O_1') \otimes (O_2 \cdot O_2')) \cdot O_3 $ \\  
\end{ruletable}

\begin{ruletable}{Rules for sets.}
    R-SET0
    & $ \mathbf{U}(\sigma) \star \mathbf{U}(\tau) \ \reduce\ \mathbf{U}(\sigma \times \tau) $
\end{ruletable}

\begin{ruletable}{Rules for sum operators.}
    R-SUM-CONST0
    & $ \sum_{x \in s} 0 \ \reduce\ 0 $ \\
    R-SUM-CONST1
    & $ \sum_{x \in s} \ZEROK(\sigma)\ \reduce\ \ZEROK(\sigma) $ \\
    R-SUM-CONST2
    & $ \sum_{x \in s} \ZEROB(\sigma)\ \reduce\ \ZEROB(\sigma) $ \\
    R-SUM-CONST3
    & $ \sum_{x \in s} \ZEROO(\sigma, \tau)\ \reduce\ \ZEROO(\sigma, \tau) $ \\
    R-SUM-CONST4
    & $ \ONEO(\sigma) \ \reduce\ \sum_{i \in \mathbf{U}(\sigma)} \ket{i} \cdot \bra{i} $
\end{ruletable}

\begin{ruletable}{Rules for eliminating $\delta_{s, t}$. These rules match the $\delta$ operator modulo the commutativity of its arguments.}
    R-SUM-ELIM0
    & $ i \text{ free in } t \Rightarrow \sum_{i \in \mathbf{U}(\sigma)} \sum_{k_1 \in s_1} \cdots \sum_{k_n \in s_n} \delta_{i, t}$ \\
    & $ \reduce\ \sum_{k_1 \in s_1} \cdots \sum_{k_n \in s_n}  1$ \\
    \\
    R-SUM-ELIM1
    & $ i \text{ free in } t \Rightarrow $ \\
    & $ \sum_{i \in \mathbf{U}(\sigma)} \sum_{k_1 \in s_1} \cdots \sum_{k_n \in s_n} (a_1 \times \cdots \times \delta_{i, t} \times \cdots \times a_n) $ \\
    & $ \reduce\ \sum_{k_1 \in s_1} \cdots \sum_{k_n \in s_n} a_1\{i/t\} \times \cdots \times a_n\{i/t\} $ \\
    \\
    R-SUM-ELIM2
    & $ i \text{ free in } t \Rightarrow \sum_{i \in \mathbf{U}(\sigma)} \sum_{k_1 \in s_1} \cdots \sum_{k_n \in s_n} (\delta_{i, t}.A) $ \\
    & $ \reduce\ \sum_{k_1 \in s_1} \cdots \sum_{k_n \in s_n} A\{i/t\} $ \\
    \\
    R-SUM-ELIM3
    & $ i \text{ free in } t \Rightarrow $ \\
    & $ \sum_{i \in \mathbf{U}(\sigma)} \sum_{k_1 \in s_1} \cdots \sum_{k_n \in s_n} (a_1 \times \cdots \times \delta_{i, t} \times \cdots \times a_n).A $ \\
    & $ \reduce\ \sum_{k_1 \in s_1} \cdots \sum_{k_n \in s_n}  (a_1\{i/t\} \times \cdots \times a_n\{i/t\}).A\{i/t\} $ \\
    \\
    R-SUM-ELIM4
    & $ \sum_{i \in M} \sum_{j \in M} \sum_{k_1 \in s_1} \cdots \sum_{k_n \in s_n}  \delta_{i, j}$ \\ 
    & $\reduce\ \sum_{j \in M} \sum_{k_1 \in s_1} \cdots \sum_{k_n \in s_n} 1 $ \\
    \\
    R-SUM-ELIM5
    & $ \sum_{i \in M} \sum_{j \in M} \sum_{k_1 \in s_1} \cdots \sum_{k_n \in s_n} (a_1 \times \dots \times \delta_{i, j} \times \cdots \times a_n) $ \\
    & $ \reduce\ \sum_{j \in M} \sum_{k_1 \in s_1} \cdots \sum_{k_n \in s_n} (a_1\{j/i\} \times \cdots \times a_n\{j/i\}) $ \\
    \\
    R-SUM-ELIM6
    & $ \sum_{i \in M} \sum_{j \in M} \sum_{k_1 \in s_1} \cdots \sum_{k_n \in s_n} (\delta_{i, j}.A) $ \\
    & $ \reduce\ \sum_{j \in M} \sum_{k_1 \in s_1} \cdots \sum_{k_n \in s_n} A\{j/i\} $ \\
    \\
    R-SUM-ELIM7
    & $ \sum_{i \in M} \sum_{j \in M} \sum_{k_1 \in s_1} \cdots \sum_{k_n \in s_n} (a_1 \times \cdots \times \delta_{i, j} \times \cdots \times a_n).A $ \\
    & $ \reduce\ \sum_{j \in M} \sum_{k_1 \in s_1} \cdots \sum_{k_n \in s_n} (a_1\{j/i\} \times \cdots \times a_n\{j/i\}).A\{j/i\} $ \\
    \\
    R-SUM-ELIM8
    & $ \sum_{i \in M} \sum_{j \in M} \sum_{k_1 \in s_1} \cdots \sum_{k_n \in s_n} ((a_1 \times \cdots \times \delta_{i, j} \times \cdots \times a_n) + $ \\
    & $ \cdots + (b_1 \times \cdots \times \delta_{i, j} \times \cdots \times b_n)).A $ \\
    & $ \reduce\ \sum_{j \in M} \sum_{k_1 \in s_1} \cdots \sum_{k_n \in s_n} ((a_1\{j/i\} \times \cdots \times a_n\{j/i\}) + $ \\
    & $ \cdots + (b_1\{j/i\} \times \cdots \times b_n\{j/i\})).A\{j/i\} $ \\
\end{ruletable}

\begin{ruletable}{Rules for pushing terms into sum operators. Because we apply type checking on variables, and stick to unique bound variables, these operations are always sound.}
    R-SUM-PUSH0
    & $ b_1 \times \cdots \times (\sum_{i \in M} a) \times \cdots \times b_n$ \\
    & $\reduce\ \sum_{i \in M} (b_1 \times \cdots \times a \times \cdots \times b_n) $ \\
    R-SUM-PUSH1
    & $ (\sum_{i \in M} a)^* \ \reduce\ \sum_{i \in M} a^* $ \\
    R-SUM-PUSH2
    & $ (\sum_{i \in M} X)^\dagger \ \reduce\ \sum_{i \in M} X^\dagger $ \\
    R-SUM-PUSH3
    & $ a.(\sum_{i \in M} X) \ \reduce\ \sum_{i \in M} (a.X) $ \\
    R-SUM-PUSH4
    & $ (\sum_{i \in M} a).X \ \reduce\ \sum_{i \in M} (a.X) $ \\
    R-SUM-PUSH5
    & $ (\sum_{i \in M} B)\cdot K \ \reduce\ \sum_{i \in M}(B \cdot K) $ \\
    R-SUM-PUSH6
    & $ (\sum_{i \in M} O)\cdot K \ \reduce\ \sum_{i \in M}(O \cdot K) $ \\
    R-SUM-PUSH7
    & $ (\sum_{i \in M} B)\cdot O \ \reduce\ \sum_{i \in M}(B \cdot O) $ \\
    R-SUM-PUSH8
    & $ (\sum_{i \in M} K)\cdot B \ \reduce\ \sum_{i \in M}(K \cdot B) $ \\
    R-SUM-PUSH9
    & $ (\sum_{i \in M} O_1)\cdot O_2 \ \reduce\ \sum_{i \in M}(O_1 \cdot O_2) $ \\
    R-SUM-PUSH10
    & $ B \cdot (\sum_{i \in M} K) \ \reduce\ \sum_{i \in M}(B \cdot K) $ \\
    R-SUM-PUSH11
    & $ O \cdot (\sum_{i \in M} K) \ \reduce\ \sum_{i \in M}(O \cdot K) $ \\
    R-SUM-PUSH12
    & $ B \cdot (\sum_{i \in M} O) \ \reduce\ \sum_{i \in M}(B \cdot O) $ \\
    R-SUM-PUSH13
    & $ K \cdot (\sum_{i \in M} B) \ \reduce\ \sum_{i \in M}(K \cdot B) $ \\
    R-SUM-PUSH14
    & $ O_1 \cdot (\sum_{i \in M} O_2) \ \reduce\ \sum_{i \in M}(O_1 \cdot O_2) $ \\
    R-SUM-PUSH15
    & $ (\sum_{i \in M} X_1) \otimes X_2 \ \reduce\ \sum_{i \in M} (X_1 \otimes X_2) $ \\
    R-SUM-PUSH16
    & $ X_1 \otimes (\sum_{i \in M} X_2) \ \reduce\ \sum_{i \in M} (X_1 \otimes X_2) $
\end{ruletable}

\begin{ruletable}{Rules for addition and index in sum.}
    R-SUM-ADDS0
    & $\sum_{i \in M}(a_1 + \cdots + a_n) \ \reduce\ (\sum_{i \in M} a_1) + \cdots + (\sum_{i \in M} a_n) $ \\
    % %
    % R-SUM-ADDS1
    % & $\sum_{i \in M}(b_1 \times \cdots \times (a_1 + \cdots + a_n) \times \cdots \times b_m) $ \\
    % & $ \reduce\ (\sum_{i \in M} (b_1 \times \cdots \times a_1 \times \cdots \times b_m) + \cdots + (\sum_{i \in M} (b_1 \times \cdots \times a_n \times \cdots \times b_m)) $ \\
    %
    R-SUM-ADD0
    & $\sum_{i \in M}(X_1 + \cdots + X_n) \ \reduce\ (\sum_{i \in M} X_1) + \cdots + (\sum_{i \in M} X_n) $ \\
    R-SUM-INDEX0
    & $ \sum_{i \in \mathbf{U}(\sigma \times \tau)} A \ \reduce\ \sum_{j \in \mathbf{U}(\sigma)} \sum_{k \in \mathbf{U}(\tau)} A\{i/(j, k)\} $ \\
    R-SUM-INDEX1
    & $ \sum_{i \in M_1 \star M_2} A \ \reduce\ \sum_{j \in M_1} \sum_{k \in M_2} A\{i/(j, k)\} $
\end{ruletable}

\begin{ruletable}{Rules for bool index.}
    R-BIT-DELTA
    & $\delta_{0, 1} \ \reduce\ 0$ \\
    R-BIT-ONEO
    & $\ONEO(\textsf{bool})\ \reduce\ \ket{0}\bra{0} + \ket{1}\bra{1} $ \\
    R-BIT-SUM
    & $\sum_{i \in \mathbf{U}(\textsf{bool})} A\ \reduce\ A\{i/0\} + A\{i/1\}$
\end{ruletable}

\begin{ruletable}{Rules about addition and sum.}
    R-MULS2
    & $b_1 \times \cdots \times (a_1 + \cdots + a_n) \times \cdots \times b_m$ \\
    & $\reduce\ (b_1 \times \cdots \times a_1 \times \cdots \times b_m) + \cdots + (b_1 \times \cdots \times a_n \times \cdots \times b_m)$ \\
    \\
    R-SUM-ADD1
    & $ Y_1 + \cdots + Y_n + \sum_{i \in M}(a+b).X$ \\
    & $ \reduce\ Y_1 + \cdots + \sum_{i \in M}(a.X) + \cdots + \sum_{i \in M}(b.X) + Y_n$ \\
    \\
    R-SUM-FACTOR
    & $X_1 + \cdots + (\sum_{k_1 \in s_1}\cdots\sum_{k_n \in s_n}A) $ \\
    & $ + (\sum_{k_1 \in s_1}\cdots\sum_{k_n \in s_n}A) + \cdots +X_n$ \\
    & $\reduce\ X_1 + \cdots + (\sum_{k_1 \in s_1}\cdots\sum_{k_n \in s_n}(1+1).A) + \cdots +X_n$ \\
    \\
    & $X_1 + \cdots + (\sum_{k_1 \in s_1}\cdots\sum_{k_n \in s_n}a.A) $ \\
    & $ + (\sum_{k_1 \in s_1}\cdots\sum_{k_n \in s_n}A) + \cdots +X_n$ \\
    & $\reduce\ X_1 + \cdots + (\sum_{k_1 \in s_1}\cdots\sum_{k_n \in s_n}(a+1).A) + \cdots +X_n$ \\
    \\
    & $X_1 + \cdots + (\sum_{k_1 \in s_1}\cdots\sum_{k_n \in s_n}a.A) $ \\
    & $ + (\sum_{k_1 \in s_1}\cdots\sum_{k_n \in s_n}b.A) + \cdots +X_n$ \\
    & $\reduce\ X_1 + \cdots + (\sum_{k_1 \in s_1}\cdots\sum_{k_n \in s_n}(a+b).A) + \cdots +X_n$ \\
\end{ruletable}

\begin{ruletable}{Rules to eliminate labels in Dirac notation.}
    R-L-EXPAND
    & $K_R \ \reduce\ \sum_{i_{r_1}\in\bU(\sigma_{r_1})}\cdots \sum_{i_{r_n}\in\bU(\sigma_{r_n})} (\<i_R|\cdot K). (|i_{r_1}\>_{r_i}\otimes\cdots\otimes|i_{r_n}\>_{r_n})$ \\
    \\
    & $B_R \ \reduce\ \sum_{i_{r_1}\in\bU(\sigma_{r_1})}\cdots \sum_{i_{r_n}\in\bU(\sigma_{r_n})} (B\cdot |i_R\>). ({}_{r_1}\<i_{r_1}|\otimes\cdots\otimes{}_{r_n}\<i_{r_n}|)$\\
    \\
    & $O_{R,R'} \ \reduce\ \sum_{i_{r_1}\in\bU(\sigma_{r_1})}\cdots \sum_{i_{r_n}\in\bU(\sigma_{r_n})}
    \sum_{i_{r'_1}\in\bU(\sigma_{r'_1})}\cdots \sum_{i_{r'_{n'}}\in\bU(\sigma_{r'_{n'}})}$ \\
    & $(\<i_R|\cdot O\cdot |i_{R'}\>).(|i_{r_1}\>_{r_i}\otimes\cdots\otimes|i_{r_n}\>_{r_n} \otimes {}_{r'_1}\<i_{r'_1}|\otimes\cdots\otimes{}_{r'_{n'}}\<i_{r'_{n'}}|)$
\end{ruletable}

\begin{ruletable}{Rules for labelled Dirac notation.}
    R-ADJDK
    & $ ({}_r\<i|)^\dagger \ \reduce\ |i\>_r$ \\
    R-ADJDB
    & $ (|i\>_r)^\dagger \ \reduce\ {}_r\<i|$ \\
    R-ADJD0
    & $ (D_1 \otimes \cdots \otimes D_n)^\dagger \ \reduce\ D_1^\dagger \otimes \cdots \otimes D_n^\dagger$ \\
    R-ADJD1
    & $ (D_1\cdot D_2)^\dagger \ \reduce\ D_2^\dagger \cdot D_1^\dagger$ \\
    R-SCRD0
    & $ D_1 \otimes \cdots \otimes (a.D_n) \otimes \cdots \otimes D_m \ \reduce\ a.(D_1 \otimes \cdots \otimes D_m) $ \\
    R-SCRD1
    & $ (a.D_1) \cdot D_2 \ \reduce\ a.(D_1 \cdot D_2) $ \\
    R-SCRD2
    & $ D_1 \cdot (a.D_2) \ \reduce\ a.(D_1 \cdot D_2) $ \\
    R-TSRD0
    & $ X_1 \otimes \cdots \otimes (D_1 + \cdots + D_n) \otimes \cdots X_m$ \\
    & $ \reduce\ X_1 \otimes \cdots D_1 \cdots \otimes X_m + \cdots + X_1 \otimes \cdots D_n  \cdots \otimes X_m $ \\
    R-DOTD0
    & $ (D_1 + \cdots + D_n) \cdot D\ \reduce\ D_1 \cdot D + \cdots + D_n \cdot D $ \\
    R-DOTD1
    & $ D \cdot (D_1 + \cdots + D_n)\ \reduce\ D \cdot D_1 + \cdots + D \cdot D_n $ \\
    R-SUM-PUSHD0
    & $ X_1 \otimes \cdots (\sum_{i \in M} D) \cdots \otimes X_2\ \reduce\ \sum_{i \in M} (X_1 \otimes \cdots D \cdots \otimes X_n) $ \\
    R-SUM-PUSHD1
    & $ (\sum_{i \in M} D_1) \cdot D_2 \ \reduce\ \sum_{i \in M} (D_1 \cdot D_2) $ \\
    R-SUM-PUSHD2
    & $ D_1 \cdot (\sum_{i \in M} D_2) \ \reduce\ \sum_{i \in M} (D_1 \cdot D_2) $
\end{ruletable}

\begin{ruletable} {Rules to simplify dot product in labelled Dirac notation.}
    R-L-SORT0
    & $ A : \DType(s_1, s_2), B : \DType(s_1', s_2'), s_2 \cap s_1'=\emptyset \Rightarrow A \cdot B \ \reduce\ A \otimes B $ \\
    R-L-SORT1
    & ${}_r\bra{i}\cdot\ket{j}_r \ \reduce\ \delta_{i, j}$ \\
    R-L-SORT2
    & ${}_r\bra{i}\cdot(Y_1 \otimes \cdots \otimes \ket{j}_r \otimes \cdots \otimes Y_m) \ \reduce\ \delta_{i, j}.(Y_1  \otimes \cdots \otimes Y_m)$ \\
    R-L-SORT3
    & $(X_1 \otimes \cdots \otimes {}_r\bra{i} \otimes \cdots \otimes X_n) \cdot \ket{j}_r \ \reduce\ \delta_{i,j}.(X_1 \otimes \cdots \otimes X_n)$ \\
    R-L-SORT4
    & $ (X_1 \otimes \cdots \otimes {}_r\bra{i} \otimes \cdots \otimes X_n) \cdot (Y_1 \otimes \cdots \otimes \ket{j}_r \otimes \cdots \otimes Y_m) $ \\
    & $\reduce\ \delta_{i,j}.(X_1 \otimes \cdots \otimes X_n) \cdot (Y_1 \otimes \cdots \otimes Y_m)$
\end{ruletable}

\subsection{Proof of Label Elimination \ref{thm: normalization}}\label{app:label_elim}

\subsubsection{Proof of soundness}

For the soundness of rules for labelled Dirac notation, we first notice that the semantics of $\cdot$ and $\otimes$ are bilinear functions, so the rules regarding linear properties such as generic rules, and (R-SCRD0) - (R-SUM-PUSHD2) in Table 20 are straightforward. We show the (selected, the rest are similar or easier) rest rules in Tables 20 and 21 are sound. Note that $\cdot$ is used in different scopes, we explicitly write $\circ$ for $\cdot$ for multiplication (composition) of linear operators.
\begin{itemize}
  \item (R-ADJDK) $\sem{({}_r\<i|)^\dagger} = \sem{{}_r\<i|}^\dagger = \<\sem{i}|^\dagger = |\sem{i}\> = \sem{|i\>_r}$.
  \item (R-ADJD0) 
  \begin{align*}
    &\sem{(D_1\otimes \cdots \otimes D_n)^\dagger}
    = \sem{(D_1\otimes \cdots \otimes D_n)}^\dagger \\
    =\ & (\sem{\Swap_{s_1,\cdots,s_n}} \circ (\sem{D_1}\otimes\cdots\otimes\sem{D_n}) \cdot \sem{\Swap_{s'_1,\cdots,s'_n}}^\dagger)^\dagger  \\ 
    =\ & \sem{\Swap_{s'_1,\cdots,s'_n}} \circ (\sem{D_1}\otimes\cdots\otimes\sem{D_n})^\dagger \circ \sem{\Swap_{s_1,\cdots,s_n}}^\dagger \\ 
    =\ & \sem{\Swap_{s'_1,\cdots,s'_n}} \circ (\sem{D_1}^\dagger\otimes\cdots\otimes\sem{D_n}^\dagger) \circ \sem{\Swap_{s_1,\cdots,s_n}}^\dagger \\ 
    =\ & \sem{\Swap_{s'_1,\cdots,s'_n}} \circ (\sem{D_1^\dagger}\otimes\cdots\otimes\sem{D_n^\dagger}) \circ \sem{\Swap_{s_1,\cdots,s_n}}^\dagger \\ 
    =\ & \sem{D_1^\dagger\otimes \cdots \otimes D_n^\dagger}.
  \end{align*}
  \item (R-ADJD1) here we use (R-ADJD0): 
  \begin{align*}
    &\sem{(D_1 \cdot D_2)^\dagger}
    = \sem{D_1 \cdot D_2}^\dagger \\
    =\ & (\sem{\cl(D_1,s_2\backslash s_1')} \circ \sem{\cl(D_2,s_1'\backslash s_2)})^\dagger  \\ 
    =\ & \sem{\cl(D_2,s_1'\backslash s_2)^\dagger} \circ \sem{\cl(D_1,s_2\backslash s_1')^\dagger}  \\ 
    =\ & \sem{(D_2 \otimes \mathbf{1}_{s_1'\backslash s_2})^\dagger} \circ \sem{(D_1\otimes \mathbf{1}_{s_2\backslash s_1'})^\dagger}  \\ 
    =\ & \sem{D_2^\dagger \otimes \mathbf{1}_{s_1'\backslash s_2}^\dagger} \circ \sem{D_1^\dagger \otimes \mathbf{1}_{s_2\backslash s_1'}^\dagger} \\ 
    =\ & \sem{D_2^\dagger \otimes \mathbf{1}_{s_1'\backslash s_2}} \circ \sem{D_1^\dagger \otimes \mathbf{1}_{s_2\backslash s_1'}} \\ 
    =\ & \sem{\cl(D_2^\dagger, s_1'\backslash s_2)} \circ \sem{\cl(D_1^\dagger, s_2\backslash s_1')} \\ 
    =\ & \sem{D_2^\dagger \circ D_1^\dagger}
  \end{align*}
  since $\Gamma \vdash D_2^\dagger : \cD(s_2',s_2)$ and $\Gamma \vdash D_1^\dagger : \cD(s_1',s_1)$, and if $s = \{r_1,\cdots,r_n\}$ orderedly, then
  \begin{align*}
    \sem{\mathbf{1}_{s}^\dagger} & = \sem{((\mathbf{1}_\OType(\sigma_{r_1}))_{r_1}\otimes \cdots \otimes (\mathbf{1}_\OType(\sigma_{r_n}))_{r_n})^\dagger } \\
    & = \sem{(\mathbf{1}_\OType(\sigma_{r_1}))_{r_1}^\dagger \otimes \cdots \otimes (\mathbf{1}_\OType(\sigma_{r_n}))_{r_n}^\dagger } \\
    & = \sem{(\mathbf{1}_\OType(\sigma_{r_1}))_{r_1} \otimes \cdots \otimes (\mathbf{1}_\OType(\sigma_{r_n}))_{r_n} } \\
    &= \sem{\mathbf{1}_{s}}
  \end{align*}
  where 
  \begin{align*}
    \sem{\mathbf{1}_\OType(\sigma_{r_i})_{r_i}^\dagger} & = 
    (\sem{\Swap_{r_i}}\circ \sem{\mathbf{1}_\OType(\sigma_{r_i})} \circ \sem{\Swap_{r_i}}^\dagger)^\dagger \\
    & = \sem{\Swap_{r_i}}\circ \mathbf{I}^\dagger \circ \sem{\Swap_{r_i}}^\dagger = \mathbf{I} \\
    & = \sem{\Swap_{r_i}}\circ \mathbf{1}_\OType(\sigma_{r_i}) \circ \sem{\Swap_{r_i}}^\dagger \\
    & = \sem{\mathbf{1}_\OType(\sigma_{r_i})_{r_i}}.
  \end{align*}
  since $\sem{\Swap}$ is a unitary operator.

  \item (R-L-SORT0) here we use the fact that labelled tensor is commutative,
  \begin{align*}
    & \sem{A\cdot B} = \sem{\cl(A, s'_1)} \circ \sem{\cl(B, s_2)} = \sem{A\otimes \mathbf{1}_{s'_1}}\circ  \sem{\mathbf{1}_{s_2}\otimes B} \\
    =\ & \sem{\Swap_{s_1,s'_1}}\circ (\sem{A}\otimes \sem{\mathbf{1}_{s'_1}})\circ \sem{\Swap_{s_2,s'_1}}^\dagger \circ \sem{\Swap_{s_2,s'_1}} \circ (\sem{\mathbf{1}_{s_2}}\otimes \sem{B}) \circ \sem{\Swap_{s_2,s'_2}} \\
    =\ & \sem{\Swap_{s_1,s'_1}}\circ (\sem{A}\otimes \mathbf{I})\circ (\mathbf{I}\otimes \sem{B}) \circ \sem{\Swap_{s_2,s'_2}} \\
    =\ & \sem{\Swap_{s_1,s'_1}}\circ ((\sem{A}\circ \mathbf{I})\otimes (\mathbf{I}\circ \sem{B})) \circ \sem{\Swap_{s_2,s'_2}} \\
    =\ & \sem{\Swap_{s_1,s'_1}}\circ (\sem{A}\otimes \sem{B}) \circ \sem{\Swap_{s_2,s'_2}} \\
    =\ & \sem{A\otimes B}.
  \end{align*}
  where we use the condition $s_2\cap s_1' = \emptyset$.

  \item (R-L-SORT1) 
  \begin{align*}
    \sem{{}_r\<i|\cdot|j\>_r} &= \sem{\cl({}_r\<i|,\emptyset)} \circ \sem{\cl(|j\>_r,\emptyset)} \\
    &= (\sem{\Swap_{\emptyset}} \circ \<\sem{i}| \circ \sem{\Swap_{r,\emptyset}}^\dagger) \circ (\sem{\Swap_{r,\emptyset}} \circ |\sem{j}\>\circ \sem{\Swap_{\emptyset}}) \\
    &= \<\sem{i}|\circ |\sem{j}\> \\
    &= \sem{\delta_{i,j}}
  \end{align*}

  \item (R-L-SORT4) we use the fact that labelled tensor is associative and commutative. Suppose $\Gamma \vdash X_1\otimes \cdots \otimes X_n : \cD(s_X, s_X')$ and $\Gamma \vdash Y_1\otimes \cdots \otimes Y_m : \cD(s_Y, s_Y')$.
  \begin{align*}
    &\sem{(X_1\otimes\cdots\otimes {}_r\<i|\otimes\cdots\otimes X_n)\cdot (Y_1\otimes\cdots\otimes |j\>_r \otimes\cdots\otimes Y_m)} \\
    =\ & \sem{\cl({}_r\<i| \otimes (X_1\otimes\cdots\otimes X_n), s_Y\backslash s_X')}\circ 
    \sem{\cl(|j\>_r \otimes (Y_1\otimes\cdots\otimes Y_m), s_X'\backslash s_Y)} \\
    =\ & \sem{{}_r\<i| \otimes ((X_1\otimes\cdots\otimes X_n) \otimes \mathbf{1}_{s_Y\backslash s_X'})}\circ 
    \sem{|j\>_r \otimes ((Y_1\otimes\cdots\otimes Y_m) \otimes \mathbf{1}_{s_X'\backslash s_Y})} \\
    =\ & [\sem{\Swap}_{s_X \cup s_Y\backslash s_X'}\circ(\<\sem{i}| \otimes \sem{(X_1\otimes\cdots\otimes X_n) \otimes \mathbf{1}_{s_Y\backslash s_X'}})\circ \sem{\Swap}_{\{r\}, s_X' \cup s_Y\backslash s_X'}^\dagger] \\
    & \circ [\sem{\Swap}_{\{r\}, s_Y \cup s_X'\backslash s_Y}\circ(|\sem{j}\> \otimes \sem{(Y_1\otimes\cdots\otimes Y_m) \otimes \mathbf{1}_{s_X'\backslash s_Y}})\circ \sem{\Swap}_{s_Y' \cup s_X'\backslash s_Y}^\dagger]\\
    =\ & \sem{\Swap}_{s_X\cup s_Y\backslash s_X'}\circ(\<\sem{i}| \otimes \sem{(X_1\otimes\cdots\otimes X_n) \otimes \mathbf{1}_{s_Y\backslash s_X'}})\circ \\
    &[\sem{\Swap}_{\{r\}, s_X'\cup s_Y}^\dagger
     \circ \sem{\Swap}_{\{r\}, s_Y \cup s_X'}] \circ \\
    &(|\sem{j}\> \otimes \sem{(Y_1\otimes\cdots\otimes Y_m) \otimes \mathbf{1}_{s_X'\backslash s_Y}})\circ \sem{\Swap}_{s_Y'\cup s_X'\backslash s_Y}^\dagger\\
    =\ & \sem{\Swap}_{s_X\cup s_Y\backslash s_X'}\circ[(\<\sem{i}| \otimes \sem{(X_1\otimes\cdots\otimes X_n) \otimes \mathbf{1}_{s_Y\backslash s_X'}})\circ \\
    &(|\sem{j}\> \otimes \sem{(Y_1\otimes\cdots\otimes Y_m) \otimes \mathbf{1}_{s_X'\backslash s_Y}})]\circ \sem{\Swap}_{s_Y'\cup s_X'\backslash s_Y}^\dagger\\
    =\ & \sem{\Swap}_{s_X\cup s_Y\backslash s_X'}\circ[\<\sem{i}|\sem{j}\> (\sem{(X_1\otimes\cdots\otimes X_n) \otimes \mathbf{1}_{s_Y\backslash s_X'}}\circ \\
    &\sem{(Y_1\otimes\cdots\otimes Y_m) \otimes \mathbf{1}_{s_X'\backslash s_Y}})]\circ \sem{\Swap}_{s_Y'\cup s_X'\backslash s_Y}^\dagger\\
    =\ & \sem{\delta_{i,j}} \sem{\Swap}_{s_X\cup s_Y\backslash s_X'}\circ\sem{(X_1\otimes\cdots\otimes X_n) \otimes \mathbf{1}_{s_Y\backslash s_X'}}\circ \sem{\Swap}_{s_X'\cup s_Y}^\dagger \circ \\
    &\sem{\Swap}_{s_Y\cup s_X'} \circ \sem{(Y_1\otimes\cdots\otimes Y_m) \otimes \mathbf{1}_{s_X'\backslash s_Y}}\circ \sem{\Swap}_{s_Y'\cup s_X'\backslash s_Y}^\dagger\\
    =\ & \sem{\delta_{i,j}} \sem{\cl(X_1\otimes\cdots\otimes X_n, s_Y\backslash s_X')}\circ 
    \sem{\cl(Y_1\otimes\cdots\otimes Y_m, s_X'\backslash s_Y)} \\
    =\ & \sem{\delta_{i,j}. (X_1\otimes\cdots\otimes X_n)\cdot (Y_1\otimes\cdots\otimes Y_m)}
  \end{align*}
\end{itemize}

For the soundness of step (3) in normalization, notice that all tensors in the form of Eqn. (2) are ordered, so the denotational semantics of LHS and RHS of Eqn. (2) are exactly the semantics of LHS and RHS of Eqn. (3).

\subsubsection{Proof of normal form}
We first show that every labelled Dirac expression can be rewritten into Eqn. (1) by step 1 and 2. It is routine to check that every rewriting rule preserves the type (i.e., $\cD(s_1,s_2)$ of the expression) and we omit this.
\begin{itemize}
  \item $D \equiv |i\>_r$ or ${}_r\<i|$. It is already in form of Eqn. (1).
  %%%%%%%%%%%%%%%%%%%%%%%%%%%%%%%%%%%%%%%%%%%%%%%%%%%%%%%%%%%%%%%%%%%%%%%%%%%%%%
  \item $D \equiv K_R$ or $B_R$ or $O_{R,R'}$. Directly by apply (R-L-EXPAND), by noticing that $\<i_R|\cdot K\>$ and $B\cdot |i_R\>$ and $\<i_R|\cdot O \cdot |i_{R'}\>$ are all scalars in plain Dirac notation (recall $|i_R\>$ and $\<i_R|$ defined in Appendix B).
  %%%%%%%%%%%%%%%%%%%%%%%%%%%%%%%%%%%%%%%%%%%%%%%%%%%%%%%%%%%%%%%%%%%%%%%%%%%%%%
  \item $D \equiv D^{\prime\dagger}$. By induction hypothesis, $D'$ is in form of Eqn. (1), so first apply (R-ADJ2) we get:
  $$ 
  (\sum_{i}\cdots\sum_{j} a_1 . (\ket{i}_{p} \otimes \cdots \otimes {}_q\<j|))^\dagger
  + \cdots +
  (\sum_{k}\cdots\sum_{l} a_m . (\ket{k}_{r} \otimes \cdots \otimes {}_s\<l|))^\dagger$$
  and then apply (R-SUM-PUSH2) until the innermost summation, we get:
  $$ 
  \sum_{i}\cdots\sum_{j} (a_1 . (\ket{i}_{p} \otimes \cdots \otimes {}_q\<j|))^\dagger
  + \cdots +
  \sum_{k}\cdots\sum_{l} (a_m . (\ket{k}_{r} \otimes \cdots \otimes {}_s\<l|))^\dagger.$$
  Finally, we sequentially perform (R-ADJ1), (R-ADJD0), (R-ADJDK) and (R-ADJDB) on every innermost expression and obtain the form of Eqn. (1):
  $$ 
  \sum_{i}\cdots\sum_{j} a_1^* . ({}_p\<i| \otimes \cdots \otimes \ket{j}_q)
  + \cdots +
  \sum_{k}\cdots\sum_{l} a_m^* . ({}_r\<k| \otimes \cdots \otimes \ket{l}_s).$$
  %%%%%%%%%%%%%%%%%%%%%%%%%%%%%%%%%%%%%%%%%%%%%%%%%%%%%%%%%%%%%%%%%%%%%%%%%%%%%%
  \item $D \equiv a.D'$. By induction hypothesis, $D'$ is in form of Eqn. (1), so first apply (R-SCR2) we get:
  $$ 
  a.(\sum_{i}\cdots\sum_{j} a_1 . (\ket{i}_{p} \otimes \cdots \otimes {}_q\<j|))
  + \cdots +
  a.(\sum_{k}\cdots\sum_{l} a_m . (\ket{k}_{r} \otimes \cdots \otimes {}_s\<l|))$$
  and then apply (R-SUM-PUSH3) until the innermost summation, we get:
  $$ 
  \sum_{i}\cdots\sum_{j} a.(a_1 . (\ket{i}_{p} \otimes \cdots \otimes {}_q\<j|))
  + \cdots +
  \sum_{k}\cdots\sum_{l} a.(a_m . (\ket{k}_{r} \otimes \cdots \otimes {}_s\<l|))$$
  and finally apply (R-SCR1) for every innermost summation, we have:
  $$ 
  \sum_{i}\cdots\sum_{j} (a \times a_1) . (\ket{i}_{p} \otimes \cdots \otimes {}_q\<j|)
  + \cdots +
  \sum_{k}\cdots\sum_{l} (a \times a_m) . (\ket{k}_{r} \otimes \cdots \otimes {}_s\<l|)$$
  which is in form of Eqn. (1).
  %%%%%%%%%%%%%%%%%%%%%%%%%%%%%%%%%%%%%%%%%%%%%%%%%%%%%%%%%%%%%%%%%%%%%%%%%%%%%%
  \item $D \equiv D_1 + \cdots + D_n$. By induction hypothesis, every $D_i$ is in form of Eqn. (1), by applying (R-FLATTEN) on every subterm, we directly rewrite $D$ into form of Eqn. (1).
  \item $D \equiv D_1\otimes \cdots \otimes D_n$. By induction hypothesis, every $D_i$ is already rewritten in the form of Eqn. (1), we first apply (R-TSRD0) for every $D_i$, thus $D$ is now of the form of additions of subterms which are in the form of 
  $$
  \sum_{i}\cdots\sum_{j} a_1 . (\ket{i}_{p} \otimes \cdots \otimes {}_q\<j|)
  \otimes \cdots \otimes
  \sum_{k}\cdots\sum_{l} a_m . (\ket{k}_{r} \otimes \cdots \otimes {}_s\<l|).
  $$
  Next, we apply (R-SUM-PUSHD0) for every subterms and we get:
  $$
  \sum_{i}\cdots\sum_{j}\cdots\sum_{k}\cdots\sum_{l} (a_1 . (\ket{i}_{p} \otimes \cdots \otimes {}_q\<j|))
  \otimes \cdots \otimes
  (a_m . (\ket{k}_{r} \otimes \cdots \otimes {}_s\<l|)).
  $$
  Then we apply (R-SCRD0) on the innermost expressions and obtain
  $$
  \sum_{i}\cdots\sum_{j}\cdots\sum_{k}\cdots\sum_{l} (a_1.( \cdots (a_m . ((\ket{i}_{p} \otimes \cdots \otimes {}_q\<j|)
  \otimes \cdots \otimes
  (\ket{k}_{r} \otimes \cdots \otimes {}_s\<l|))) \cdots ))
  $$
  and finally (R-SCR1) on the innermost expressions to get
  $$
  \sum_{i}\cdots\sum_{j}\cdots\sum_{k}\cdots\sum_{l} (a_1\times \cdots \times a_m) . ((\ket{i}_{p} \otimes \cdots \otimes {}_q\<j|)
  \otimes \cdots \otimes
  (\ket{k}_{r} \otimes \cdots \otimes {}_s\<l|)).
  $$
  \item $D \equiv D_1 \cdot D_2$. By induction hypothesis, $D_1$ and $D_2$ are already rewritten in the form of Eqn. (1), we first apply (R-DOTD0) and then (R-DOTD1), which lead $D$ to the form of additions of subterms which are in the form of:
  $$
  (\sum_{i}\cdots\sum_{j} a_1 . (\ket{i}_{p} \otimes \cdots \otimes {}_q\<j|))
  \cdot  
  (\sum_{k}\cdots\sum_{l} a_2 . (\ket{k}_{r} \otimes \cdots \otimes {}_s\<l|)).
  $$
  Next, by applying (R-SUM-PUSHD1) and (R-SUM-PUSHD2) for every summation, we get the subterms as
  $$
  \sum_{i}\cdots\sum_{j}\sum_{k}\cdots\sum_{l} (a_1 . (\ket{i}_{p} \otimes \cdots \otimes {}_q\<j|))
  \cdot  
  (a_2 . (\ket{k}_{r} \otimes \cdots \otimes {}_s\<l|)).
  $$
  Then apply (R-SCRD1) and (R-SCRD2) on the innermost expression to obtain:
  $$
  \sum_{i}\cdots\sum_{j}\sum_{k}\cdots\sum_{l} (a_1 . (a_2. ((\ket{i}_{p} \otimes \cdots \otimes {}_q\<j|)
  \cdot (\ket{k}_{r} \otimes \cdots \otimes {}_s\<l|)))).
  $$
  Now, we gradually apply (R-L-SORT4)\footnote{the final rule in Table 21 which has a typo of naming, i.e., it should be (R-L-SORT4) instead of (R-L-SORT1)} (or (R-L-SORT1) or (R-L-SORT2) or (R-L-SORT3) depends on the form of left part / right part of the $\cdot$) to eliminate \textbf{all} bras (in the LHS of $\cdot$) and kets (in the RHS of $\cdot$) with same labels. This will always terminate as we only have the finite number of kets and bras. Then the innermost expression is of form:
  \begin{align*}
    &\sum_{i}\cdots\sum_{j}\sum_{k}\cdots\sum_{l} (a_1 . ( a_2. (\delta_{?,?}. (\cdots (\delta_{?,?}. ((\ket{i_1}_{p_1} \otimes \cdots \otimes \ket{i_n}_{p_n} \otimes \\
    &{}_{q_1}\<j_1| \otimes \cdots \otimes {}_{q_m}\<j_m|)
  \cdot (\ket{k_1}_{r_1} \otimes \cdots \otimes \ket{k_{n'}}_{r_{n'}} \otimes {}_{s_1}\<l_1| \otimes \cdots \otimes {}_{s_{m'}}\<l_{m'}|))))))).
  \end{align*}
  where $?$s are some indexes (but for simplicity, we do not explicitly give the names), and we also use the fact that labelled tensor is commutative and associative and then reordering the kets and bras by kets first.

  We can further apply (R-L-SORT0) to translate $\cdot$ to $\otimes$, which is guaranteed by: LHS of $\cdot$ has type $\cD(\{p_1,\cdots,p_n\},\{q_1,\cdots,q_m\})$ and RHS of $\cdot$ has type $\cD(\{r_1,\cdots,r_{n'}\}, \\ \{s_1,\cdots,s_{m'}\})$, and $\{q_1,\cdots,q_m\}\cap \{r_1,\cdots,r_{n'}\} = \emptyset$ (otherwise, suppose $q_x = r_y = t$, then ${}_{q_x}\<j_x|$ and $|k_y\>_{r_y}$ are pairs that can be eliminated by applying (R-L-SORT4) which contradicts to that all pairs have been eliminated).
  
  We finally apply (R-L-SCR1) for the innermost expression and get the form of Eqn. (1):
  \begin{align*}
    &\sum_{i}\cdots\sum_{j}\sum_{k}\cdots\sum_{l} (a_1 \times a_2 \times \delta_{?,?} \times \cdots \times \delta_{?,?}). ((\ket{i_1}_{p_1} \otimes \cdots \otimes \ket{i_n}_{p_n} \otimes \\
    &{}_{q_1}\<j_1| \otimes \cdots \otimes {}_{q_m}\<j_m|)
  \otimes (\ket{k_1}_{r_1} \otimes \cdots \otimes \ket{k_{n'}}_{r_{n'}} \otimes {}_{s_1}\<l_1| \otimes \cdots \otimes {}_{s_{m'}}\<l_{m'}|)).
  \end{align*}
  
  \item $D \equiv \sum_s f$. By induction hypothesis, for every $i\in s$, $f(i)$ is already rewritten in the form of Eqn. (1). By applying (R-SUM-ADD0) we directly translate $D$ to the form of Eqn. (1).
\end{itemize}

The step (3) is straightforward since: 1. reordering always succeeds and terminates, 2. since $D$ is well-typed and after applying rewriting rules it is still well-typed, every subterm of additions has the same type, this ensures that every subterm of the form Eqn. (2) has the same ordered labels.
Finally, notice that all tensors in Eqn. (2) are ordered, so the denotational semantics of LHS and RHS of Eqn. (2) are exactly the semantics of LHS and RHS of Eqn. (3).

\section{Efficient algorithm for proving equivalence}
\label{sec: decide}

Now we analyse the axioms in $E$ to understand the difficulty and solution for normalization. 
For $\alpha$-equivalence, we want to rule out the influence of bound variable names. Therefore we use de Bruijn notation~\cite{deBruijn1972lambda}, which replaces the name with the distance from the lambda abstraction to the variable. For instance, the nominal lambda abstraction \( \lambda x. x \) is transformed into \( \lambda . \$0 \), while \( \lambda x. \lambda y. (x\ (y\ x)) \) is transformed into \( \lambda.\lambda. (\$1\ (\$0\ \$1)) \).

The remaining axioms, such as AC-equivalence and SUM-SWAP, assert equivalence under permutations. A standard approach for proving such equivalences is to normalize terms by sorting in a predefined order. For example, given the dictionary order \( a < b < c \), the term \( b + c + a \) (and any other AC-equivalent term) is normalized into \( a + b + c \). However, in our setting, two intertwined difficulties arise: how to assign an order to all terms in the language, and how to simultaneously sort for both axioms.

Consider the following two equivalent terms:
\[
\sum_{i \in s_1} \sum_{j \in s_2} \bra{i} A \ket{j} \times \bra{j} B \ket{i}
= 
\sum_{i \in s_2} \sum_{j \in s_1} \bra{i} B \ket{j} \times \bra{j} A \ket{i}
\]
While these two terms are equivalent, directly sorting the elements of scalar multiplication using lexical order does not yield the same form.

To address this issue, we propose an algorithm to sort in two steps. The key observation is that in a successive sum expression \( \sum_{i \in s_1} \cdots \sum_{j \in s_n} A \), the names and order of the bound variables \( i, \dots, j \) can be freely permuted. Therefore, a good idea is to normalize AC-equivalence first, where all bound variables are treated uniformly. Afterwards, the order of summation can then be determined based on the position of the bound variables.

% \begin{align*}
%     \sum_{i \in s_1} \sum_{j \in s_2} \bra{i} A \ket{j} \times \bra{j} B \ket{i} 
%     \qquad
%     \sum_{i \in s_2} \sum_{j \in s_1} \bra{i} B \ket{j} \times \bra{j} A \ket{i} \\
%     \sum_{i \in s_1} \sum_{j \in s_2} \bra{i} A \ket{j} \times \bra{j} B \ket{i} 
%     \qquad 
%     \sum_{i \in s_2} \sum_{j \in s_1} \bra{j} A \ket{i} \times \bra{i} B \ket{j} \\
%     \sum_{i \in s_1} \sum_{j \in s_2} \bra{i} A \ket{j} \times \bra{j} B \ket{i} \\
%     \sum_{s_1} \sum_{s_2} \bra{\$1} A \ket{\$0} \times \bra{\$0} B \ket{\$1}
% \end{align*}

In the example above, we first ignore the bound variables and sort the sum body into \( \bra{\bullet} A \ket{\bullet} \times \bra{\bullet} B \ket{\bullet} \). Then, we swap the summations such that the bound variable at the first \( \bullet \) position appears at the outermost position. The results will have the same de Bruijn normal form, namely \( \sum_{s_1} \sum_{s_2} \bra{\$1} A \ket{\$0} \times \bra{\$0} B \ket{\$1} \).

To describe the algorithm in the following, we introduce two key notations. For a term \( e = f(a_1, a_2, \dots, a_n) \), \( \textrm{head}(e) \) denotes the function symbol \( f \), while \( \textrm{arg}(e, i) \) refers to the \( i \)-th argument \( a_i \) of the term. In this context, variables and constants are treated as functions with zero arguments.
\begin{definition}[Order Without Bound Variables]
Let \( \mathcal{B} \) represent the set of bound variables, with the assumption that all bound variables are unique. We also assume that a total order exists over all symbols. The relation \( e_1 =_\mathcal{B} e_2 \) holds if:
\begin{itemize}
    \item \( \textrm{head}(e_1) = \textrm{head}(e_2) \), and for all \( i \), \( \textrm{arg}(e_1, i) =_\mathcal{B} \textrm{arg}(e_2, i) \), or
    \item \( e_1 \in \mathcal{B} \) and \(e_2 \in \mathcal{B}\).
\end{itemize}

The relation \( e_1 <_\mathcal{B} e_2 \) holds between two terms if:
\begin{itemize}
    \item $e_1 \notin \mathcal{B}$ and $e_2 \in \mathcal{B}$, or
    \item $head(e_1) < head(e_2)$, or
    \item $head(e_1) = head(e_2)$, and there exists $n$ with $arg(e_1, n) <_\mathcal{B} arg(e_2, n)$, where $arg(e_1, i) =_\mathcal{B} arg(e_2, i)$ for all $i < n$.
\end{itemize}
\end{definition}
It can be shown that \( e_1 =_\mathcal{B} e_2 \) if and only if neither \( e_1 <_\mathcal{B} e_2 \) nor \( e_2 <_\mathcal{B} e_1 \) holds. The purpose of this ordering is to compare function symbols in a top-down manner while ignoring bound variables. This order enables normalization of terms for AC equivalence.
\begin{definition}[Sort Transformation]
    For a term $e$ with bound variable set $\mathcal{B}$,
    The sort transformation is defined in~\Cref{alg: sort}.
\end{definition}

\begin{algorithm}
    \caption{Sort Transformation}
    \label{alg: sort}
    \begin{algorithmic}[1]
        \Procedure{Sort}{$e, \mathcal{B}$}
            \If{$e \equiv \lambda x : T . e'$}
                \State \Return $\lambda x : T . \textsc{Sort}(e')$
            \ElsIf{$e \equiv \lambda x : \Index. e'$}
                \State \Return $\lambda x : \Index . \textsc{Sort}(e')$
            \ElsIf{$e \equiv f(a_1, \cdots, a_n)$}
                \State $ls := \textsc{Sort}(a_1), \cdots, \textsc{Sort}(a_n)$
                \State $ls := ls$ sorted by $<_\mathcal{B}$
                \State \Return $f(ls)$
            \EndIf
        \EndProcedure
    \end{algorithmic}
\end{algorithm}

After sorting, the next step is the \textit{swap transformation}, which arranges successive summations based on the order of bound variables.
\begin{definition}[Swap Transformation]
For a term \( e \) with a sorting result \( \textsc{Sort}(e) \), the swap transformation proceeds by ordering all bound variables according to their first appearances, except in function definitions \( \lambda x \). The swap transformation then reorders the successive summations accordingly.
\end{definition}

Take the term $\sum_{i \in s_2} \sum_{j \in s_1} \bra{i} B \ket{j} \times \bra{j} A \ket{i}$ as an example. Its bound variable set $\mathcal{B} = \{i, j\}$. Assume we have $A < B$, then the sorting result will be $\sum_{i \in s_2} \sum_{j \in s_1} \bra{j} A \ket{i} \times \bra{i} B \ket{j}$. Then we set the order for bound variables to be such that $j < i$ because $j$ appears first in the body. Using the swap transformation, the sorted result will be $\sum_{j \in s_1} \sum_{i \in s_2} \bra{j} A \ket{i} \times \bra{i} B \ket{j}$.

% The order depends on there occurances in the last sorting result. If no occurance, then the order will depend on the set (for sum) and the type (for lambda abstraction only).

%%%%%%%%%%%%%%%%%%%%%%%

% The idea is to assign an order to terms, which is independent on the bound variables. Because we can have terms with nested AC symbols.

% Lastly, we can prove that the equivalence established by this normalization procedure is sound with respect to the semantics.

% \begin{theorem}[Soundness]
%     For any well-formed context \( \Gamma \) and well-typed expressions \( e_1 \) and \( e_2 \), if  $e_1$ and $e_2$ have the same normal form, then \( \sem{e_1} = \sem{e_2} \).
% \end{theorem}

% \begin{proof}
%     The soundness of the term rewriting procedure follows from the fact that each rewriting rule preserves equivalence. Furthermore, the operations in the sort and swap transformations respect the AC-equivalence and SUM-SWAP axioms. Finally, the de Bruijn normalization ensures soundness for \( \alpha \)-equivalence.
% \end{proof}

\section{Examples for labelled Dirac notation}
\newcommand{\tr}{\mathrm{tr}}
\label{sec: examples for labelled}
\begin{itemize}
    \item (LDN-1) \( \ket{s}_Q \otimes \ket{t}_R = \ket{(s, t)}_{(Q, R)} \)
    \item (LDN-2) \( {O_1}_{Q} \cdot {O_2}_{(Q,R)} = ((O_1 \otimes \mathbf{1}_\mathcal{O}) \cdot O_2)_{(Q,R)} \)
    \item (LDN-3) \( M_{r_1} \sum_{i} \ket{(i, i)}_{(r_1, r_2)} = M^T_{r_2} \sum_{i}\ket{(i, i)}_{(r_1, r_2)} \)
    \item (LDN-4) \( \bra{\Phi}_{(x, y)} M_y \cdot N_x \ket{\Phi}_{(x, y)} = \mathrm{tr}(M^T N) \), where $\ket{\Phi} = \sum_{i} \ket{(i, i)}$
    \item (LDN-5) \( \sum_{i} {}_x\bra{i} O_x \ket{i}_x = \mathrm{tr}(O) \)
    \item (LDN-6) \( \sum_{j} {}_y\bra{j} (\sum_{i} {}_x\bra{i} O_{(x,y)} \ket{j}_x) \ket{j}_y = \mathrm{tr}(O) \)
    \item (LDN-7) \( \sum_{i} {}_{(x, y)}\bra{i} O_{(x,y)} \ket{i}_{(x, y)} = \mathrm{tr}(O) \)
    \item (LDN-8) \( \mathrm{tr}_x(\mathrm{tr}_y(O_{((y,z),x)})) = \mathrm{tr}_y(\mathrm{tr}_x(O_{((y,z),x)})) \)
    \item (LDN-9) \( \mathrm{tr}_x(\mathrm{tr}_y(O_{((y,z),x)})) = \mathrm{tr}_{(x, y)}(O_{((y,z),x)}) \)
    \item (LDN-10) \begin{align*}
        & \tr_{((a',(b,b')),c')}\Big[\tr_{r}\Big(U_{(r,(a,b))} \cdot\Big(|s\>_r\<s|\otimes 
        \Big[V_{((a',(b,b')),c')}\cdot 
        \big(|\phi\>_{(a,a')}\<\phi|\otimes \\
        &\qquad |\psi\>_{((b,b'),(c,c'))}\<\psi|\big)
        \cdot V^\dagger_{((a',(b,b')),c')}\Big]
        \Big) \cdot U^\dagger_{(r,(a,b))} \Big)\Big] \\
        =\ &\tr_{(((r,a'),(b,b')),c')}\Big[\big(U_{(r,(a,b))}\cdot V_{((a',(b,b')),c')}\cdot(|s\>_r\otimes |\phi\>_{(a,a')}\otimes |\psi\>_{((b,b'),(c,c'))})\big)\cdot \\
        &\qquad \big(U_{(r,(a,b))}\cdot V_{((a',(b,b')),c')}\cdot(|s\>_r\otimes |\phi\>_{(a,a')}\otimes |\psi\>_{((b,b'),(c,c'))})\big)^\dagger\Big]
    \end{align*}
    \item (LDN-11)    
        \begin{align*}
            \mbox{set}\qquad &U \triangleq \sum_i|i\>\<i|\otimes  P_i \qquad V \triangleq \sum_i|i\>\<i|\otimes Q_i \\
            \mbox{show}\qquad &U_{(a,b)}\cdot W_{(b,c)}\cdot V_{(a,c)} = 
            \sum_i |i\>_a\<i|\otimes \big((P_i)_c\cdot W_{(b,c)}\cdot (Q_i)_c\big)
        \end{align*}
    \item (LDN-12) \( \ket{i}_{a;b}\bra{j} \cdot C_{(b,c)} \cdot D_{(c,d)} = {}_b\bra{j} \cdot C_{(b,c)} \cdot D_{(c,d)} \cdot \ket{i}_{a}\)
    \item (LDN-13) \( (A_{(a,b)} \otimes B_{(c,d)} \otimes C_{(e,f)}) \cdot (D_{(b,c)} \otimes E_{(d,e)}) \cdot (F_{(a,b)} \otimes G_{(c,d)} \otimes H_{(e,f)}) = (A_{(a,b)} \otimes C_{(e,f)}) \cdot (B_{(c,d)} \cdot D_{(b,c)}) \cdot (E_{(d,e)} \cdot G_{(c,d)}) \cdot (F_{(a,b)} \otimes H_{(e,f)})\)
    \item (LDN-14) \( \textsf{CNOT}_{rq}\ket{\textrm{GHZ}}_{pqr} = (\textsf{CNOT}\ket{00})_{rq}\ket{0}_p + (\textsf{CNOT}\ket{11})_{rq}\ket{1}_p \)
    \item (LDN-15) \( \textsf{CNOT}_{pq}\ket{\textrm{GHZ}}_{pqr} = (\textsf{CNOT}\ket{00})_{pq}\ket{0}_p + (\textsf{CNOT}\ket{11})_{pq}\ket{1}_p \)
    \item (LDN-16) \begin{align*}
        \mbox{set}\qquad &\ket{\textrm{GHZ}}\triangleq\sum_i \ket{iii}\bra{iii} \quad M\triangleq\sum_{ij}\ket{ij}\bra{ij}\otimes U_{ij} \quad N \triangleq \sum_i \ket{i}\bra{i} \otimes U_{ii}\\
        \mbox{show}\qquad &M_{prq} \ket{\textrm{GHZ}}_{prq} = N_{rq} \ket{\textrm{GHZ}}_{pqr}
    \end{align*}
    \item (LDN-17) \begin{align*}
        \mbox{set}\qquad &\ket{\textrm{GHZ}}\triangleq\sum_i \ket{iii}\bra{iii} \quad M\triangleq\sum_{ij}\ket{ij}\bra{ij}\otimes U_{ij} \quad N \triangleq \sum_i \ket{i}\bra{i} \otimes U_{ii}\\
        \mbox{show}\qquad &N_{rq} \ket{\textrm{GHZ}}_{pqr} = N_{pq}\ket{\textrm{GHZ}}_{pqr}
    \end{align*}
    \item (LDN-18) \( -\ket{0}_q\ket{+-}_{q_1, q_2} = X_{q_2} \ket{0}_q\ket{+-}_{q_1, q_2} \)
    % \item \( \ket{1}_q \ket{--}_{(q1, q2)} = \textsf{CNOT}_{(q1, q2)} \ket{0}_q \ket{+-}_{(q1, q2)} \)
\end{itemize}